\documentclass[11pt,aps,preprint,prd,showpacs,superscriptaddress,amssymb,amsmath,floatfix,nofootinbib]{revtex4-2}

\usepackage{graphicx}
\usepackage{subfigure}
\usepackage{hyperref}
\usepackage{multirow}
\usepackage[dvipsnames]{xcolor}
\allowdisplaybreaks[1]
\usepackage[utf8]{inputenc}
\usepackage{color}
\usepackage{tikz}
\begin{document}

\def\simge{
    \mathrel{\rlap{\raise 0.511ex
        \hbox{$>$}}{\lower 0.511ex \hbox{$\sim$}}}}
\def\simle{
    \mathrel{\rlap{\raise 0.511ex 
        \hbox{$<$}}{\lower 0.511ex \hbox{$\sim$}}}}
\let\ifcomments\iftrue
\def\commentsoff{\global\let\ifcomments\iffalse}

\let\commentsize\small
\def\tinycomments{\global\let\commentsize\footnotesize}

\def\ask#1{%
\ifcomments%
$\langle\!\langle${\commentsize #1}$\rangle\!\rangle$%
\else%
\fi%
}


\newcommand{\1}{\mbox{1}\hspace{-0.25em}\mbox{l}}
\newcommand{\del}{\partial}
\newcommand{\td}{{\rm{d}}}
\newcommand{\e}{{\rm{e}}}
\newcommand{\img}{{\rm{i}}}
\newcommand{\Tr}{{\rm{Tr}}}
\definecolor{ao(english)}{rgb}{0.0, 0.5, 0.0}
\def\as#1{{\color{blue}{#1}}}
\def\slashchar#1{\setbox0=\hbox{$#1$} 
\dimen0=\wd0 
\setbox1=\hbox{/} \dimen1=\wd1 
\ifdim\dimen0>\dimen1 
\rlap{\hbox to \dimen0{\hfil/\hfil}} 
#1 
\else 
\rlap{\hbox to \dimen1{\hfil$#1$\hfil}} 
/ 
\fi}

\newcommand*\circled[1]{\tikz[baseline=(char.base)]{
            \node[shape=circle,draw,inner sep=0.3pt] (char) {#1};}}
\newcommand*\circledsingle[1]{\tikz[baseline=(char.base)]{
            \node[shape=circle,draw,inner sep=1.8pt] (char) {#1};}}

\title{
$\Delta I = 3/2$ and $\Delta I = 1/2$ channels of $K\to\pi\pi$ decay at the physical point with periodic boundary conditions
}

\newcommand{\Columbia}{
Physics Department,
Columbia University,
New York, NY 10027, USA
}
\newcommand{\UConn}{
Physics Department,
University of Connecticut,
Storrs, CT 06269, USA
}
\newcommand{\BNL}{
Physics Department,
Brookhaven National Laboratory,
Upton, NY 11973, USA
}
\newcommand{\RBRC}{
RIKEN-BNL Research Center,
Brookhaven National Laboratory,
Upton, NY 11973, USA
}
\newcommand{\CSIBNL}{
Computational Science Initiative,
Brookhaven National Laboratory,
Upton, NY 11973, USA
}
\newcommand{\Bern}{
Albert Einstein Center,
Institute for Theoretical Physics,
University of Bern,
CH-3012 Bern, Switzerland
}
\newcommand{\Regensburg}{
Fakult\"at f\"ur Physik, Universit\"at Regensburg,
Universit\"atsstra{\ss}e 31,
93040 Regensburg, Germany
}
\newcommand{\UCB}{
Department of Physics,
University of California,
Berkeley, CA 94720, USA
}
\newcommand{\LBNL}{
Nuclear Science Division,
Lawrence Berkeley National Laboratory,
Berkeley, CA 94720, USA
}
\newcommand{\Milano}{
Dipartimento di Fisica,
Universit\'a di Milano-Bicocca,
Piazza della Scienza 3,
I-20126 Milano, Italy
}
\newcommand{\INFN}{
INFN, Sezione di Milano-Bicocca,
Piazza della Scienza 3,
I-20126 Milano, Italy
}

\author{Thomas~Blum}
\affiliation{\UConn}
\affiliation{\RBRC}

\author{Peter~A.~Boyle}
\affiliation{\BNL}


\author{Daniel~Hoying}
\affiliation{\Bern}

\author{Taku~Izubuchi}
\affiliation{\RBRC}
\affiliation{\BNL}

\author{Luchang~Jin}
\affiliation{\UConn}
\affiliation{\RBRC}

\author{Chulwoo~Jung}
\affiliation{\BNL}

\author{Christopher~Kelly}
\affiliation{\CSIBNL}

\author{Christoph~Lehner}
\affiliation{\Regensburg}


\author{Amarjit~Soni}
\affiliation{\BNL}

\author{Masaaki~Tomii}
\affiliation{\UConn}


\collaboration{RBC \& UKQCD Collaborations}

\begin{abstract}
We present a lattice calculation of the $K\to\pi\pi$ matrix elements and amplitudes with both the $\Delta I = 3/2$ and 1/2 channels and $\varepsilon'$, the measure of direct $CP$ violation.  
We use periodic boundary conditions (PBC), where the correct kinematics of $K\to\pi\pi$ can be achieved via an excited two-pion final state.  To overcome the difficulty associated with the extraction of excited states, our previous work~\cite{Bai:2015nea,RBC:2020kdj} successfully employed G-parity boundary conditions, where pions are forced to have non-zero momentum enabling the $I=0$ two-pion ground state to express the on-shell kinematics of the $K\to\pi\pi$ decay.
Here instead we overcome the problem using the variational method which allows us to resolve the two-pion spectrum and matrix elements up to the relevant energy where the decay amplitude is on-shell. 

In this paper we report an exploratory calculation of $K\to\pi\pi$ decay amplitudes and $\varepsilon'$ using PBC on a coarser lattice size of $24^3\times64$ with inverse lattice spacing $a^{-1}=1.023$~GeV and the physical pion and kaon masses.  The results are promising enough to motivate us to continue our measurements on finer lattice ensembles in order to improve the precision in the near future. 
\end{abstract}

\maketitle

\clearpage

\section{Introduction}
\label{sec:intro}

$K\to\pi\pi$ decay is an ideal probe to search for new physics beyond the Standard Model (SM) as it contains charge-parity ($CP$)-violating decay processes. 
The amount of $CP$ violation in the SM is believed to be too small to explain the dominance of matter over antimatter in the current universe.  
Particularly $\varepsilon'$, the measure of direct $CP$ violation in $K\to\pi\pi$ decay, is very sensitive to potential new sources of $CP$ violation.
The experimental measurements were performed by the NA48~\cite{NA48:2002tmj} and KTeV~\cite{Worcester:2009qt,KTeV:2010sng} collaborations and their average is quoted as Re$(\varepsilon'/\varepsilon)_{\rm exp}=16.6(2.3)\times10^{-4}$~\cite{ParticleDataGroup:2022pth}, where $\varepsilon$ is the measure of indirect $CP$ violation $|\varepsilon| = 2.228(11)\times10^{-3}$.

Since the decay processes receive considerable nonperturbative QCD effects, lattice QCD should play a key role in giving a SM prediction of $\varepsilon'/\varepsilon$ and, in fact, the RBC and UKQCD collaborations have achieved the first {\it ab initio} SM prediction of this quantity~\cite{Bai:2015nea,RBC:2020kdj}.  
Our most recent result~\cite{RBC:2020kdj} is
Re$(\varepsilon'/\varepsilon)_{2020}=21.7(2.6)(6.2)(5.0)\times10^{-4}$~\cite{RBC:2020kdj}.
The first two errors are statistical and systematic, respectively, but excluding the systematic error due to electromagnetic and isospin-breaking corrections, which is listed as the third error.
While consistent with the experimental result, the relatively large uncertainty motivates further refinement of the calculation especially since the current error in the lattice calculation is a lot bigger than  in the two decades old experimental determination.

In the isospin limit $\varepsilon'/\varepsilon$ is determined by
\begin{equation}
\frac{\varepsilon'}{\varepsilon}
= \frac{\img\omega\e^{\img(\delta_2-\delta_0)}}{\sqrt2\varepsilon}
\left[
\frac{\mbox{Im}(A_2)}{\mbox{Re}(A_2)}-\frac{\mbox{Im}(A_0)}{\mbox{Re}(A_0)}
\right],
\label{eq:epsilonPrime}
\end{equation}
where $\delta_I$ is the $S$-wave two-pion scattering phase shift with isospin $I$ at the energy of the kaon mass, and $A_I = \langle(\pi\pi)_I|H_W|K^0\rangle$ is the $K\to\pi\pi$ decay amplitude defined with the weak Hamiltonian $H_W$, kaon initial state $|K^0\rangle$ and two-pion final state $\langle(\pi\pi)_I|$ with the definite isospin $I$.  The normalization of these states is clarified and the relations between the $K\to\pi\pi$ amplitudes and decay rates are explicitly given in Appendix~\ref{sec:normalization}.
In Eq.~\eqref{eq:epsilonPrime} we also define $\omega={\rm Re}(A_2)/{\rm Re}(A_0)$.
We employ the three-flavor effective weak Hamiltonian with the form~\cite{Buras:1993dy,Buchalla:1995vs}
\begin{equation}
H_W = \frac{G_F}{\sqrt2}V_{us}^*V_{ud}\sum_i [z_i(\mu)+\tau y_i(\mu)]Q_i(\mu),
\label{eq:Hw}
\end{equation}
where we define the Fermi constant $G_F$, Cabibbo-Kobayashi-Maskawa (CKM) matrix elements $V_{q'q}$ connecting up-type ($q'$) and down-type ($q$) quarks, their ratio $\tau = -V_{ts}^*V_{td}/V_{us}^*V_{ud}$, the Wilson coefficients $z_i(\mu)$ and $y_i(\mu)$ that encompass the effects of heavy particle fields that are integrated out, and the effective four-quark operators $Q_i(\mu)$ renormalized in the same scheme and scale, $\mu$, as the Wilson coefficients. 
The effective operators $Q_i$ relevant for $K\to\pi\pi$ have the strangeness-changing nature, $\Delta S=1$, and the dominant contributions are associated with the four-quark electroweak operators given in Eqs.~\eqref{eq:Q1}--\eqref{eq:Q10}.
The matrix elements $\langle(\pi\pi)_I|Q_i|K^0\rangle$ depend on nonperturbative QCD properties in the low-energy regime. 
Lattice QCD is the only known approach to computing non-perturbative physics that is systematically-improvable, $i.e.$ for which all of the systematic errors can be quantified and improved with sufficient computational effort.

One long-standing obstacle in this subject, originally described in Ref.~\cite{Maiani:1990ca}, is that it is not straightforward to extract the unique energy-conserving on-shell matrix elements from Euclidean correlation functions at large time separations.  In fact, in a periodic box and in the rest frame, the ground two-pion state in the rest frame comprises two pions at rest and has an energy near twice the pion mass, $2m_\pi$, which is not equal to the initial kaon state energy $m_K$.  While using periodic boundary conditions (PBC) and extracting signals of the ground states in the rest frame is the most common approach to lattice calculations, the $K\to\pi\pi$ matrix elements obtained in this way do not correspond to the energy-conserving on-shell matrix elements.
In early attempts to compute $K\to\pi\pi$ on the lattice, chiral perturbation theory (ChPT) was utilized to relate the on-shell $K\to\pi\pi$ matrix elements to some other matrix elements with unphysical kinematics that were accessible to numerical calculations at that time~\cite{Bernard:1985wf,JLQCD:1997kmy,CP-PACS:2001jkd,RBC:2001pmy,Lin:2002nq,Laiho:2002jq,Laiho:2003uy}. 

More recently, lattice calculations of $K\to\pi\pi$ matrix elements with (nearly) on-shell kinematics have employed a different approach.
The $\Delta I = 3/2$ process with the $I=2$ two-pion final state was computed with physical pion and kaon masses at a finite lattice spacing~\cite{Blum:2011ng,Blum:2012uk} and in the continuum limit~\cite{Blum:2015ywa}.  In these works, the matrix elements of $K^+\to\pi^+\pi^+$, which are related to $A_2$ by the Wigner-Eckart theorem and isospin symmetry, were computed by imposing anti-periodic boundary conditions (APBC) for spatial directions on the down quark field.
This results in the $\pi^+$ states also satisfying APBC in those directions, and therefore having momenta discretized in odd-integer multiples of $\pi/L$, where $L$ is the lattice size.
As a result, by tuning $L$, the energy of the $\pi^+\pi^+$ ground state matching the kaon mass $m_K$ was realized. 

Calculation of the $\Delta I = 1/2$ process with the $I=0$ two-pion final state is much more complicated due to the presence of many more diagrams, including noisy, disconnected contributions, possible mixing of the quark bilinear operators with the effective 
four-quark 
operators causing a power divergence, and so on.  In addition, the APBC procedure is not applicable for this isospin channel because it only makes $\pi^+$ and $\pi^-$ anti-periodic but $\pi^0$, which is also essential for the $\Delta I=1/2$ channel, is still periodic.  Importantly, it also breaks isospin. For $\Delta I=3/2$, this is circumvented because $\pi^+\pi^+$ cannot mix with other two-pion states due to charge conservation. For $\Delta I=1/2$ there is no way to avoid this issue.  
For our previous calculations of the I=0 decay, we resolved this difficulty through the use of G-parity boundary conditions (GPBC)~\cite{Wiese:1991ku,Christ:2019sah}, which employ a combined charge-conjugation and isospin rotation to induce APBC on both charged and neutral pion states, while preserving the isospin symmetry.
With this approach we were successfully able to calculate the $\Delta I = 1/2$ process and $\varepsilon'$.

Besides the works by the RBC and UKQCD collaborations summarized above, a moving frame was employed to realize the physical kinematics with the ground two-pion final state by Ishizuka {\it et al.}~\cite{Ishizuka:2018qbn}.
They reported the $\Delta I = 1/2$ and $3/2$ channels of $K\to\pi\pi$ amplitudes near on-shell and Re($\varepsilon'/\varepsilon$) computed with Wilson fermions at unphysical pion and kaon masses.
They utilized $CPS$ symmetry~\cite{Bernard:1987pr,Donini:1999sf,Ishizuka:2015oja}, which is the symmetry under $CP$ transformation followed by an interchange of the strange quark ($S$) with the down quark, to ensure the absence of four-quark operator mixing with wrong chirality even with the Wilson fermion action.

In this work, we aim to avoid the complexities of manipulating the boundary conditions and of moving frames, by employing PBC on an appropriately sized lattice such that the on-shell decay can be obtained via an excited two-pion state with energy $E_{\pi\pi}\approx m_K$.
We extract signals from an excited two-pion state with the energy near the kaon mass, $E_{\pi\pi}\approx m_K$ using the generalized eigenvalue problem (GEVP) method~\cite{Luscher:1990ck,Blossier:2009kd,Bulava:2011yz}, which provides a convenient method for isolating the contributions of individual low-lying states to Euclidean correlation functions.
The GEVP method has been used for several calculations of matrix elements, for example for nucleon structure~\cite{Owen:2012ts,Yoon:2016dij,Dragos:2016rtx,Karpie:2021pap,HadStruc:2021wmh,HadStruc:2022yaw,Barca:2022uhi}, $B$ physics~\cite{Bernardoni:2014kla,Blossier:2016vgh,Bahr:2019eom,Gerardin:2021jch}, light-meson radiative transitions~\cite{Shultz:2015pfa,Owen:2015fra} and form factors~\cite{Owen:2015gva}.
We can utilize this method not only for removing excited-state contamination from the ground-state signal but also for extracting signals from low-lying excited states.
The latter is an important goal of this work and we have reported our successful extractions of a few excited two-pion states in the companion two-pion scattering paper~\cite{RBC:2023xqv}.
The method also enables us to extract these signals at relatively short Euclidean times, where the signals of excited states could be still resolved, with the truncated excited-state contamination under control.

Our previous GPBC work~\cite{RBC:2020kdj} also employed multiple two-pion interpolation operators to remove the excited-state contamination using multi-state fits.  
In our first GPBC calculation of $K\to\pi\pi$~\cite{Bai:2015nea}, we only used a single $\pi\pi$-like operator, a product of two single-pion operators projected to $I=0$.
In Ref.~\cite{RBC:2020kdj} we additionally introduced an iso-singlet scalar bilinear operator, which we call a $\sigma$-like operator, and another $\pi\pi$-like operator with the same total momentum but different constituent pion momenta.  We then observed a significant change in the two-pion phase shift, $K\to\pi\pi$ matrix elements and $\varepsilon'/\varepsilon$
relative to our earlier calculation~\cite{Bai:2015nea}, which we attribute to excited-state contamination that was not formerly resolvable from the rapidly-growing statistical noise when measured with a single operator and lower statistics.  The $\sigma$-like operator in particular was shown to play a significant role in removing the excited-state contamination.
For the three-operator basis employed in Ref.~\cite{RBC:2021acc}, this fit-based approach was found to be equivalent in its resolution as the GEVP, but, unlike GEVP, also offered the flexibility to use a different number of states than operators to describe the two-pion correlation function in the region in which the excited-state contamination cannot be resolved from the noise. This ultimately proved important to obtain a result with minimal excited-state contamination. In our companion paper~\cite{RBC:2023xqv} we describe a "re-basing" strategy that also allows the GEVP approach to consider a lower number of state than operators. We believe that the GEVP will offer improved stability over a fit-based approach for larger numbers of operators and states, for which the covariance matrix may become ill conditioned.

While computing $\varepsilon'/\varepsilon$ with a different setup is itself interesting as there have been very few lattice results despite its phenomenological importance, we expect some further benefits of using PBC for $K\to\pi\pi$ calculation.
To discuss this we here remark on two major systematic errors on $\varepsilon'$ estimated in our previous GPBC work~\cite{RBC:2020kdj}: the finite lattice spacing error, and electromagnetic and isospin-breaking corrections.
The first resulted from computing $\varepsilon'$ on a single, rather coarse lattice spacing of 1.38 GeV. Work is underway to repeat the GPBC calculation on two finer ensembles in order to take the continuum limit and remove this error~\cite{Blum:2022wsz,Kelly:2022bpr}.  The necessity of generating ensembles for the specific purpose of $K\to\pi\pi$ calculation is requires an extra computational cost.
With PBC, on the other hand, we have already generated finer ensembles~\cite{RBC:2014ntl,Jung:2022sjp} with domain wall fermions at physical masses up to the inverse lattice spacing of 2.69~GeV, which have been used for other various projects.  
We can use these ensembles for $K\to\pi\pi$ calculation once the approach in the present work is found to be feasible.
One potential obstacle to using these ensembles is that, since these ensembles were generated without tuning the volume for $K\to\pi\pi$ calculation, physical kinematics cannot be precisely achieved with a two-pion state allowed in the given volume.

Even though electromagnetic and isospin-breaking corrections are typically of order $O(1\%)$, they are significant for $\varepsilon'/\varepsilon$ because of the $\Delta I =1/2$ rule.  The $\Delta I =1/2$ process is enhanced relative to $\Delta I =3/2$ (${\rm Re}(A_0)/{\rm Re}(A_2)\approx22.5$)
because Re($A_2$) is suppressed by a factor of 10 due to non-perturbative dynamics~\cite{RBC:2020kdj} and a further factor of two from higher-energy kinematics.
This results in a strong suppression of $\varepsilon'$ through the coefficient $\omega=\mbox{Re}(A_2)/\mbox{Re}(A_0)$ in Eq.~\eqref{eq:epsilonPrime} and a corresponding $20\times$ enhancement in the relative size of electromagnetic and isospin-breaking effects.
These effects were estimated to be $O(20\%)$ based on ChPT and the large-$N_c$ expansion of QCD~\cite{Cirigliano:2019cpi}, and while a direct prediction from Lattice QCD would certainly be useful, it is still inaccessible due to lack of a full formalism.
This requires an extension of the formalism given by L\"uscher~\cite{Luscher:1990ux} and Lellouch-L\"uscher~\cite{Lellouch:2000pv} that deals with two-hadron systems in finite volumes.
While there are related studies on-going~\cite{Christ:2017pze,Cai:2018why,Christ:2021guf}, we expect PBC is more suitable than GPBC to accomplish it because GPBC mixes the up and down quarks violating charge conservation.

In this paper we present our first PBC results for the $\Delta I = 3/2$ and $\Delta I = 1/2$ channels of $K\to\pi\pi$ amplitudes and Re($\varepsilon'/\varepsilon$) on a $24^3\times64$ ensemble with $2+1$-flavors of domain wall fermions with physical pion and kaon masses and an inverse lattice spacing of $a^{-1}=1.023(2)$~GeV.  We employ the all-to-all (A2A) propagator method~\cite{Foley:2005ac} with 2,000 low modes for the light quark and spin-color-time diluted random noise for both the light and strange quarks.  The all-mode averaging (AMA) technique~\cite{Blum:2012uh,Shintani:2014vja} is also employed to accelerate the sampling.  With the A2A procedure, AMA is implemented by reducing the number of configurations for exact calculations rather than the number of propagator sources~\cite{RBC:2023xqv}. 
A preparatory study of two-pion scattering with the same setup and GEVP was recently reported~\cite{RBC:2023xqv}. 
For two-pion interpolation operators we employ four 
$\pi\pi$-like operators, products of two single pion operators, with different relative momenta for both the $I=2$ and $I=0$ channels and additionally one $\sigma$-like scalar quark-bilinear operator for the $I=0$ channel. 
We employ a single kaon operator expecting the contamination from kaon excited states is less significant.
We apply the Lellouch-L\"uscher method~\cite{Lellouch:2000pv} to relate the two-pion state in the finite box with that in infinite volume.
We use the RI-SMOM renormalization procedure~\cite{Aoki:2007xm} with step scaling~\cite{Arthur:2010ht} up to the renormalization scale of 4~GeV.
In general these details are similar to the previous GPBC calculation~\cite{RBC:2020kdj}.  
We employ the same M\"obius domain wall formalism and Iwasaki+DSDR gauge action as the GPBC calculation of the $I=0$ matrix element, and our physical volume is the same to within a percent.
Besides the lattice spacing, the choice of boundary condition and two-pion operators, and the inclusion of the $\Delta I=3/2$ matrix element, the calculation is largely the same as the GPBC measurement, differing only in minor details such as the number of low-eigenmodes, the modifications to the A2A approach required in the G-parity case to treat the explicit flavor structure, and the use of cost-reduction techniques such as AMA and zM\"obius ($c.f.$, below).
However, the differing finite-volume effects resulting from the change in boundary conditions result in an $I=0$ two-pion energy that does not as closely match the kaon energy, requiring an interpolation to on-shell kinematics.
We use the lattice results for the matrix elements with the ground and first-excited two-pion final states for the interpolation and estimate the corresponding systematic error.

\begin{table}[tbp]
\centering
\begin{tabular}{lcc}
\hline
Quantity & This work & Experiment\\
\hline
\hline
{\rm Re}$(A_2)$ & $1.74(15)(48)\times10^{-8}~{\rm GeV}$ & $1.479(4)\times10^{-8}$~GeV\\
{\rm Im}$(A_2)$ & $-5.91(13)(1.75)\times10^{-13}~{\rm GeV}$ & \ldots\\
{\rm Re}$(A_0)$ & $2.84(57)(87)\times10^{-7}~{\rm GeV}$ & $3.3201(18)\times10^{-7}$~GeV\\
{\rm Im}$(A_0)$ & $-8.7(1.2)(2.6)\times10^{-11}~{\rm GeV}$ & \ldots\\
Re($A_0$)/Re($A_2$) & $16.3(3.7)(6.7)$ & 22.45(6)\\
$\omega=\mbox{Re}(A_2)/\mbox{Re}(A_0)$ & 0.061(14)(25) & 0.04454(12)\\
Re($\varepsilon'/\varepsilon$) & $29.4(5.2)(11.1)(5.0)\times10^{-4}$ & $16.6(2.3)\times10^{-4}$\\
\hline
\end{tabular}
\caption{A summary of the primary results of this work shown in the middle column. The values in parentheses give the statistical and systematic errors, respectively. For the last entry the systematic error associated with electromagnetic and isospin breaking effects is listed separately as the third error, which we inherit from the estimation in Ref.~\cite{RBC:2020kdj} based on the large-$N_c$ expansion of QCD and ChPT~\cite{Cirigliano:2019cpi}.  The corresponding experimental values are shown in the right column if applicable.}
\label{tab:primalresult}
\end{table}

For the convenience of the reader we summarize the primary results of this work in Table~\ref{tab:primalresult}.  We estimate various systematic errors.  The systematic error due to electromagnetic and isospin-breaking corrections is listed for Re($\varepsilon'/\varepsilon$) as the third error.  We inherit most of the systematic errors from Ref.~\cite{RBC:2020kdj} with a few exceptions, which need a new estimation for the setup in the present work and are discussed in Section~\ref{sec:AI}.  Table~\ref{tab:primalresult} also shows the corresponding experimental values for comparison, except Im$(A_I)$, which are not accessible from experiments.

The paper is organized as follows.
Section~\ref{sec:ensemble} briefly explains the lattice ensemble of gauge fields used in this study and measurement details.
Section~\ref{sec:2pt} gives results for two-point functions including the kaon mass, a brief summary of the two-pion spectrum study in Ref.~\cite{RBC:2023xqv}, and the corresponding Lellouch-L\"uscher factors.
In Section~\ref{sec:3pt} we present the $K\to\pi\pi$ three-point functions and the results for the matrix elements obtained by combining the three-point correlation functions with the results given in Section~\ref{sec:2pt}.
Section~\ref{sec:NPR} is devoted to the operator renormalization procedure and results.
In Section~\ref{sec:AandEpslon'} we present the remainder of analysis to obtain the $K\to\pi\pi$ amplitudes and Re($\varepsilon'/\varepsilon$), including the interpolation of the renormalized matrix elements to physical kinematics.
Section~\ref{sec:conclusion} concludes the present work and discusses future prospects.

\section{Lattice ensemble and overall measurement procedure}
\label{sec:ensemble}

In this paper we present our lattice calculation carried out on a $24^3\times 64$ lattice ensemble.
We employ Iwasaki $+$ DSDR (dislocation suppressing determinant ratio) gauge action~\cite{RBC:2012cbl} and $2+1$-flavor M\"obius domain wall fermions~\cite{Brower:2004xi,Brower:2012vk} with the extent for the fifth direction $L_s = 24$, the M\"obius scale $b+c = 4$, $b-c=1$ and the domain-wall height $M_5 = 1.8$.
We choose $\beta = 1.633$, which corresponds to the inverse lattice spacing $a^{-1} = 1.023(2)$~GeV and hence spatial extent $L = 4.639(9)$~fm and time extent $L_T=12.34(2)$~fm.
We tune the input light quark mass to $am_l = 0.00107$ and the strange quark mass to $am_s = 0.0850$ so that the pion and kaon masses have (nearly) their physical values.
The pion mass is $am_\pi = 0.13944(17)$ and the kaon mass is, as quoted in the following subsection, $am_K = 0.50189(36)$.  See Ref.~\cite{Tu:2020vpn} for more details on the ensemble.

We use the all-to-all (A2A) quark propagator method \cite{Foley:2005ac}, which combines exact low-mode solutions with a stochastic approximation to the high-mode contribution, with low-mode deflation for acceleration of the conjugate gradient (CG) inversions.
We calculate 2,000 low modes for the light quarks via the local coherent Lanczos algorithm~\cite{Clark:2017wom} with the zM\"obius action~\cite{Mcglynn:2015uwh,Abramczyk:2017oxr} with $L_s=12$, while the strange quark propagators are calculated without low-modes.  The high(all)-mode contributions to light (strange) quark propagators are computed with one random source for each spin, color, and time-slice (spin-color-time dilution). This requires $64 \times 12 =768$ CG inversions per configuration for each of the light and strange quarks.
Since different random sources are generated for different configurations, 

We perform measurements on 258 configurations, which are separated by 10 or 20 molecular dynamics time units, with which we do not see apparent autocorrelation in the correlation functions discussed below.
We implement the all-mode averaging (AMA) technique \cite{Blum:2012uh,Shintani:2014vja} to save computational time.
Since the A2A prescription as used in this calculation naturally requires sampling on all time slices, we perform an exact calculation on a small subset of configurations with the same A2A contraction strategy as for the approximate (sloppy) calculations and combine the results using the superjackknife approach. A detailed description was given in Ref.~\cite{RBC:2023xqv}. In this work we have performed the exact calculations on 14 configurations.
For the approximate part, the CG is stopped at 400 iterations and the light quark propagators are calculated with the zM\"obius fermion action~\cite{Mcglynn:2015uwh,Abramczyk:2017oxr}, which well approximates quantities obtained with the M\"obius action and tighter CG stopping residual, but with smaller $L_s=12$ and hence with smaller cost. 

The $\chi^2$ fits presented in the paper are all performed with a fixed covariance matrix for all (super)jackknife samples and the errors are estimated by the standard jackknife method.

\section{Two-point functions}
\label{sec:2pt}

\subsection{Kaon correlation function}
\label{sec:kaon2pt}

We calculate the two-point function of kaon interpolation operators in the rest frame
\begin{equation}
    C^K(t) = 
    \left\langle O^K(t) O^K(0)^\dag\right\rangle.
    \label{eq:kaon2pt}
\end{equation}
Here the bracket represents ensemble average and the kaon operator is defined by
\begin{equation}
    O^K(t) = \sum_{\vec x,\vec y}
    f_r(||\vec x-\vec y||)
    \bar d(t,\vec x) \img\gamma_5 s(t,\vec y),
\end{equation}
where a $1S$ hydrogen-like wave function smearing
\begin{equation}
    f_r(||\vec x-\vec y||) = \exp(-||\vec x-\vec y||/r)
\end{equation}
is employed with the radius $r = 2.0a$ and with Coulomb gauge fixing.  Here we define the periodic modulus $||\vec x -\vec y||$, the length of the shortest straight path from $\vec y$ to $\vec x$ in the periodic box.

To calculate the kaon two-point function we take advantage of A2A propagators which allow us to calculate the two-point function for every time translation of the source and sink operators and thereby perform an average to improve the statistical precision.

The kaon two-point function at sufficiently large values of $t$ and $L_T-t$ behaves as
\begin{equation}
    C^K(t) \rightarrow A_KA_K^*\left(\e^{-m_Kt}+\e^{-m_K(L_T-t)}\right)
    \label{eq:fitfunc_kaon}
\end{equation}
with the ground-state kaon mass $m_K$ and $A_K = \langle 0|O^K|K(m_K)\rangle$.

\begin{figure}[tbp]
\begin{center}
\includegraphics[width=105mm]{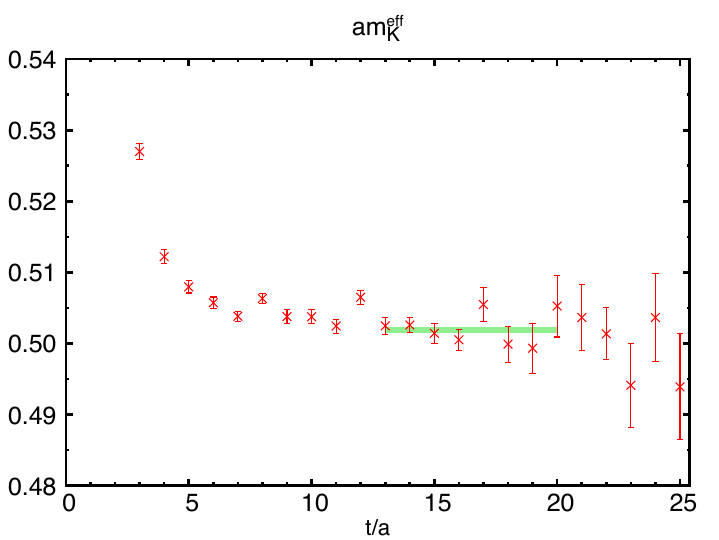}
\caption{
Effective kaon mass and a band of fit result with the fit function in Eq.~\eqref{eq:fitfunc_kaon}.  The fit range is $13\le t/a\le20$.
}
\label{fig:kaonEfm}
\end{center}
\end{figure}

We define an effective kaon mass based on the first term in Eq.~\eqref{eq:fitfunc_kaon},
\begin{equation}
    m_K^{\rm eff}(t) = -\ln\frac{C^K(t)}{C^K(t-1)},
    \label{eq:kaonEfm}
\end{equation}
which is valid for time slices well below $L_T/2 = 32a$.
Figure~\ref{fig:kaonEfm} shows the effective kaon mass along with the result from a fit to  Eq.~\eqref{eq:fitfunc_kaon} using a fixed covariance matrix with range $13\le t/a \le 20$.  The fit result reads
\begin{align*}
    am_K &= 0.50189(36),\\
    A_KA_K^*  &= 1.5604(76) \times 10^6,\\
    \chi^2/d.o.f. &= 0.55,
\end{align*}
which corresponds to about 513 MeV.

\subsection{Two-pion correlation functions and energies}
\label{sec:pipi2pt}

In this subsection we summarize the results for the two-pion correlators and energies obtained in our companion paper, Ref.~\cite{RBC:2023xqv}.

To construct two-pion operators, we first define single pion operators with specific momenta, $(0,0,0)$, $(0,0,\frac{2\pi}{L})$, $(0,\frac{2\pi}{L},\frac{2\pi}{L})$, $(\frac{2\pi}{L},\frac{2\pi}{L},\frac{2\pi}{L})$ and permutations with respect to cubic symmetry.
As for the kaon, Coulomb gauge fixed hydrogen-like wave functions are used for single pion interpolating operators but with radius $r = 1.5a$, which well matches with the smearing radius for single pion operators used in the GPBC calculation in physical units~\cite{RBC:2020kdj}.
Next two-pion operators are formed by multiplying two single-pion operators together with opposite momenta so that the total momentum is zero.  Since the SO(3) symmetry of angular momentum is broken down to a discrete symmetry in a finite box, we use the simplest irreducible representation that couples to the $S$-wave state, the A1 representation. 
In addition we perform isospin projection of the two-pion operators to the $I=2$ and $I=0$ representations.
We separate the two single-pion operators by 3 time-slices in order to reduce the vacuum state contribution to the $I=0$ channel~\cite{Bai:2015nea,RBC:2020kdj}.  The same separation in physical units were employed in the GPBC calculation~\cite{RBC:2020kdj}.

For the $I=0$ channel, we also introduce a $\sigma$-like operator, an iso-singlet scalar bilinear of the up and down quarks since it was previously found to play an important role in controlling contamination from excited states \cite{Briceno:2016mjc,RBC:2020kdj}.
The $\sigma$-like operator is also smeared in the same way as the single pion operators.

Thus, we employ four two-pion operators for the $I=2$ channel and five for the $I=0$ channel.  In order to distinguish the $\sigma$-like operator from the other two-pion operators we refer to the latter as $\pi\pi$-like operators.  It is also convenient to denote them as $\pi\pi(000)$, $\pi\pi(001)$, $\pi\pi(011)$, $\pi\pi(111)$ and $\sigma$, where the three-digit number in the parentheses represents the three-dimensional momentum, in units of $2\pi/L$, of a single pion interpolation operator.
The single pion and $\sigma$ operators, which are dimension-3, are summed over three-dimensional space to project onto a certain momentum and are therefore dimensionless.

We compute the following two-point correlation functions
\begin{equation}
    C_{ab}^{{\rm 2pt,}I}(t) = 
    \left\langle O_a^{I}(t) O_b^{I}(-\Delta_b)^\dag\right\rangle
    - \left\langle O_a^{I}\right\rangle 
    \left \langle {O_b^{I}}^\dag\right\rangle,
    \label{eq:pipi2pt}
\end{equation}
where $O_a^I$ is a two-pion operator labeled by the operator index $a(=1,2,\ldots,N)$ and isospin $I$.  $\Delta_b$ is 0 when $b$ corresponds to the $\sigma$ operator and otherwise $3a$ so that the time variable $t$ indicates the separation between the ``inner", or nearer, pions or $\sigma$ operator of the source and sink. The second term on the RHS corresponds to the vacuum subtraction relevant for the $I=0$ channel.
In what follows we omit the superscript $I$ for simplicity.

These two-pion correlation functions behave as
\begin{align}
C_{ab}^{\rm2pt}(t)
& = \sum_n A_{n,a}A_{n,b}^*\e^{-E_n t}
+ \mbox{thermal effects},
\label{eq:state_expansion}
\end{align}
where $E_n$ ($n=0,1,\ldots$) is the energy of the $n$-th two-pion state of the corresponding isospin channel and $A_{n,a} = \langle 0| O_a | \pi\pi(E_n)\rangle$.
Here the thermal effect is significant, and its leading contribution is an around-the-world propagation effect of the ground-state pion in the time direction which is independent of $t$.
In order to eliminate this contribution we perform a subtraction and obtain the following behavior:
\begin{align}
    C_{ab}^{\rm2pt}(t,\delta_t) &= C_{ab}^{\rm2pt}(t) - C_{ab}^{\rm2pt}(t+\delta_t)
    \notag\\
    &= \sum_n \left(1 - \e^{-E_n\delta_t}\right)
    A_{n,a}A_{n,b}^*\e^{-E_n t} + \ldots
    \label{eq:state_expansion_subt}
\end{align}
with an arbitrary time shift $\delta_t$.
The ellipsis represents time-dependent thermal effects, which are invisible in the results shown below and in Ref.~\cite{RBC:2023xqv} with the given statistical precision and therefore neglected in this work.
While this subtraction cancels the vacuum subtraction term in Eq.~\eqref{eq:pipi2pt}, we find a minor statistical improvement by continuing to apply the vacuum subtraction in a time-dependent manner, $i.e.$ by replacing the second term on the right hand side of Eq.~\eqref{eq:pipi2pt} with $\left\langle O_a(t)\right\rangle \left\langle O_b(-\Delta_b)^\dag\right\rangle$, where the vacuum expectation values are taken at a specific time slice indicated in the parentheses.  The time-translation average is taken after the modified vacuum subtraction.

To decompose the correlation functions into contributions from individual states we employ the variational method \cite{Luscher:1990ck}, where we solve the Generalized Eigenvalue Problem (GEVP) for a given $N\times N$ correlator matrix, $C^{\rm2pt}(t,\delta_t)$,
\begin{equation}
C^{\rm2pt}(t,\delta_t)V_n(t,t_0,\delta_t) = \lambda_n(t,t_0,\delta_t) C^{\rm2pt}(t_0,\delta_t)V_n(t,t_0,\delta_t).
\label{eq:GEVP}
\end{equation}
From the original operators and the eigenvectors
$V_{n,a}(t,t_0,\delta_t)$ we can construct an operator,
\begin{equation}
    {\cal O}_n^{(t,t_0,\delta_t)} = 
    \sum_a V_{n,a}(t,t_0,\delta_t)O_a,
\end{equation}
which couples well to the state labeled by $n$ but not with any of the other $N$ low-lying states.  We can thus calculate the state-specific correlation functions:
\begin{align}
    C_n^{\rm2pt}(t,\delta_t,t',t_0) 
    &= V_n(t',t_0,\delta_t)^\dag C(t,\delta_t) V_n(t',t_0,\delta_t)
    \notag\\
    &= \left(1 - \e^{-E_n\delta_t}\right) B_n^{(t',t_0,\delta_t)} {B_n^{(t',t_0,\delta_t)}}^* \e^{-E_nt} + 
    \ldots,
    \label{eq:diagCorrpipi}
\end{align}
where the state index $n$ is not summed over, and we define
\begin{equation}
    B_n^{(t,t_0,\delta_t)} =  
    \left\langle 0\big| {\cal O}_n^{(t,t_0,\delta_t)} \big| \pi\pi(E_n)\right\rangle.
    \label{eq:Bcoef}
\end{equation}
The ellipsis in Eq.~\eqref{eq:diagCorrpipi} represents minor contributions from higher states with energies larger than $E_{N-1}$ and the remaining thermal effects.

The (generalized) eigenvalues $\lambda_n(t,t_0,\delta_t)$ obtained by solving the GEVP \eqref{eq:GEVP} provide the effective two-pion energies 
\begin{equation}
    E^{\rm eff}_n(t,t_0,\delta_t) = \ln{\frac{\lambda_n(t,t_0,\delta_t)}{\lambda_n(t+a,t_0,\delta_t)}},
    \label{eq:def_Eeff}
\end{equation}
which should be independent of $t,t_0$ and $\delta_t$ at sufficiently large time separations ($t$) where the higher-state contamination is invisible.  Ref.~\cite{Blossier:2009kd} demonstrated that the higher-state contamination in the energy of the $n$-th state defined by \eqref{eq:def_Eeff} is $O(\e^{-(E_{N+1}-E_n)t})$ in the region $t_0\ge t/2$.  In this work we choose $t_0 = t-a$, which satisfies the inequality for $t_0\ge a$.  In Ref.~\cite{RBC:2023xqv} we tuned the value of $\delta_t$ and found the statistical precision can be optimized by choosing $\delta_t = 5a$ for $I=2$ and $\delta_t = 8a$ for $I=0$, with which we quote the results throughout this paper. 

In practice, there is a significant correlation between two-pion correlation functions with and without interactions between two pions. 
We utilize this correlation to improve the statistical and systematic precision of the two-pion energies.
We compute the difference between the fully-interacting and non-interacting two-pion energies
\begin{equation}
\Delta E^{\rm eff}_n(t,t_0,\delta_t)
= E^{\rm eff}_n(t,t_0,\delta_t) - E_n^{\rm0,eff}(t,t_0,\delta_t),
\label{eq:Delta_Eeff}
\end{equation}
where the non-interacting two-pion effective energy $E_n^{\rm0,eff}(t,t_0,\delta_t)$ is calculated by the same procedure as the interacting one but using non-interacting two-pion correlators, $i.e.,$ a product of two single-pion correlators, ensemble-averaged separately, with pion operators placed at the same time slices as the ones for the interacting two-pion correlators.
Then we add back the non-interacting two-pion energy obtained by using the continuum dispersion relation to obtain the improved effective energy,
\begin{equation}
    E_n^{\rm eff\,\prime}(t,t_0,\delta_t)
    = \Delta E^{\rm eff}_n(t,t_0,\delta_t) + 2\sqrt{m_\pi^2 + \frac{4n\pi^2}{L^2}}.
    \label{eq:DR}
\end{equation}

\begin{figure}[tbp]
\begin{center}
\begin{tabular}{c}
\includegraphics[width=80mm]{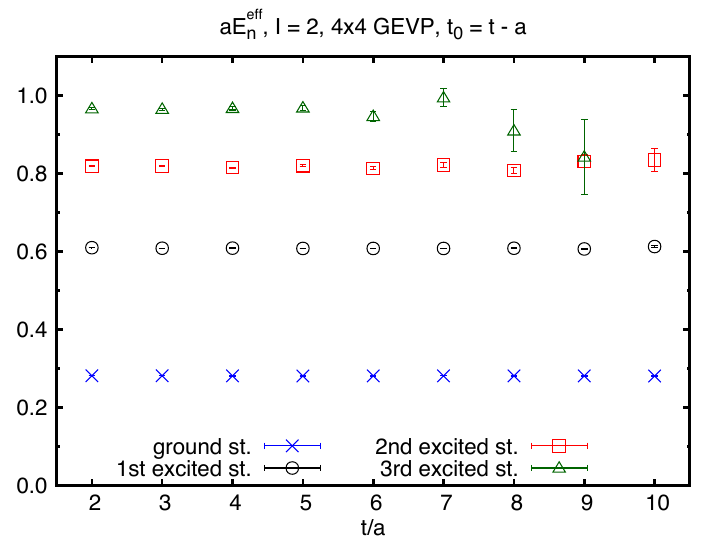}
\end{tabular}
\hfill
\begin{tabular}{c}
\includegraphics[width=80mm]{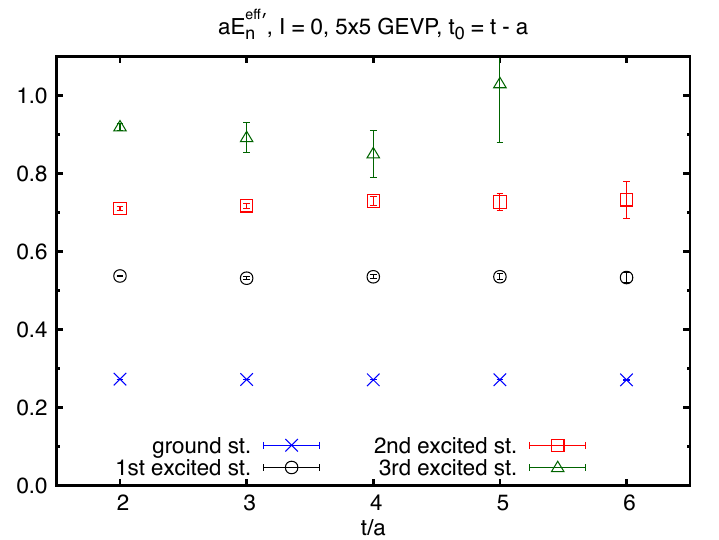}
\end{tabular}
\caption{Effective two-pion energies for $I=2$ (left) and $I=0$ (right) obtained from GEVP analysis and the improvement~\eqref{eq:DR} using the continuum dispersion relation.  Here we solve GEVP using the four two-pion operators for $I=2$ and five operators including the $\sigma$-like operator for $I=0$.  The results are shown in lattice units. 
}
\label{fig:efms0_24c}
\end{center}
\end{figure}

Figure~\ref{fig:efms0_24c} shows the improved effective two-pion energies for the $I=2$ and $I=0$ channels.
We omit the $I=0$ fourth excited state energy because it is unresolved.
The figure indicates that we can well extract the signal from the four lowest states of the $I=2$ channel and three lowest ones of the $I=0$ channel.

\begin{table}[tp]
    \centering
    \begin{tabular}{ccc}
    \hline
        fit range & $aE_n$ & $F$  \\
        \hline
        4--10 & 0.28128(34) & 32.09(15) \\
        5--9 & 0.60789(31) & 28.686(22) \\
        3--9 & 0.81743(56)  & 26.242(15) \\
        \hline
    \end{tabular}
    \caption{Results for the $I=2$ two-pion energies obtained from a constant fit to the effective two-pion energies shown with the corresponding fit range and corresponding Lelloch-L\"uscher factors.}
    \label{tab:pipiI2}
\end{table}

\begin{table}[tp]
    \centering
    \begin{tabular}{ccc}
    \hline
        fit range & $aE_n$ & $F$  \\
        \hline
        4--8 & 0.27060(40) & 38.38(46) \\
        3--6 & 0.5349(34) & 29.184(68) \\
        3--5 & 0.7242(65) & 31.56(33) \\
        \hline
    \end{tabular}
    \caption{The same as Table~\ref{tab:pipiI2} but for the $I=0$ channel.}
    \label{tab:pipiI0}
\end{table}

We perform constant fits to each two-pion effective energy considering the correlation among the data points at different time slices.
The results are summarized in Tables~\ref{tab:pipiI2} and \ref{tab:pipiI0} for the $I=2$ and $I=0$ channels, respectively.
In the $I=2$ channel, the first-excited state energy is closest to the kaon mass, $am_K = 0.50189(36)$, but is 21\% larger, hence a somewhat sizeable interpolation will be required to obtain physical kinematics as we will discuss in Section~\ref{sec:onshellME}. For the $I=0$ channel the difference is only 5\%, requiring a much more modest interpolation to be performed.

As noted in the beginning of this subsection more detailed and sophisticated discussion on two-pion correlators and energies was presented in Ref.~\cite{RBC:2023xqv}.

\subsection{Lellouch-L\"uscher factor}
\label{sec:LL}

Because of the interaction between two pions in a finite box, the normalization for the two-pion states is not the same as in the infinite-volume limit.
We need to correctly normalize the two-pion states in order to obtain the $K\to\pi\pi$ amplitudes in infinite volume.
The prescription to determine the normalization factor $F$, the Lellouch-L\"uscher (LL) factor, was provided in Ref.~\cite{Lellouch:2000pv}.
With a given two-pion energy $E_{\pi\pi}$ in a finite box, the normalization factor can be determined as
\begin{equation}
    F^2 = \frac{4\pi m_K E_{\pi\pi}^2}{k^3}
    \left(k\frac{\td\delta_0}{\td k} + q\frac{\td\phi}{\td q}\right),
    \label{eq:LLfactor}
\end{equation}
where 
\begin{equation}
    k = \sqrt{\left(\frac{E_{\pi\pi}}{2}\right)^2-m_\pi^2},\ \ \ \ 
    q = \frac{Lk}{2\pi}
\end{equation}
and $\delta_0$ is the two-pion phase shift and $\phi$ is a known function,
\begin{equation}
    \tan\phi(q) = -\frac{\pi^{3/2}q}{Z_{00}(1;q^2)}
\end{equation}
for periodic boundary conditions defined with the L\"uscher zeta function,
\begin{equation}
    Z_{00}(s;q^2) = \frac{1}{\sqrt{4\pi}}\sum_{\vec n}\left(|\vec n|^2-q^2\right)^{-s}.
    \label{eq:luscher zeta}
\end{equation}
Eq.~\eqref{eq:luscher zeta} is valid for ${\rm Re}(s)\ge3/2$. For ${\rm Re}(s)<3/2$ its analytic continuation is used.  We employ the efficient numerical implementation given in Ref.~\cite{Yamazaki:2004qb} to evaluate it at $s=1$.
In this work, the phase shift term, the first term in the parentheses on the right hand side of Eq.~\eqref{eq:LLfactor}, is calculated via dispersion theory based on the Roy equation \cite{Roy:1971tc} with inputs from chiral perturbation theory (ChPT) and experimental data, which is consistent with results from lattice QCD using the L\"uscher finite volume method~\cite{Luscher:1990ck}, both near the two-pion threshold ~\cite{Beane:2007xs,Fu:2017apw,Culver:2019qtx,FlavourLatticeAveragingGroupFLAG:2021npn} and at larger two-pion energies~\cite{NPLQCD:2011htk,Fischer:2020jzp,RBC:2021acc,RBC:2023xqv,Bruno:2023pde}.  In practice, we use Eqs.~(17.1)--(17.3) of Ref.~\cite{Colangelo:2001df} with the pion mass on our lattice ensemble to compute the phase shift.

The formula~\eqref{eq:LLfactor} is valid up to the exponentially suppressed corrections~\cite{Bedaque:2006yi}, $O(\e^{-m_\pi L})$, in the elastic region, $2m_\pi < E_{\pi\pi} < 4m_\pi$ in the case of the rest frame.
In the inelastic region above the four-pion threshold, $E_{\pi\pi} \ge 4m_\pi$, and there are extra states that are not accounted for by the L\"uscher prescription.
The systematic effect slightly above the inelastic threshold is expected to be small since the effects of four-pion states appear at next-to-next-to-leading order (NNLO) in ChPT~\cite{NPLQCD:2011htk}.
In addition, some pion scattering studies have found that the systematic effects appear to be sub-statistical slightly above the inelastic threshold~\cite{NPLQCD:2011htk,Fischer:2020jzp,RBC:2023xqv,Bruno:2023pde}.  Since the LL factor~\eqref{eq:LLfactor} is derived using the L\"uscher formula for the phase shift, the systematic error on the LL factor slightly above the inelastic threshold is expected to be small.
In this section, we calculate the LL factor up to the second excited state with an energy slightly above 800~MeV and 700~MeV for $I=2$ and $I=0$, respectively.  For calculations of $K\to\pi\pi$ amplitudes in Section~\ref{sec:onshellME} we only use the ground and first-excited two-pion states for the interpolation to obtain the physical kinematics.  Out of these states, only the $I=2$ first excited state is above (but close to) the threshold. 

Since the $I=0$ two-pion ground state energy in the rest frame is smaller than $2m_\pi$ due to the attractive interaction in this channel, the corresponding value of $k^2$ is negative, and the formulae above needs to be analytically continued.  While this kind of analytic continuation for $I=0$ was implicitly performed for the calculation of the scattering length by a number of works as emphasized in Ref.~\cite{Fu:2017apw}, here we give the formulae for the LL factor below the two-pion threshold. 
\begin{align}
F^2 &= -\frac{4\pi m_KE_{\pi\pi}^2}{{k'}^3}
\left(k'\frac{\td\delta_0'}{\td k'} + q'\frac{\td\phi'}{\td q'}\right),\\
k' &= \sqrt{m_\pi^2-\left(\frac{E_{\pi\pi}}{2}\right)^2},\ \ \ \ 
q' = \frac{Lk'}{2\pi},\\
\tanh\phi'(q') &= -\frac{\pi^{3/2}q'}{Z_{00}(1;-{q'}^2)},
\end{align}
where $\delta_0'(k')$ is obtained by replacing $\tan$ with $\tanh$, $k^2$ with $-{k'}^2$ and the kinematic factor $\sqrt{1-4m_\pi^2/s}$ with $2k'/E_{\pi\pi}$, for the case of rest frame, in Eq.~(17.1) of Ref.~\cite{Colangelo:2001df} ($k$ in our convention is denoted by ``$q$'' in ~\cite{Colangelo:2001df}).

While this is primarily a prescription for accounting for a finite-volume two-pion state, it also contains some extra factors such as $m_K$ and $E_{\pi\pi}$ due to the difference in the state normalization; states on the lattice are normalized to unity, whereas a relativistic normalization is employed in infinite volume.
The factor $m_K$ is associated with the difference in the convention of the kaon state normalization.
Thus the LL factor \eqref{eq:LLfactor} gives the relation between the $K\to\pi\pi$ amplitudes on finite lattice with those in infinite volume: $A_I = FA_I^{FV}$.

The results for the LL factor $F$ for the lowest three energy states are summarized in Tables~\ref{tab:pipiI2} and \ref{tab:pipiI0} for $I=2$ and $I=0$, respectively, along with the corresponding two-pion energies $E_n$ discussed in the previous subsection.

\section{$K\to\pi\pi$ three-point functions and matrix elements}
\label{sec:3pt}

\subsection{Four-quark operators}
\label{sec:4qOps}

$K\to\pi\pi$ decay is well described by the $\Delta S = 1$ weak Hamiltonian in Eq.~\eqref{eq:Hw}.
In this work, the $\Delta S = 1$ four-quark operators in the three-flavor theory are employed as effective operators $Q_i$, which we define following the convention in 
Refs.~\cite{Buras:1993dy,Buchalla:1995vs},
\begin{align}
    Q_1 &= 
    \bar s_\alpha \gamma_\mu(1-\gamma_5)u_\beta\cdot
    \bar u_\beta \gamma_\mu(1-\gamma_5)d_\alpha,
    \label{eq:Q1}
    \\
    Q_2 &= 
    \bar s_\alpha \gamma_\mu(1-\gamma_5)u_\alpha\cdot
    \bar u_\beta \gamma_\mu(1-\gamma_5)d_\beta,
    \label{eq:Q2}
    \\
    Q_3 &= 
    \bar s_\alpha \gamma_\mu(1-\gamma_5)d_\alpha\sum_q
    \bar q_\beta \gamma_\mu(1-\gamma_5)q_\beta,
    \\
    Q_4 &= 
    \bar s_\alpha \gamma_\mu(1-\gamma_5)d_\beta\sum_q
    \bar q_\beta \gamma_\mu(1-\gamma_5)q_\alpha,
    \\
    Q_5 &= 
    \bar s_\alpha \gamma_\mu(1-\gamma_5)d_\alpha\sum_q
    \bar q_\beta \gamma_\mu(1+\gamma_5)q_\beta,
    \\
    Q_6 &= 
    \bar s_\alpha \gamma_\mu(1-\gamma_5)d_\beta\sum_q
    \bar q_\beta \gamma_\mu(1+\gamma_5)q_\alpha,
    \\
    Q_7 &= \frac{3}{2}
    \bar s_\alpha \gamma_\mu(1-\gamma_5)d_\alpha\sum_qe_q
    \bar q_\beta \gamma_\mu(1+\gamma_5)q_\beta,
    \\
    Q_8 &= \frac{3}{2}
    \bar s_\alpha \gamma_\mu(1-\gamma_5)d_\beta\sum_qe_q
    \bar q_\beta \gamma_\mu(1+\gamma_5)q_\alpha,
    \\
    Q_9 &= \frac{3}{2}
    \bar s_\alpha \gamma_\mu(1-\gamma_5)d_\alpha\sum_qe_q
    \bar q_\beta \gamma_\mu(1-\gamma_5)q_\beta,
    \\
    Q_{10} &= \frac{3}{2}
    \bar s_\alpha \gamma_\mu(1-\gamma_5)d_\beta\sum_qe_q
    \bar q_\beta \gamma_\mu(1-\gamma_5)q_\alpha,
    \label{eq:Q10}
\end{align}
where the sums over $q$ run for all the active quarks: up, down and strange in three-flavor theory.  
The sum over the Lorentz index $\mu$ is implicitly taken for each operator.
The color indices are explicitly shown by $\alpha$ and $\beta$, while the spin indices are omitted as they are always contracted in the trivial manner. The electric charge of a quark $q$ is expressed by $e_q$ for the electroweak penguin operators $Q_{7\mathchar`-10}$.  Here, the current-current operators, $Q_1$ and $Q_2$, dominate the physics of the real parts of the amplitudes; the QCD penguin operators, $Q_{3\mathchar`-6}$, dominate that of Im$(A_0)$; and the electroweak penguin operators, $Q_{7\mathchar`-10}$, that of Im$(A_2)$. 
Note, while the lattice calculation does not include electromagnetic effects, we do include the effective operators resulting from short-distance photonic propagation due to the significant role they play in the $I=2$ channel decay.

As is well known, a lattice calculation preserves all dimension-4 Fierz relations, while these are broken in dimensional regularization approaches to perturbation theory for which the dimension-dependence of $\gamma_5$ breaks certain Fierz relations leading.  Fierz symmetry gives rise to three relations among the four-quark operators.  We therefore use them to reduce the operator basis to seven operators, $\{Q_j'\}_{j = 1,2,3,5,6,7,8}$ defined in \cite{RBC:2001pmy}, in our lattice calculation, which we call the chiral basis, as well as the ten-operator basis above. 
The linear independence of the chiral basis is convenient to renormalize the four-quark operators as it requires only a minimum number of independent renormalization conditions, and the inverse renormalization matrix needed for step scaling is well defined.
It should also be noted that each operator in the chiral basis transforms as a specific representation of ${\rm SU}(3)_L\times {\rm SU}(3)_R$ chiral symmetry so that the mixing among the operators is minimal, while the current-current ($Q_{1,2}$) and electroweak penguin ($Q_{7\mathchar`-10}$) operators are composed of multiple representations.  This property is convenient especially when fermions with good chiral symmetry such as domain wall fermions are employed.
The basis enlargement from the chiral basis to the ten-operator basis after renormalization only has nontriviality in the perturbative matching from a nonperturbative scheme to $\rm\overline{MS}$, which was well discussed in Ref.~\cite{Lehner:2011fz} and is taken into account in Section~\ref{sec:AandEpslon'}.

Alternative definitions of $Q_1$ and $Q_2$ can be formed by applying the Fierz identities~\cite{RBC:2001pmy,Lehner:2011fz}, $\widetilde Q_1 = \bar s_\alpha\gamma_\mu(1-\gamma_5)d_\alpha\cdot \bar u_\beta \gamma_\mu(1-\gamma_5)u_\beta$ and $\widetilde Q_2 = \bar s_\alpha \gamma_\mu(1-\gamma_5)u_\beta\cdot\bar u_\beta \gamma_\mu(1-\gamma_5)d_\alpha$.  These definitions give rise to identical numerical results on the lattice and simplify the structure of the contractions by making all four-quark operators with an odd (even) index have a color-diagonal (color-mixed) structure.  However, the Wilson coefficients in the $\rm\overline{MS}$ scheme depend on the definitions of $Q_1$ and $Q_2$ since dimensional regularization is employed.  We use the definitions in Eqs.~\eqref{eq:Q1} and \eqref{eq:Q2} for computing the Wilson coefficients based on the formulae given in Refs.~\cite{Buras:1993dy,Buchalla:1995vs}.

While these operators are all relevant for the $\Delta I = 1/2$ channel, the QCD penguin operators $Q_{3\mathchar`-6}$ and four operators $Q_{2,3,5,6}'$ in the chiral basis are purely in the $(8,1)$ representation and do not contribute to the $\Delta I = 3/2$ channel.  As a result the number of independent operators is three for this channel. 
Our earlier works on the $\Delta I = 3/2$ channel \cite{Blum:2011ng,Blum:2012uk,Blum:2015ywa} employed a three-operator basis that is purely made of $\Delta I = 3/2$ operators, $Q_{(27,1)}$, $Q_{(8,8)}$ and $Q_{(8,8)\rm mix}$.  In this work we employ the ten operators in Eqs.~\eqref{eq:Q1}--\eqref{eq:Q10} and the chiral basis of the seven operators for both the $\Delta I = 3/2$ and $\Delta I = 1/2$ channels to apply the same numerical analysis, although the matrix elements of the QCD penguin operators are always zero for the $\Delta I =3/2$ channel. 

\subsection{$K\to\pi\pi$ three-point functions}
\label{sec:K2pipi3pt}

With the kaon and two-pion interpolation operators described in Sec~\ref{sec:2pt} and local four-quark operators in Eqs.~\eqref{eq:Q1}--\eqref{eq:Q10}, we compute the $K\to\pi\pi$ three-point functions
\begin{equation}
C_{ai}^{\rm3pt}(t_1,t_2)
= \left\langle O_a(t_1+t_2) Q_i(t_1) O^K(0)^\dag\right\rangle.
\label{eq:k2pipi3pt}
\end{equation}
Again the isospin index $I$ is suppressed for simplicity and we discuss the three-point function in general for both the $\Delta I = 3/2$ and $1/2$ channels corresponding the $I=2$ and 0 two-pion operators on the right hand side, respectively. The subscript $a$ labels a two-pion operator, including the quark bilinear $\sigma$.
As mentioned in the previous section the source and sink operators are dimensionless as each bilinear of these operators is summed over three-dimensional space.
The dimension-6 operator $Q_i$ is also summed over three-dimensional space for the measurements and thus these correlation functions are dimension-3.
The time indices $t_1$ and $t_2$ stand for the time separations between the four-quark and kaon operators, and between the two-pion and four-quark operators, respectively.
We calculate the $K\to\pi\pi$ three-point functions with several time separations between the kaon source and two-pion sink operators. We choose $(t_1 + t_2)/a = 6,7,9,10,11,13$ in this work.
Counting the parity of the two-pion and kaon operators one can recognize that the three-point functions are contributed by the parity-odd part of the four-quark operators.  The parity-even part of the four-quark operators only increases statistical error on the three-point functions and is therefore excluded from the measurements.

\begin{figure}[tbp]
\begin{center}
\begin{tabular}{c}
\includegraphics[width=80mm]{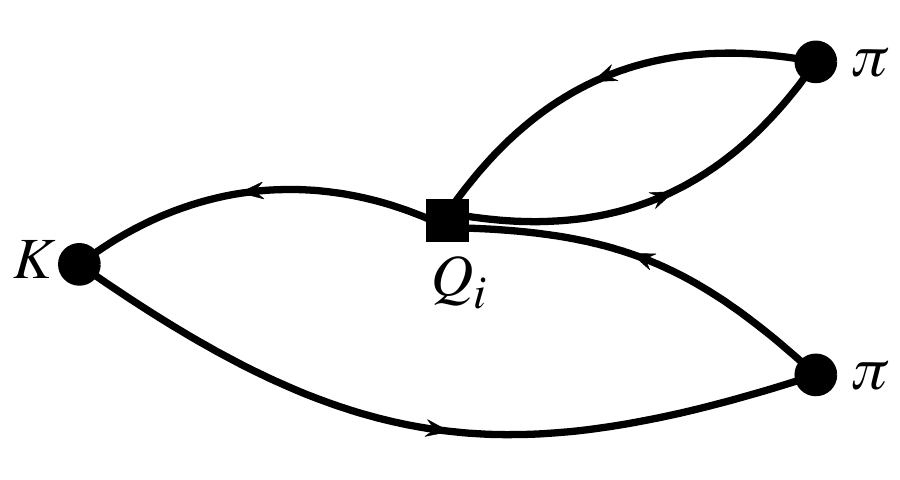}
\\
{\it type1}
\end{tabular}
\hfill
\begin{tabular}{c}
\includegraphics[width=80mm]{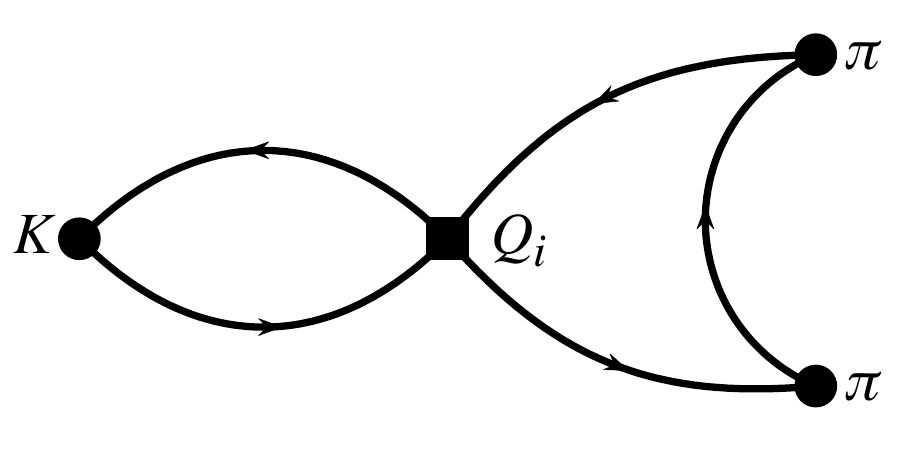}
\\
{\it type2}
\end{tabular}
\begin{tabular}{c}
\includegraphics[width=80mm]{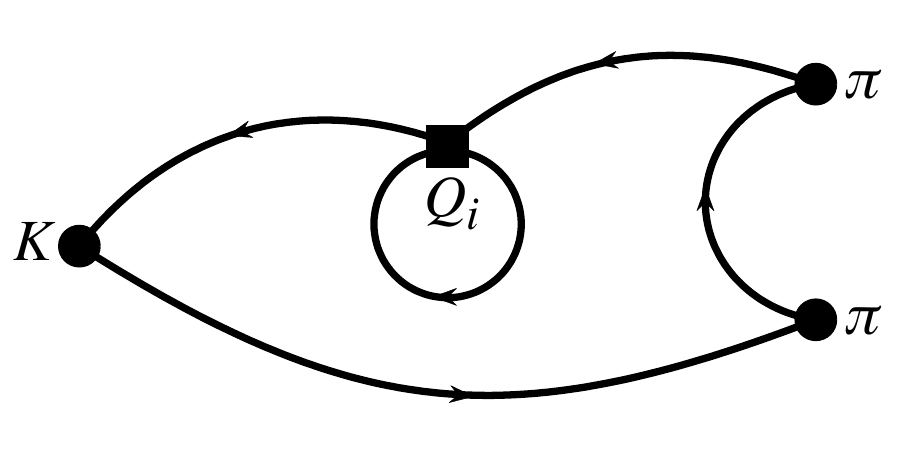}
\\
{\it type3}
\end{tabular}
\hfill
\begin{tabular}{c}
\includegraphics[width=80mm]{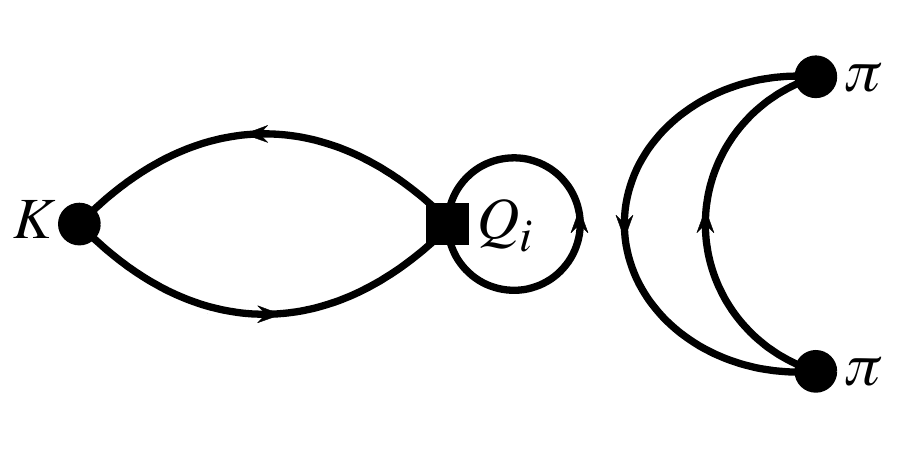}
\\
{\it type4}
\end{tabular}
\caption{
Diagrams for the $K\to\pi\pi$ three-point functions with four-quark and $\pi\pi$-like sink operators.  A line connecting the kaon ($K$) and the four-quark operator ($Q_i$) is a strange-quark propagator.  The quark loop in $type3$ and $type4$ diagrams is either a light- or strange-quark propagator.  All other lines represent a light-quark propagator.
}
\label{fig:diagrams_k2pipi}
\end{center}
\end{figure}

\begin{figure}[tbp]
\begin{center}
\begin{tabular}{c}
\includegraphics[width=80mm]{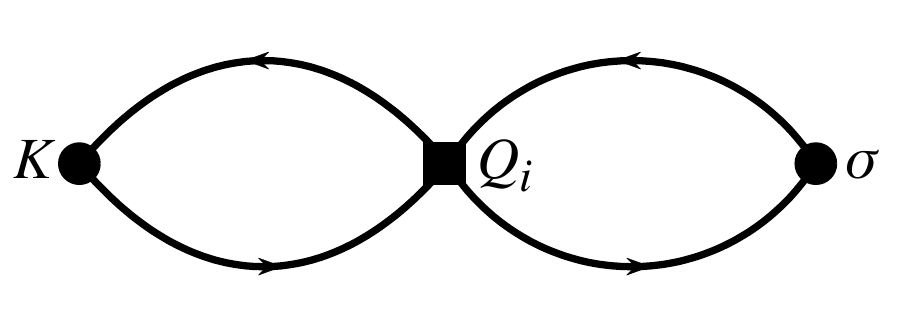}
\\
{\it type2}
\end{tabular}
\\
\begin{tabular}{c}
\includegraphics[width=80mm]{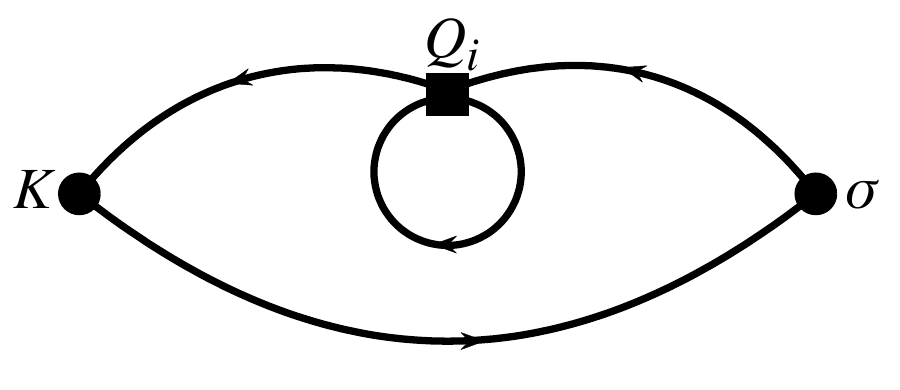}
\\
{\it type3}
\end{tabular}
\hfill
\begin{tabular}{c}
\includegraphics[width=80mm]{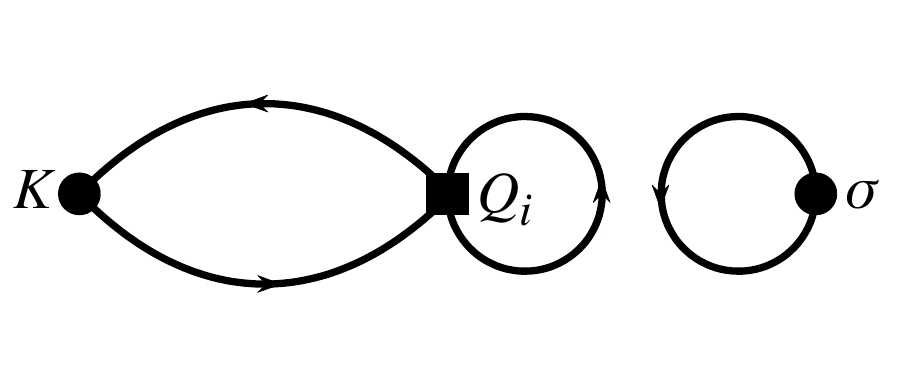}
\\
{\it type4}
\end{tabular}
\caption{
Same as Figure~\ref{fig:diagrams_k2pipi} but with a $\sigma$-like sink operator.
}
\label{fig:diagrams_k2sigma}
\end{center}
\end{figure}

The Wick contractions for these correlation functions yield four classes of diagram, which are summarized in Figure~\ref{fig:diagrams_k2pipi} for the case where $O_a$ is a $\pi\pi$-like operator and Figure~\ref{fig:diagrams_k2sigma} for the case where $O_a$ is a $\sigma$-like operator.
The $\Delta I=3/2$ channel contains contributions from only {\it type1} diagrams, whereas the $\Delta I = 1/2$ channel receives contributions from all the diagrams.
While the contraction formulae with a $\pi\pi$-like sink operator for PBC were given in Ref.~\cite{Blum:2011pu}, the formulae with a $\sigma$-like operator were not presented.
We summarize the contraction formulae for both $K\to\pi\pi$ three-point functions including those with a $\sigma$-like operator in Appendix~\ref{sec:contractions}.

Since {\it type3} and {\it type4} diagrams, which are needed for the $\Delta I = 1/2$ channel, include a quark loop, the correlation functions in this channel contain a power divergence $\sim a^{-2}$ that needs to be removed.
For the rest of the article, we distinguish the subtracted ($Q_i$) and unsubtracted ($\widehat Q_i$) four-quark operators and relate them by
\begin{equation}
    Q_i = \widehat Q_i - \alpha_i \bar s\gamma_5d,
\end{equation}
where we determine the subtraction coefficients $\alpha_i$ by imposing the following condition on the two-point functions:
\begin{equation}
    \left\langle Q_i(t_1) O^K(0)^\dag\right\rangle = 0.
    \label{eq:subt cond}
\end{equation}
While the coefficients $\alpha_i$ do not have to depend on the time separation $t_1$ between the kaon and four-quark operators, we choose this approach because it is found to offer a minor statistical improvement~\cite{RBC:2020kdj}.
The contribution of this pseudoscalar operator vanishes for on-shell matrix elements, but the power divergence still afflicts the measurements described here because the kinematics are not energy-conserving.
It should also be noted that the subtraction condition \eqref{eq:subt cond} ensures the absence of the vacuum contribution to the $K\to\pi\pi$ three-point function
\begin{equation}
    \langle0|O_a|0\rangle\left\langle 0\big|Q_i(t_1) O^K(0)^\dag\big|0\right\rangle = 0
\end{equation}
as long as we neglect thermal effects like $\sim\left\langle \pi\big|Q_i(t_1) O^K(0)^\dag\big|\pi\right\rangle\e^{-m_\pi(L_T-t_1)}$, which is negligible as seen below.
Therefore we do not perform a subtraction of this vacuum effect.
From the condition \eqref{eq:subt cond} we obtain
\begin{equation}
    \alpha_i(t_1)
    = \frac{\left\langle\widehat Q_i(t_1) O^K(0)^\dag\right\rangle}
    {\left\langle \bar s\gamma_5d(t_1) O^K(0)^\dag\right\rangle}.
    \label{eq:alpha_i}
\end{equation}
These correlation functions are averaged over all time translations with the A2A quark propagators.

\begin{figure}[tbp]
\begin{center}
\begin{tabular}{c}
\includegraphics[width=80mm]{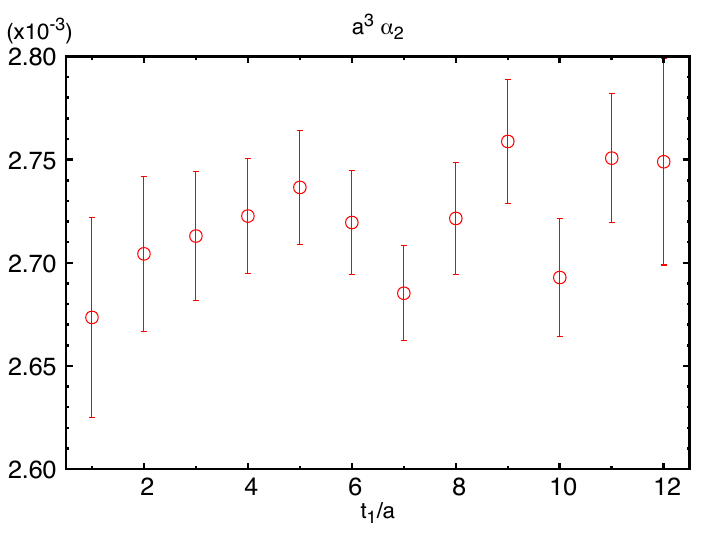}
\end{tabular}
\hfill
\begin{tabular}{c}
\includegraphics[width=80mm]{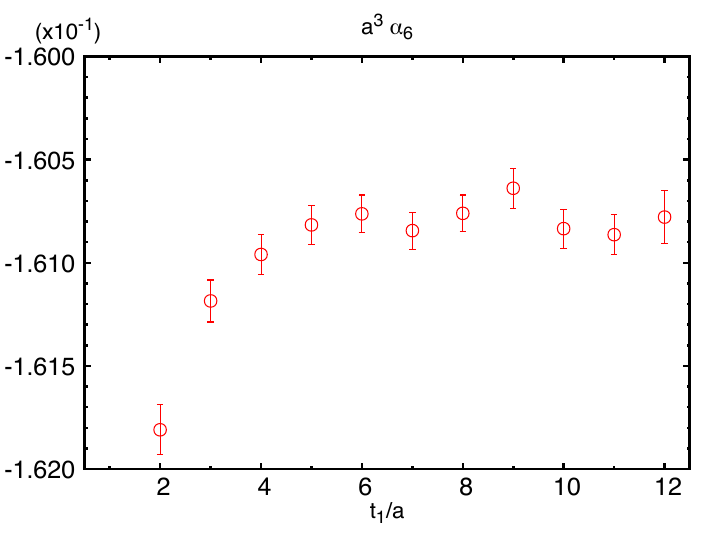}
\end{tabular}
\caption{Subtraction coefficients $\alpha_2$ (left) and $\alpha_6$ (right) obtained by Eq.~\eqref{eq:alpha_i}.
}
\label{fig:alpha_i}
\end{center}
\end{figure}

Figure~\ref{fig:alpha_i} shows the results for $\alpha_2$ and $\alpha_6$.
The stable plateau seen up to $t_1 = 12a$ indicates that the thermal effect, which contributes to the numerator of Eq.~\eqref{eq:alpha_i} in the form $\sim\left\langle \pi\big|Q_i(t_1) O^K(0)^\dag\big|\pi\right\rangle\e^{-m_\pi(L_T-t_1)}$, is not significant.
While the values of $\alpha_i$ on the plateau correspond to the subtraction condition $\langle0|Q_i|K\rangle = 0$, the other data points also remove the power divergence from the quark loop of {\it type3} and {\it type4} diagrams but lead to different values of $K\to\pi\pi$ matrix elements with energy-non-conserving kinematics.  
In this work, the $\Delta I = 1/2$ matrix elements are determined from the region of large enough $t_1$ where $\langle0|Q_i|K\rangle = 0$ is satisfied.

\begin{figure}[tbp]
\begin{center}
\begin{tabular}{c}
\includegraphics[width=80mm]{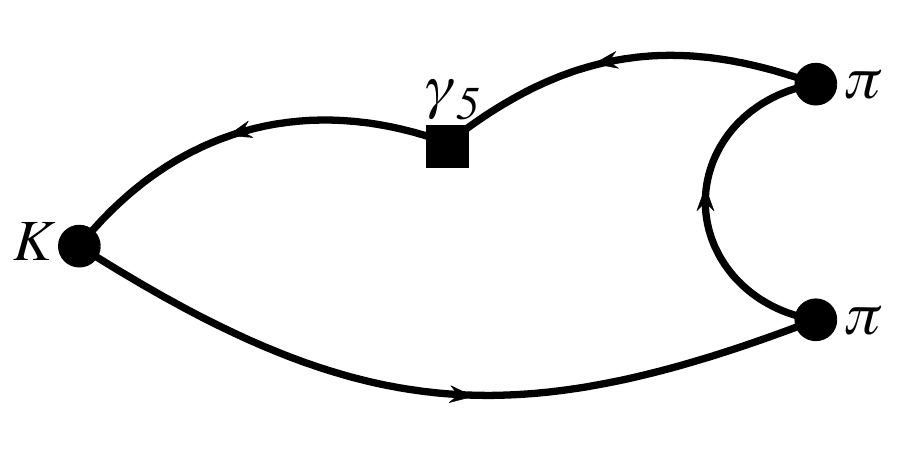}
\\
{\it type3}
\end{tabular}
\hfill
\begin{tabular}{c}
\includegraphics[width=80mm]{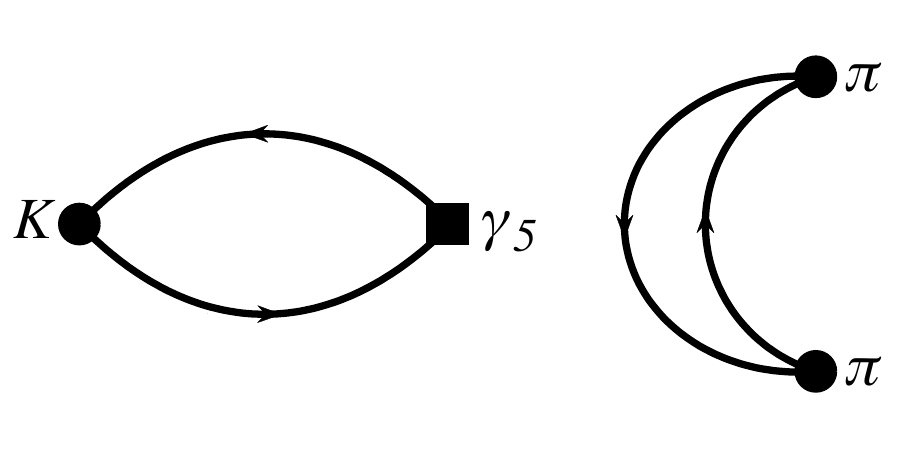}
\\
{\it type4}
\end{tabular}
\caption{
Diagrams for the $K\to\pi\pi$ three-point functions with the bilinear and a $\pi\pi$-like sink operators needed for the subtraction of power divergence arising from the loop diagrams in Figure~\ref{fig:diagrams_k2pipi}.
A line connecting the kaon ($K$) and bilinear operator ($\gamma_5$) represents a strange-quark propagator, while the all other lines correspond to a light-quark propagator.
}
\label{fig:diagrams_k2pipi_2q}
\end{center}
\end{figure}

\begin{figure}[tbp]
\begin{center}
\begin{tabular}{c}
\includegraphics[width=80mm]{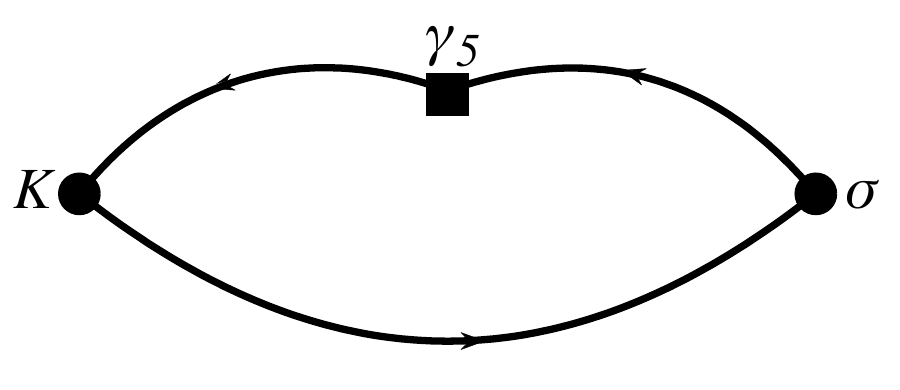}
\\
{\it type3}
\end{tabular}
\hfill
\begin{tabular}{c}
\includegraphics[width=80mm]{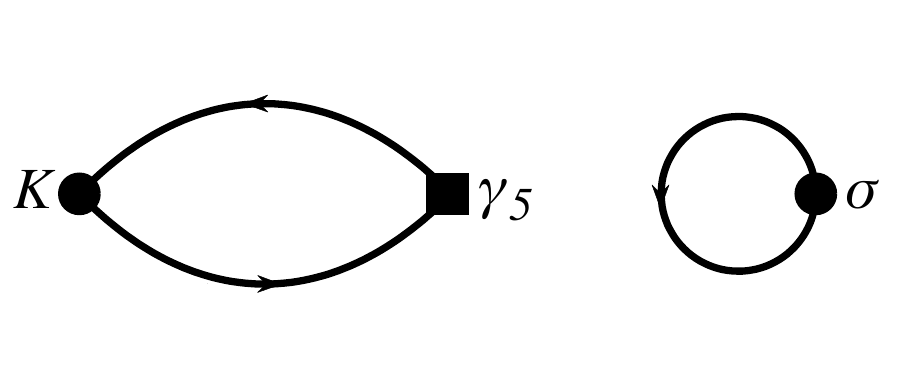}
\\
{\it type4}
\end{tabular}
\caption{
Same as Figure~\ref{fig:diagrams_k2pipi_2q} but for the subtraction of power divergence arising from the loop diagrams in Figure~\ref{fig:diagrams_k2sigma} with a $\sigma$-like sink operator.
}
\label{fig:diagrams_k2sigma_2q}
\end{center}
\end{figure}

To implement this subtraction for the $K\to\pi\pi$ three-point functions we need to calculate additional diagrams, which are summarized in Figures~\ref{fig:diagrams_k2pipi_2q} and \ref{fig:diagrams_k2sigma_2q} for $\pi\pi$-like and $\sigma$-like sink operators, respectively.
The contraction formulae for the correlation functions are also summarized in Appendix~\ref{sec:contractions}.

While the A2A quark propagator method allows us to take the average of three-point functions over all space-time translations, the cost of performing contractions on every space-time translation is comparable to, or even larger than, the cost of generating the A2A quark propagators.
However the diagrams that require a large fraction of the contraction cost are {\it type1} and {\it type2}, which are fully connected and can be calculated precisely with relatively fewer measurements.
In our previous GPBC calculation~\cite{RBC:2020kdj} we calculated {\it type1} and {\it type2} diagrams for every translation of eight time-slices and {\it type3} and {\it type4} diagrams every time-slice with a full three-dimensional volume average for all diagrams.
We observed that the {\it type4} diagram still dominated the statistical error while the cost for {\it type1} and {\it type2} diagrams was significant.

In this work we reduce the number of measurements of {\it type1} and {\it type2} diagrams in the spatial directions as well as for the time direction by calculating these diagrams on a uniform grid of $8^3$ sites, for eight time translations for the four-quark operator, while {\it type3} and {\it type4} diagrams are calculated for all space-time translations.

\begin{figure}[tbp]
\begin{center}
\begin{tabular}{c}
\includegraphics[width=80mm]{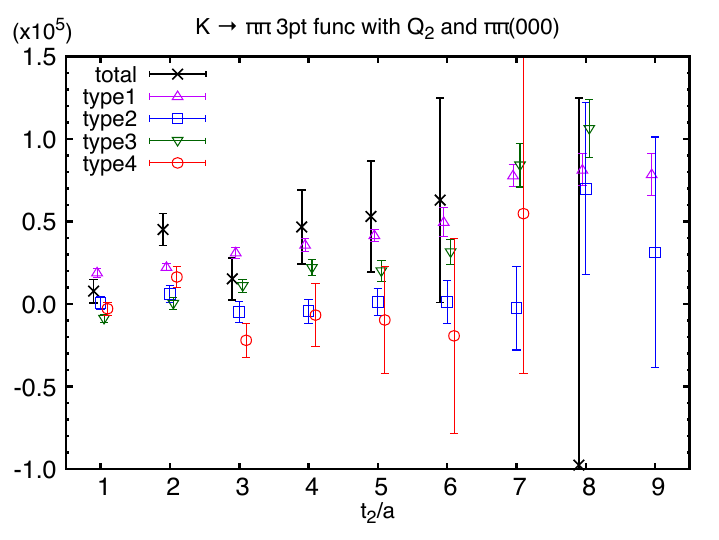}
\end{tabular}
\hfill
\begin{tabular}{c}
\includegraphics[width=80mm]{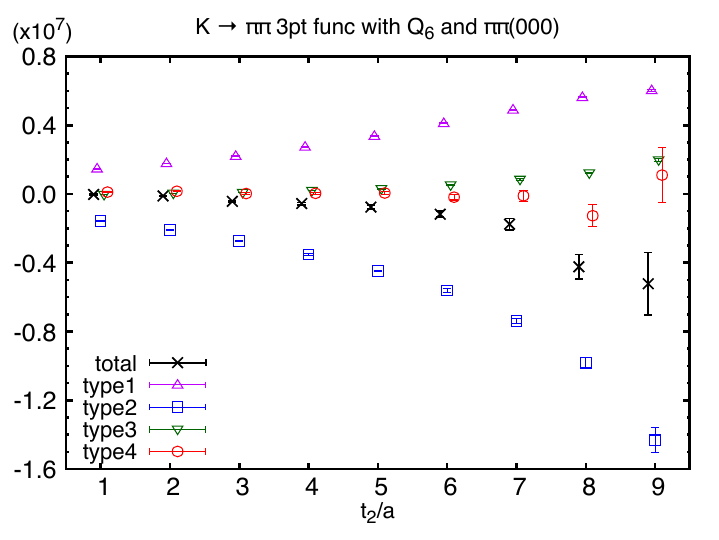}
\end{tabular}
\begin{tabular}{c}
\includegraphics[width=80mm]{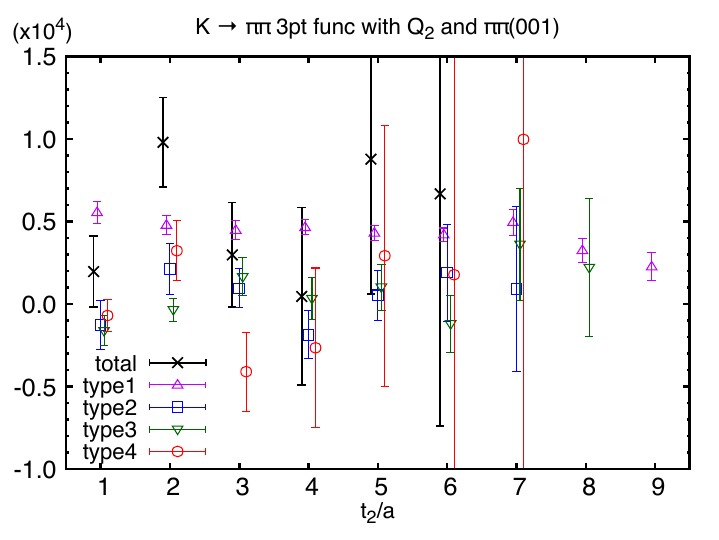}
\end{tabular}
\hfill
\begin{tabular}{c}
\includegraphics[width=80mm]{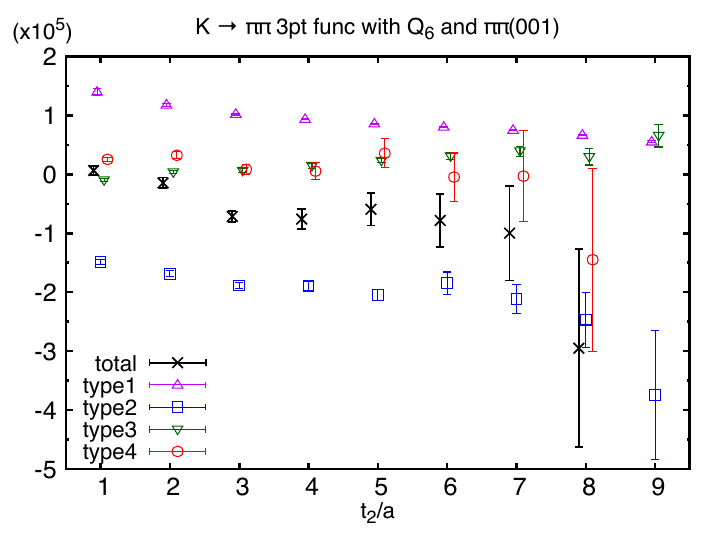}
\end{tabular}
\begin{tabular}{c}
\includegraphics[width=80mm]{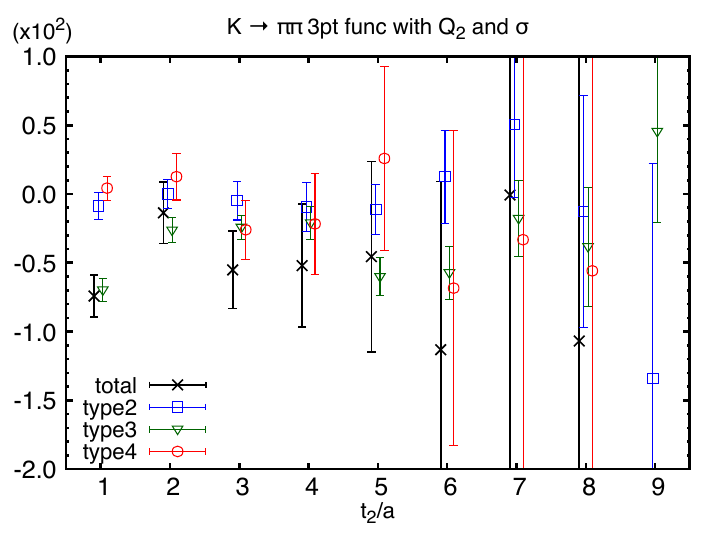}
\end{tabular}
\hfill
\begin{tabular}{c}
\includegraphics[width=80mm]{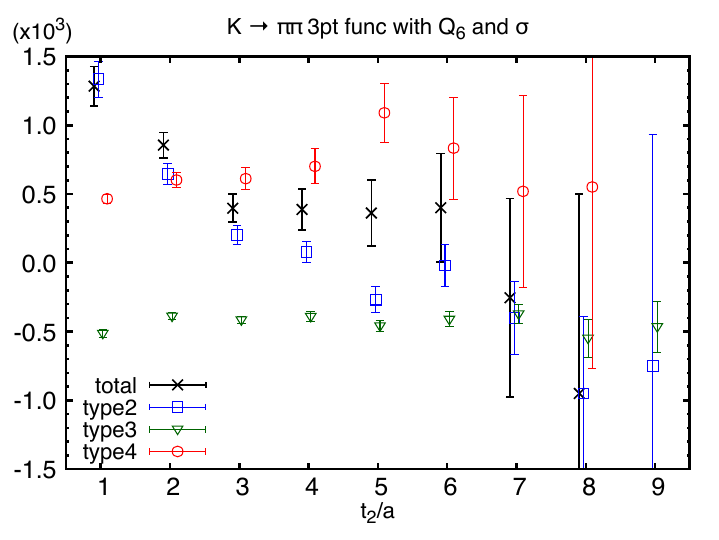}
\end{tabular}
\caption{Breakdown of the $\Delta I = 1/2$ channel of $K\to\pi\pi$ three-point functions at $t_1 + t_2 = 10a$ into contributions from each diagram.  Results with $Q_2$ (left) and $Q_6$ (right) operators, and with $\pi\pi(000)$ (upper), $\pi\pi(001)$ (middle) and $\sigma$ (lower) are shown in lattice units.  The power divergence of {\it type3} and {\it type4} diagrams is removed. 
}
\label{fig:3ptfunc}
\end{center}
\end{figure}

Figure~\ref{fig:3ptfunc} shows the breakdown of the $\Delta I=1/2$ $K\to\pi\pi$ three-point functions into contributions from each diagram at $t_1 + t_2 = 10a$.
We show the results with $Q_2$ and $Q_6$ operators, which provide the dominant contribution to Re($A_0$) and Im($A_0$), respectively, and with the $\pi\pi(000)$, $\pi\pi(001)$ and $\sigma$ operators, which strongly couple with either the ground or first-excited two-pion state.
While the {\it type4} diagrams, which are disconnected, are expected to dominate the statistical error, that is not always the case at small time separations $t_2 \le 3a$ between the two-pion and four-quark operators, where the errors on the {\it type2} diagrams are comparable or even larger that on {\it type4}. 
In addition to the intrinsic signal restoration of disconnected diagrams at short times, this should also be because of the reduction in the number of measurements for the {\it type1} and {\it type2} diagrams.
Note that we do not extract the matrix elements at such short time separations, where the contamination from higher excited states is still significant. 
At larger time separations, $t_2>3a$, on the other hand, the {\it type4} diagram dominates the statistical error.
This indicates that the cost reduction with less number of measurements of {\it type1} and {\it type2} diagrams does not significantly impact the total statistical precision of the correlation functions at time separations where the $K\to\pi\pi$ matrix elements are extracted.

\subsection{$K\to\pi\pi$ matrix elements}
\label{sec:Meff}

Using the eigenvectors obtained by solving the GEVP~\eqref{eq:GEVP}, we can extract the three-point functions with the contribution from a specific two-pion state labeled by $n$:
\begin{align}
    C_{n,i}^{\rm3pt}(t_1,t_2,\delta_t,t,t_0)
    &= \sum_a V_{n,a}(t,t_0,\delta_t)C_{ai}^{\rm3pt}(t_1,t_2)
    \notag\\
    &= B_n^{(t,t_0,\delta_t)} M_{n,i} A_K^* \e^{-m_Kt_1-E_nt_2} + \ldots,
\end{align}
where the ellipsis represents the contamination from excited states and potential thermal effects, and
\begin{equation}
    M_{n,i} = \left\langle \pi\pi(E_n) | Q_i | K\right\rangle.
\end{equation}
We define the effective matrix elements~\cite{Bulava:2011yz},
\begin{equation}
    M_{n,i}^{\rm eff}(t_1,t_2,\delta_t,t,t_0)
    = C_{n,i}^{\rm3pt}(t_1,t_2,\delta_t,t,t_0) R^K(t_1)R_n(t_2,\delta_t,t,t_0),
\end{equation}
with
\begin{align}
    R^K(t_1) &= \e^{m_K^{\rm eff}(t_1)t_1/2} \left[C^K(t_1)\right]^{-1/2},
    \\
    R_n(t_2,\delta_t,t,t_0)
    &= \left(1 - \e^{-E_n^{\rm eff}(t,t_0,\delta_t)\delta_t}\right)^{1/2}\e^{E^{\rm eff}_n(t,t_0,\delta_t)t_2/2}
    \left[C_n^{\rm2pt}(t_2,\delta_t,t,t_0)\right]^{-1/2}.
\end{align}
The factor $(1 - \e^{-E_n^{\rm eff}(t,t_0,\delta_t)\delta_t})^{1/2}$ is associated with the matrix subtraction of two-pion correlation functions in Eq.\eqref{eq:diagCorrpipi}.  Here we use the effective two-pion energy $E_n^{\rm eff}$ defined in Eq.~\eqref{eq:def_Eeff} rather than the improved one $E_n^{\rm eff\prime}$ given in Eqs.~\eqref{eq:DR} to compensate the exponential time dependence of the three-point functions. 
Following the discussion in Ref.~\cite{Bulava:2011yz}, we choose $t_0 = t - a = t_2$ and then one can reduce the number of time arguments and simplify the state-projected correlation functions,
\begin{equation}
    C_{n,i}^{\rm3pt}(t_1,t_2,\delta_t) = C_{n,i}^{\rm3pt}(t_1,t_2,\delta_t,t_2+a,t_2),
\end{equation}
and the effective matrix elements
\begin{equation}
    M_{n,i}^{\rm eff}(t_1,t_2,\delta_t) = M_{n,i}^{\rm eff}(t_1,t_2,\delta_t,t_2+a,t_2).
\end{equation}
We also limit our discussion to $\delta_t = 5a$ and $8a$ for the $\Delta I = 3/2$ and $1/2$ channels, respectively
These are the best choices, according to the discussion in our two-pion scattering companion paper~\cite{RBC:2023xqv}.  We do not expect any significant error reduction from further tuning of $\delta_t$ as the main source of statistical errors is the $K\to\pi\pi$ three-point functions themselves, which are independent of $\delta_t$.
Most of the results for the matrix elements we show below are obtained by using the four two-pion operators $\pi\pi(000)$, $\pi\pi(001)$, $\pi\pi(011)$ and $\pi\pi(111)$ for the $\Delta I = 3/2$ channel and the five operators including the additional $\sigma$-like operator for the $\Delta I = 1/2$ channel.  Sets with fewer two-pion operators are also employed when discussing systematic effects due to excited two-pion states.

\begin{figure}[tbp]
\begin{center}
\begin{tabular}{c}
\includegraphics[width=80mm]{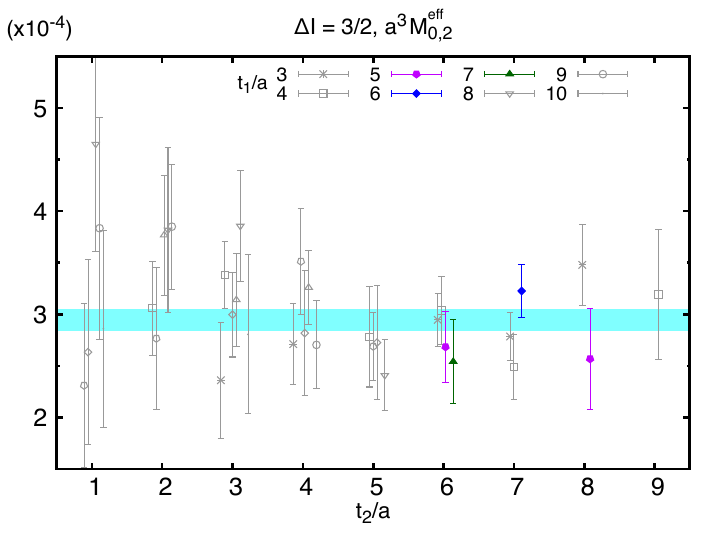}
\end{tabular}
\hfill
\begin{tabular}{c}
\includegraphics[width=80mm]{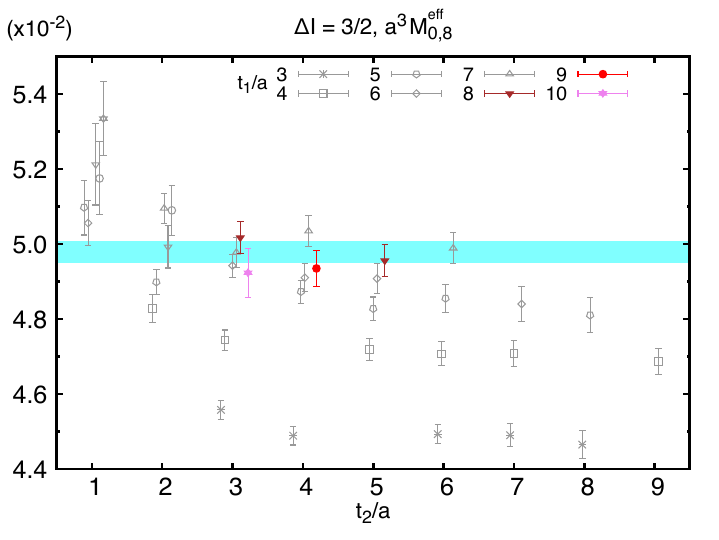}
\end{tabular}
\caption{Effective matrix elements of $\Delta I=3/2$ $K\to\pi\pi$ decay with the ground two-pion state for $Q_2$~(left) and $Q_8$ (right) plotted in lattice units.  The band represents the result for constant fit with the fit range $t_1\ge5a$ and $t_2\ge6a$ for $Q_2$ and $t_1\ge8a$ and $t_2\ge3a$ for $Q_8$.  The data points in the fit range are plotted as filled colored points, while unfilled gray points are out of the fit range.  
}
\label{fig:efmlm2_n0_fit}
\end{center}
\end{figure}

\begin{figure}[tbp]
\begin{center}
\begin{tabular}{c}
\includegraphics[width=80mm]{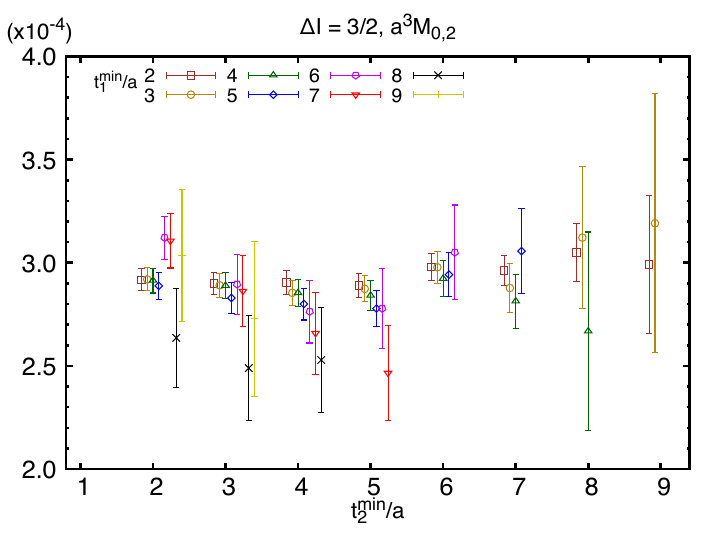}
\end{tabular}
\hfill
\begin{tabular}{c}
\includegraphics[width=80mm]{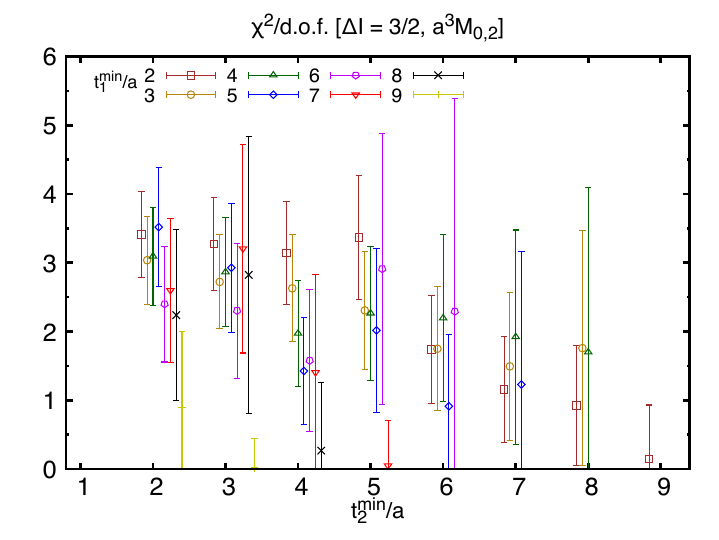}
\end{tabular}
\begin{tabular}{c}
\includegraphics[width=80mm]{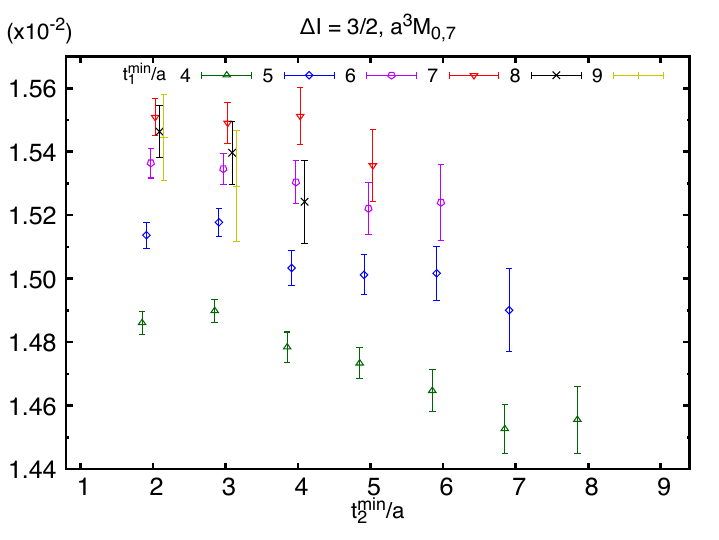}
\end{tabular}
\hfill
\begin{tabular}{c}
\includegraphics[width=80mm]{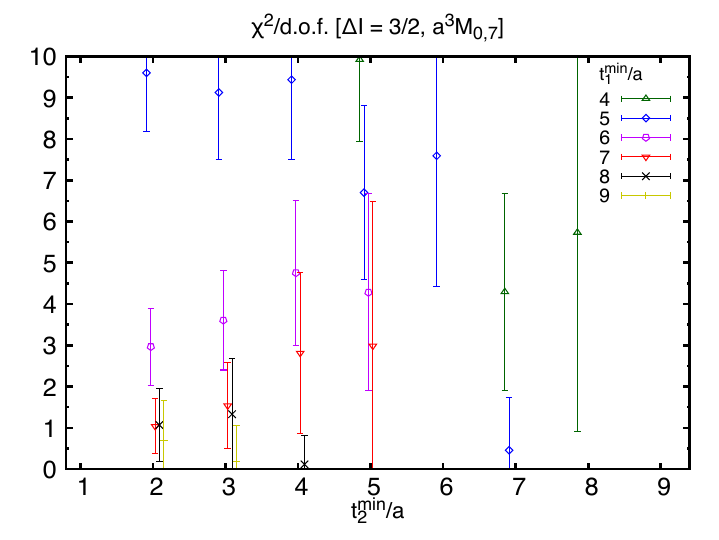}
\end{tabular}
\begin{tabular}{c}
\includegraphics[width=80mm]{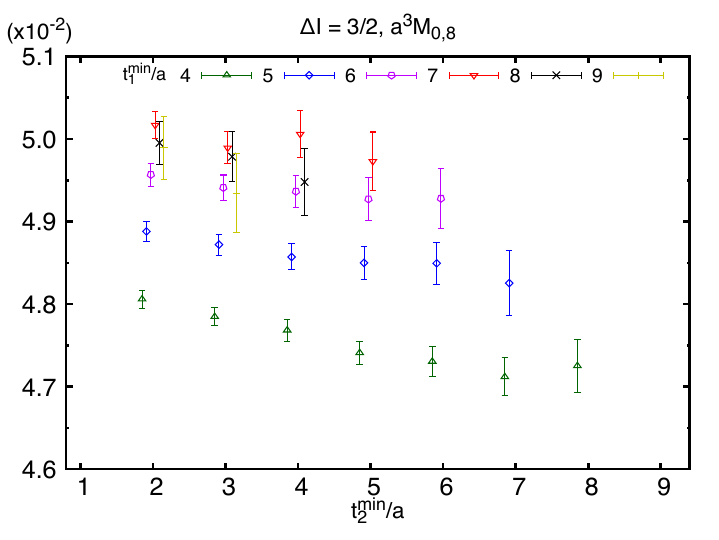}
\end{tabular}
\hfill
\begin{tabular}{c}
\includegraphics[width=80mm]{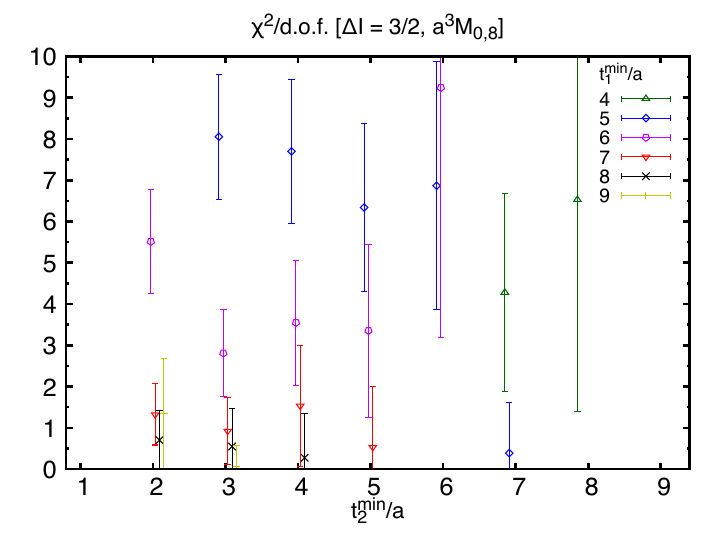}
\end{tabular}
\caption{Results for the $\Delta I = 3/2$ matrix elements (left) $M_{0,2}$ (upper), $M_{0,7}$ (middle) and $M_{0,8}$ (lower) with the ground two-pion final state obtained by correlated constant fits with various fit ranges indicated by $(t_1^{\rm min},t_2^{\rm min})$.  The corresponding values of $\chi^2/d.o.f.$ are shown on the right panels.  
}
\label{fig:range_dps_I2_n0}
\end{center}
\end{figure}

Figure~\ref{fig:efmlm2_n0_fit} shows the $\Delta I = 3/2$ effective $K\to\pi\pi$ matrix elements $M_{0,i}^{\rm eff}$ with the ground two-pion final state and with the four-quark operators labeled by $i=2$ and $i=8$.  
We do not see significant dependence on $t_2$.  The lack of $t_2$-dependence at small separations $t_2$ between the four-quark and two-pion operators implies small contamination from two-pion excited states excluded from the GEVP.  Similarly, the plateau at large $t_2$ with a fixed separation $t_1$ between the kaon and four-quark operators, indicates that the thermal effects are negligible.
Throughout the paper, we do not see significant thermal effects.
The band represents the result for a correlated constant fit with the range explained in the caption.
To visually distinguish the fit range, filled colored symbols  denote data points that are used in the fit while unfilled gray points are not.

Figure~\ref{fig:range_dps_I2_n0} shows the results for a correlated constant fit to the effective matrix elements $M_{0,2}^{\rm eff}$, $M_{0,7}^{\rm eff}$ and $M_{0,8}^{\rm eff}$ with various fit ranges $t_1\ge t_1^{\rm min}$ and $t_2\ge t_2^{\rm min}$ plotted in lattice units along with the corresponding values of $\chi^2/d.o.f.$
Despite the wide plateau for $M_{0,2}^{\rm eff}$ appearing in the left panel of Figure~\ref{fig:efmlm2_n0_fit}, Figure~\ref{fig:range_dps_I2_n0} indicates that enlarging the (correlated) fit range of $t_1$ significantly decreases the statistical error while increasing the value of  $\chi^2/d.o.f.$ This may indicate that the results with wider fit ranges could still receive significant excited-state contamination.
In order to minimize such systematic errors, only the four data points that satisfy $t_1\ge5a$ and $t_2\ge6a$ are used for the fit to quote the final result.
On the other hand $M_{0,8}^{\rm eff}$ in Figure~\ref{fig:efmlm2_n0_fit} has a significant dependence on the time separation $t_1$ between the kaon and four-quark operators, while much smaller dependence on the separation $t_2$ between the pion and four-quark operators is observed for each value of $t_1$.
The group of points with $t_1 = a$ deviates significantly from groups with $t_1 > a$, and smaller deviations are observed for larger values of $t_1$.
This indicates that the contamination from kaon excited states is quite significant. 
For $t_1 + t_2 = 13a$ the consecutive data points at $(t_1/a,t_2/a) = (10,3)$, $(9,4)$ and $(8,5)$ appear consistent within statistical precision.
We therefore choose a fit range that satisfies 
$t_1\ge8a$ and $t_2\ge3a$.
The same trend is observed in Figure~\ref{fig:range_dps_I2_n0} for $M_{0,7}$ and therefore we choose the same fit range for $M_{0,7}$ and $M_{0,8}$.

\begin{figure}[tbp]
\begin{center}
\begin{tabular}{c}
\includegraphics[width=80mm]{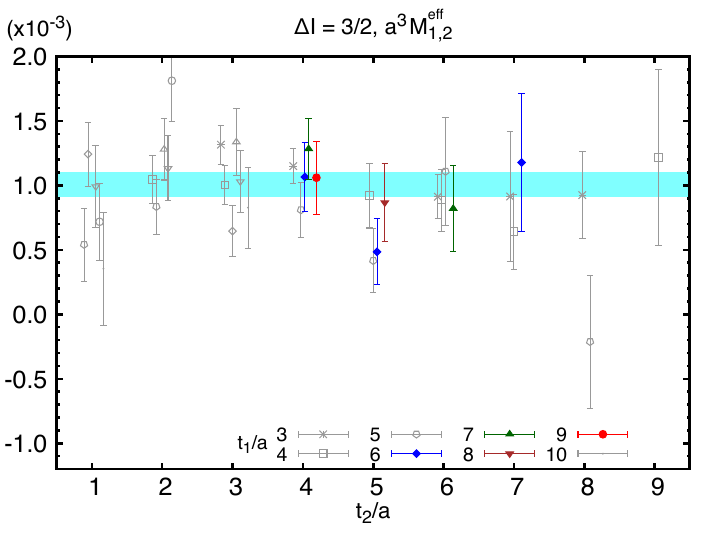}
\end{tabular}
\hfill
\begin{tabular}{c}
\includegraphics[width=80mm]{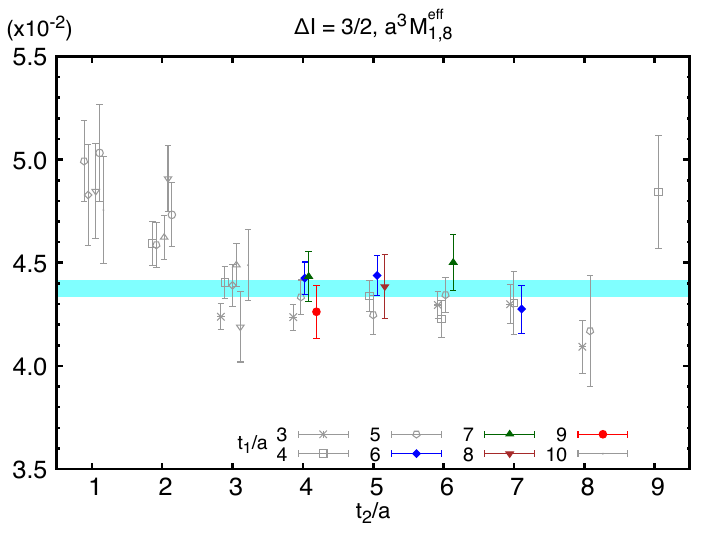}
\end{tabular}
\caption{Effective matrix elements of $\Delta I=3/2$ $K\to\pi\pi$ decay with the first-excited $\pi\pi$ state for $Q_2$~(left) and $Q_8$ (right).  The band represents the result for constant fit with the fit range $t_1\ge6a$ and $t_2\ge5a$ for $Q_2$ and $t_1\ge6a$ and $t_2\ge4a$ for $Q_8$.  The data points in the fit range are plotted as filled colored points, while unfilled gray points are out of the fit range.  
}
\label{fig:efmlm2_n1_fit}
\end{center}
\end{figure}

\begin{figure}[tbp]
\begin{center}
\begin{tabular}{c}
\includegraphics[width=80mm]{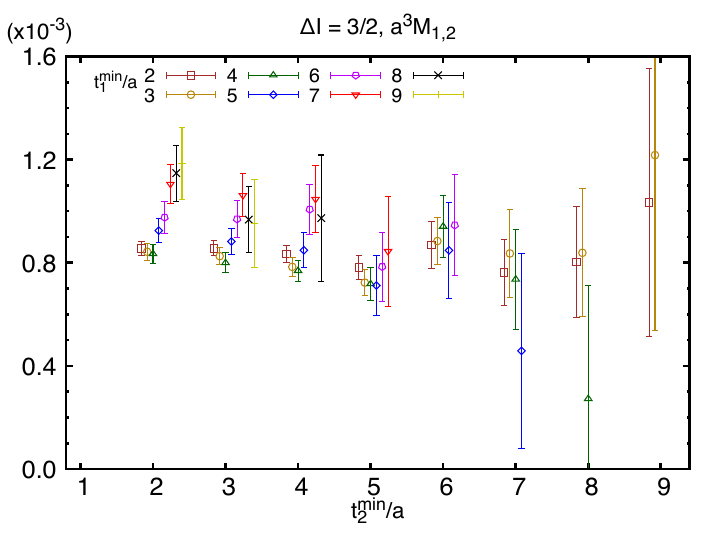}
\end{tabular}
\hfill
\begin{tabular}{c}
\includegraphics[width=80mm]{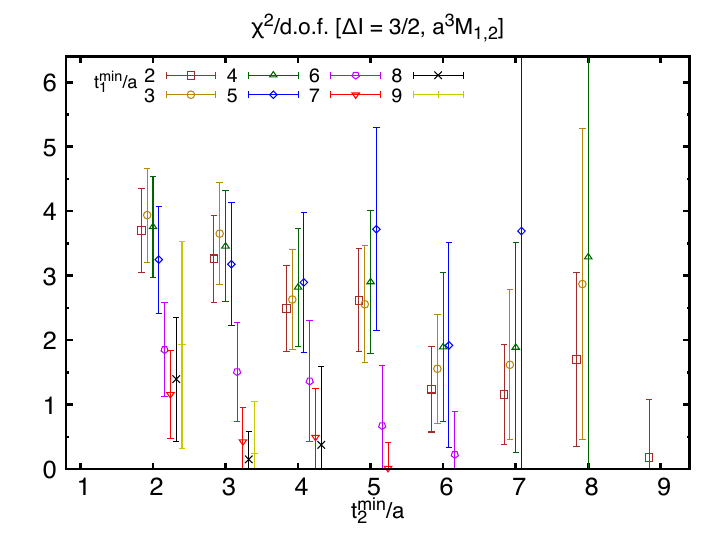}
\end{tabular}
\begin{tabular}{c}
\includegraphics[width=80mm]{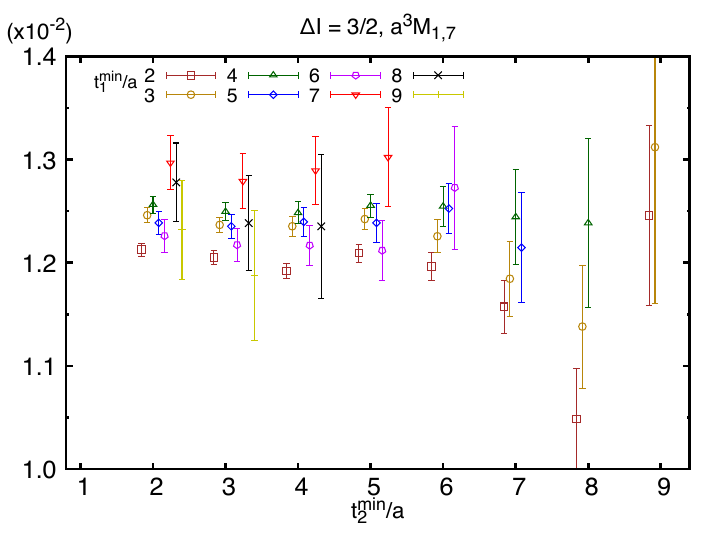}
\end{tabular}
\hfill
\begin{tabular}{c}
\includegraphics[width=80mm]{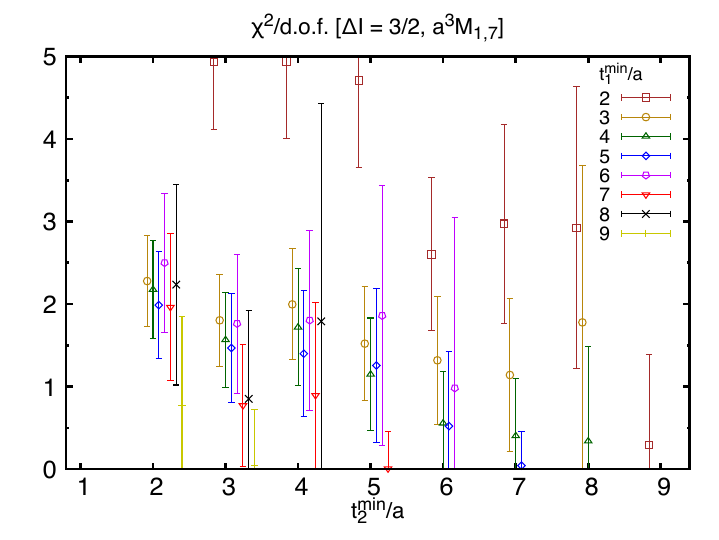}
\end{tabular}
\begin{tabular}{c}
\includegraphics[width=80mm]{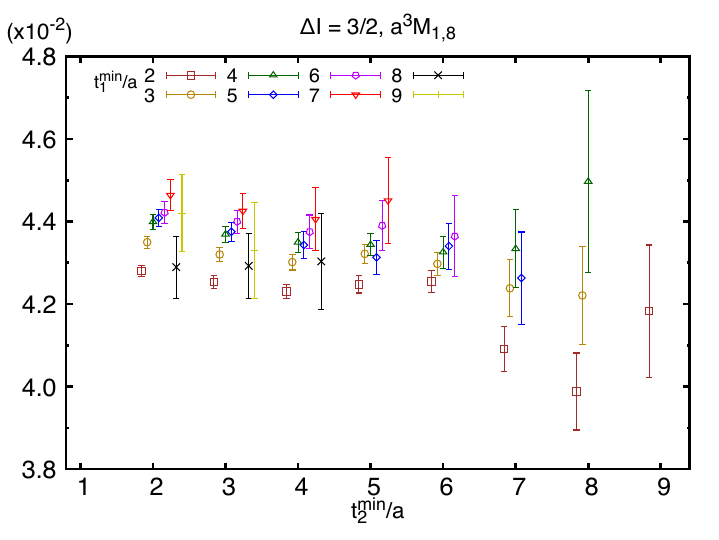}
\end{tabular}
\hfill
\begin{tabular}{c}
\includegraphics[width=80mm]{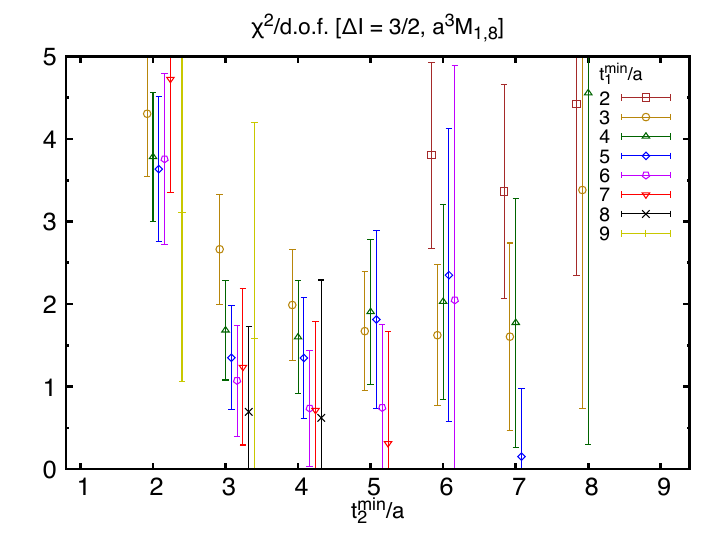}
\end{tabular}
\caption{Same as Figure~\ref{fig:range_dps_I2_n0} but results with the first-excited two-pion final state.
}
\label{fig:range_dps_I2_n1}
\end{center}
\end{figure}

Figure~\ref{fig:efmlm2_n1_fit} shows the $\Delta I = 3/2$ effective $K\to\pi\pi$ matrix elements $M_{1,i}^{\rm eff}$ with the first-excited two-pion state and with the four-quark operators labeled by $i=2$ and $i=8$ along with the result (band) for a correlated constant fit with the range indicated in the caption.
The fit results for $M_{1,2}$, $M_{1,7}$ and $M_{1,8}$ with  various fit ranges and the corresponding values of $\chi^2/d.o.f.$ are shown in Figure~\ref{fig:range_dps_I2_n1}.
Similar to the matrix elements with the ground two-pion final state, the fit range for $M_{1,2}$ needs to be limited to have a reasonably small value of $\chi^2/d.o.f.$ despite an apparent wider plateau observed in the left panel of Figure~\ref{fig:efmlm2_n1_fit}.
The contamination from excited kaon states in $M_{1,8}^{\rm eff}$ appears less than that in $M_{0,8}^{\rm eff}$.

\begin{figure}[tbp]
\begin{center}
\begin{tabular}{c}
\includegraphics[width=80mm]{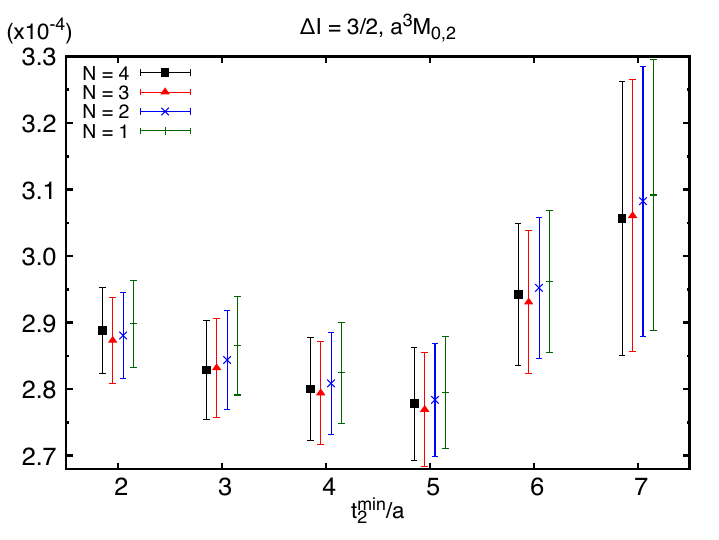}
\end{tabular}
\hfill
\begin{tabular}{c}
\includegraphics[width=80mm]{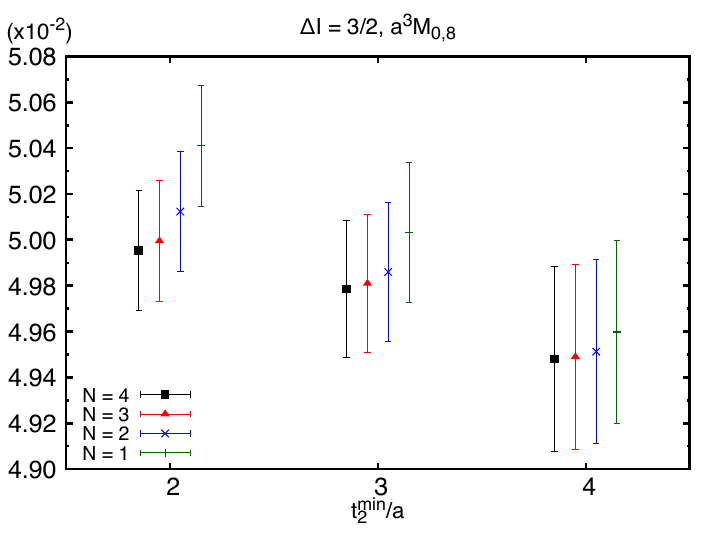}
\end{tabular}
\begin{tabular}{c}
\includegraphics[width=80mm]{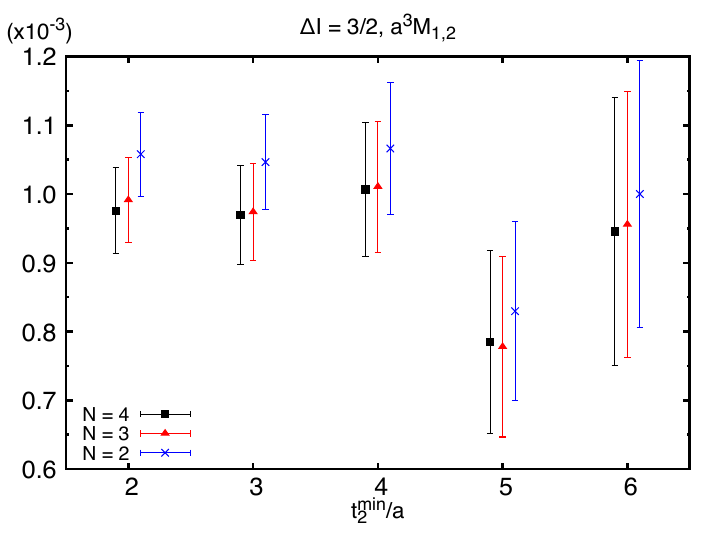}
\end{tabular}
\hfill
\begin{tabular}{c}
\includegraphics[width=80mm]{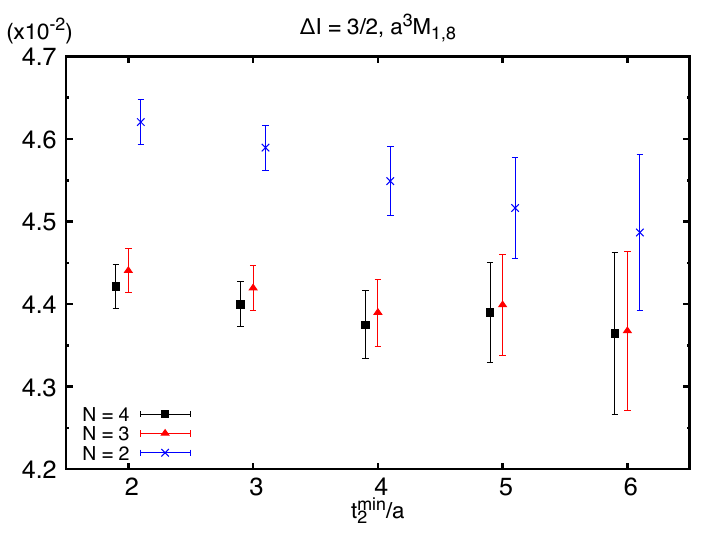}
\end{tabular}
\caption{Fit results for $\Delta I = 3/2$ matrix elements with different GEVP sizes $N=2,3$ and 4.  Results for $M_{0,2}$ (upper/left) at $t_1^{\rm min}=5a$, $M_{0,8}$ (upper/right) at $t_1^{\rm min}=8a$, $M_{1,2}$ (lower/right) and $M_{1,8}$ (lower/left) at $t_1^{\rm min}=6a$ are shown.  The operators with one of the smallest $N$ relative momenta are included in each analysis.  The results with a single two-pion operator $\pi\pi(000)$ without GEVP ($N=1$) are also shown in the upper panels. 
}
\label{fig:gevp_dps_I2}
\end{center}
\end{figure}

To investigate the contamination from neglected excited two-pion states, we perform the same analyses with fewer two-pion operators.  Figure~\ref{fig:gevp_dps_I2} shows the fit results for the $\Delta I = 3/2$ matrix elements with various GEVP sizes $N$ for the ground and first-excited two-pion final states at fixed value of $t_1^{\rm min}$ specified in the caption.
We observe somewhat noticeable deviation at large $t_2^{\rm min}$ such as the difference of $M_{1,2}$ between $t_2^{\rm min}=4a$ and $5a$.  While 1--2$\sigma$ deviation can be caused by either or both of statistical and systematic effects and it is not trivial to distinguish, here we demonstrate that such `$t_2^{\rm min}$-dependence' is not due to two-pion excited states but likely caused by a statistical fluctuation.
Note that the results with $N=3$ and 4 are mostly consistent, while the GEVP up to $N=4$ successfully decomposes $N$ two-pion states for the $I=2$ channel up to $t_0(=t_2) \sim 6a$ as seen in Figure~\ref{fig:efms0_24c}. 
Because of the consistency between $N=3$ and 4 elsewhere and the well resolved signal of the $N=4$ state, it is natural to expect the contamination from excited two-pion states is negligible and some deviation of the data points seen at large $t_2^{\rm min}$ should be due to a statistical fluctuation.

Since we did not perform measurements with multiple kaon operators, we are not able to investigate potential contamination from excited kaon states in the same way as for excited two-pion states.  In Figure~\ref{fig:range_dps_I2_n1} we see fluctuations with varying $t_1^{\rm min}$ for the matrix elements with the $I=2$ two-pion first-excited state.  However these fluctuations are of marginal statistical significance and occur in both directions.  These may not necessarily be due to kaon excited states but could also be due to limited statistics similar to the aforementioned example with two-pion states.  
Most of the fit results for the matrix elements at fixed $t_2^{\rm min}$
in Figures~\ref{fig:range_dps_I2_n0} and \ref{fig:range_dps_I2_n1}
display	smooth behavior	at small values of $t_1^{\rm min}$ consistent with excited state contamination which should smoothly decrease to zero,
but transition to fluctuating at a larger $t_1^{\rm min}$ more consistent with statistical fluctuation.
Excited-state contamination is unlikely to be the cause of these fluctuations at larger $t_1$, and we conclude these are merely statistical fluctuations.

\begin{table*}[tbp]
\centering
\begin{tabular}{|c|cc|cc|cc|}
\hline
$i$ & $(t_1^{min},t_2^{min})$ & $a^3M_{0,i}$ & $(t_1^{min},t_2^{min})$ & $a^3M_{1,i}$ & $(t_1^{min},t_2^{min})$ & $a^3M_{2,i}$\\
\hline
1 & $(5a,6a)$ & $0.0002885(93)$[0.9] & $(6a,4a)$ & $0.001005(97)$[1.4] & $(5a,4a)$ & $0.00209(27)$[1.0]\\
2 & $(5a,6a)$ & $0.000294(11)$[0.9] & $(6a,4a)$ & $0.001006(97)$[1.4] & $(5a,4a)$ & $0.00210(26)$[1.0]\\
7 & $(8a,3a)$ & $0.015397(99)$[1.3] & $(6a,4a)$ & $0.01217(19)$[1.8] & $(5a,4a)$ & $0.01319(61)$[1.0]\\
8 & $(8a,3a)$ & $0.04979(30)$[0.5] & $(6a,4a)$ & $0.04375(41)$[0.7] & $(5a,4a)$ & $0.0479(10)$[1.8]\\
9 & $(5a,6a)$ & $0.000433(14)$[0.9] & $(6a,4a)$ & $0.00151(15)$[1.4] & $(5a,4a)$ & $0.00314(40)$[1.0]\\
10 & $(5a,6a)$ & $0.000441(16)$[0.9] & $(6a,4a)$ & $0.00151(15)$[1.4] & $(5a,4a)$ & $0.00315(40)$[1.0]\\
\hline
\end{tabular}
\caption{Fit results for the $\Delta I = 3/2$ channel of the $K\to\pi\pi$ matrix elements with the ground ($M_{0,i}$), first-excited ($M_{1,i}$) and second-excited ($M_{2,i}$) two-pion states and with the four-quark operators relevant for this isospin channel. The fit range is indicated by $(t_1^{min},t_2^{min})$, which means data points that satisfy $t_1\ge t_1^{min}$ and $t_2\ge t_2^{min}$ are used for the fit.  For the ground-state matrix elements we choose different fit ranges depending on four-quark 
operator.  The values of $\chi^2/d.o.f.$ are shown in the square brackets.}
\label{tab:fitres_ME_I2}
\end{table*}

Table~\ref{tab:fitres_ME_I2} summarizes the fit results for the $\Delta I = 3/2$ channel of $K\to\pi\pi$ matrix elements $M_{n,i}$ with $n=0,1,2$ and the current-current and electroweak penguin four-quark operators. 
Because of the significantly different trends of the excited-state contamination in the matrix elements with the ground two-pion state ($n=0$) between $i=2$ and $i=7,8$ described above, we quote the results with different fit ranges depending on the four-quark operator for the ground two-pion final state.  For excited two-pion states, it is not necessary to tune the fit range depending on four-quark operator, and we choose a common fit range for each final state
for simplicity.
The results for $i=1,2,9$ and 10 are linearly dependent for this isospin channel.  The difference between $M_{n,1}$ and $M_{n,2}$ is caused by the slight violation of the Fierz symmetry due to the use of the stochastic A2A method, which approximates the quark propagators in the given gauge configuration.  Since three-point functions before and after Fierz transformation are calculated by taking the spin and color contractions quite differently, the difference between the exact and approximated quark propagators can cause the difference between the three-point functions and hence the corresponding matrix elements.
The quark propagator becomes exact in the large-hit or large-ensemble limit.
Both results are thus correct within the statistical error, $e.g.$ we observe the jackknife average of the difference $a^3(M_{0,1}-M_{0,2}) = -5.7(5.7)\times10^{-6}$ and $a^3(M_{1,1}-M_{1,2}) = -1.6(1.5)\times10^{-6}$.  We gain a minor improvement in the statistical precision by taking the average.  However, the matrix elements with $i=9$ and 10 are exactly the same as those with $i=1$ and 2, respectively, up to the overall factor of $3/2$ as the identical contractions are used for computing the three-point functions for these pairs (see Appendix~\ref{sec:contractions}).

\begin{figure}[tbp]
\begin{center}
\begin{tabular}{c}
\includegraphics[width=80mm]{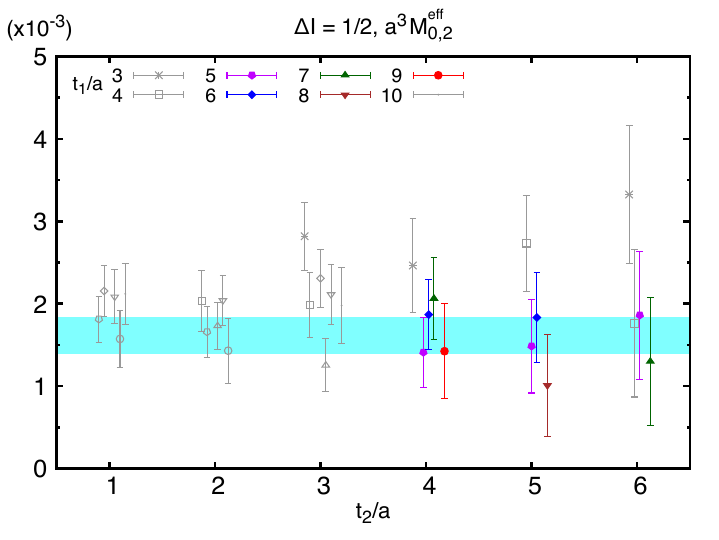}
\end{tabular}
\hfill
\begin{tabular}{c}
\includegraphics[width=80mm]{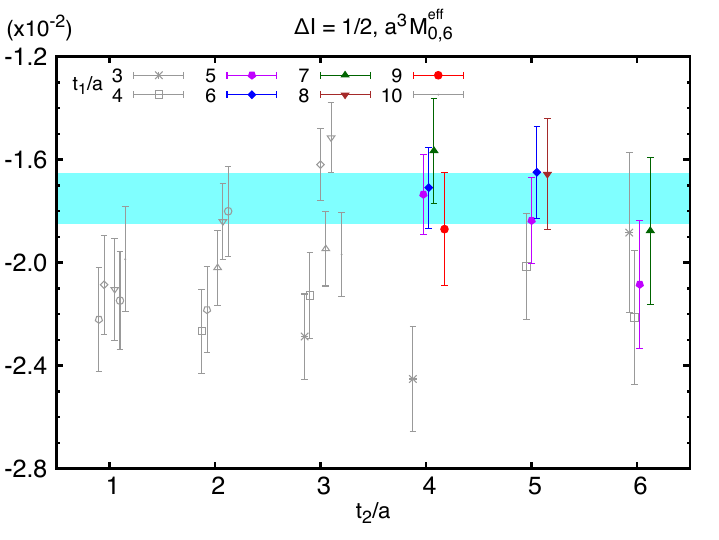}
\end{tabular}
\caption{Effective matrix elements of $\Delta I=1/2$ $K\to\pi\pi$ decay with the ground two-pion state for $Q_2$~(left) and $Q_6$~(right).  The band represents the result for a correlated constant fit with the fit range  $t_1\ge5a$ and $t_2\ge4a$ for both $Q_2$ and  $Q_6$.  The data points in the fit range are plotted as filled colored points, while unfilled gray points are out of the fit range. 
}
\label{fig:efmlm0_n0_fit}
\end{center}
\end{figure}

\begin{figure}[tbp]
\begin{center}
\begin{tabular}{c}
\includegraphics[width=80mm]{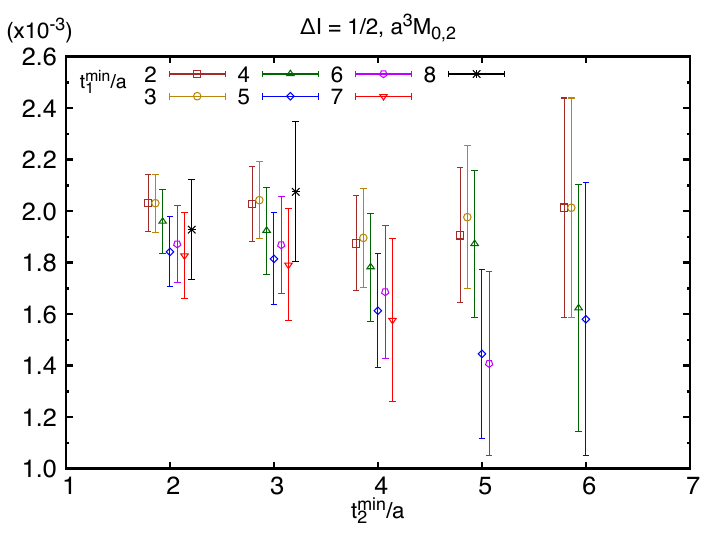}
\end{tabular}
\hfill
\begin{tabular}{c}
\includegraphics[width=80mm]{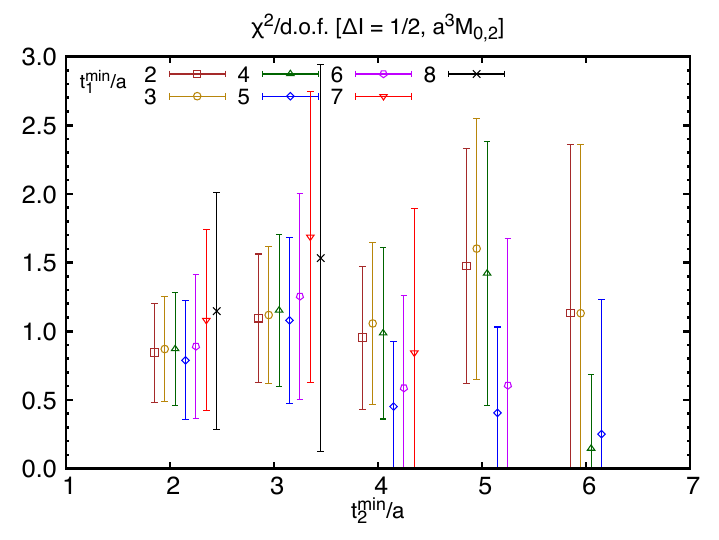}
\end{tabular}
\begin{tabular}{c}
\includegraphics[width=80mm]{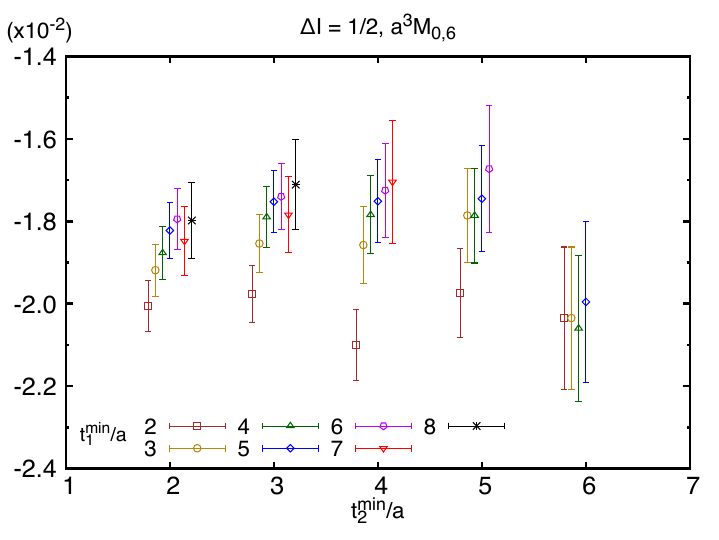}
\end{tabular}
\hfill
\begin{tabular}{c}
\includegraphics[width=80mm]{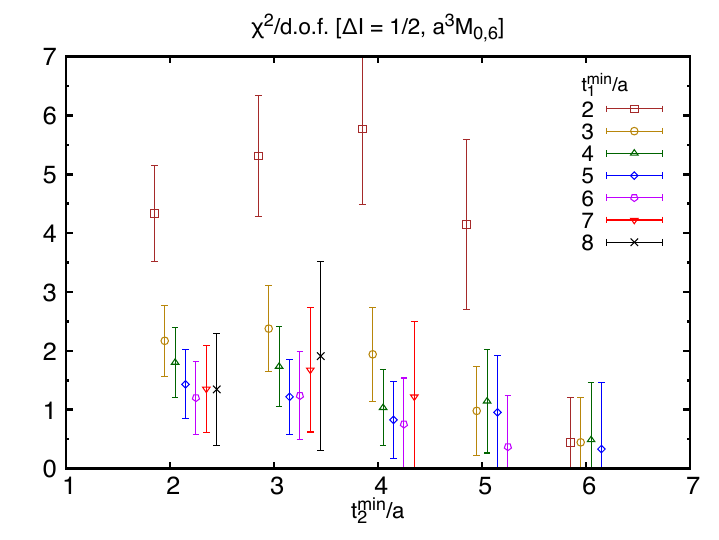}
\end{tabular}
\caption{Same as figure~\ref{fig:range_dps_I2_n0} but the results for the $\Delta I = 1/2$ matrix elements $M_{0,2}$ (upper) and $M_{0,6}$ (lower) with the ground two-pion state.}
\label{fig:range_dps_I0_n0}
\end{center}
\end{figure}

Figure~\ref{fig:efmlm0_n0_fit} shows the $\Delta I = 1/2$ effective $K\to\pi\pi$ matrix elements $M_{0,i}^{\rm eff}$ with the ground two-pion state and the four-quark operators labeled by $i=2$ and $i=6$ along with the result for a correlated constant fit with the range $(t_1^{\rm min},t_2^{\rm min})=(5a,4a)$.
The fit results with other various fit ranges and the corresponding values of $\chi^2/d.o.f.$ are shown in Figure~\ref{fig:range_dps_I0_n0}.
Unlike the case of the $\Delta I = 3/2$ channel the values of $\chi^2/d.o.f.$ are reasonably small for most of the fit ranges and less dependent on fit range.  In addition we do not see a significant difference in the trend of the fit range dependence for different four-quark operators.  For simplicity we choose a common fit range for all four-quark operators when we quote the final results.

\begin{figure}[tbp]
\begin{center}
\begin{tabular}{c}
\includegraphics[width=80mm]{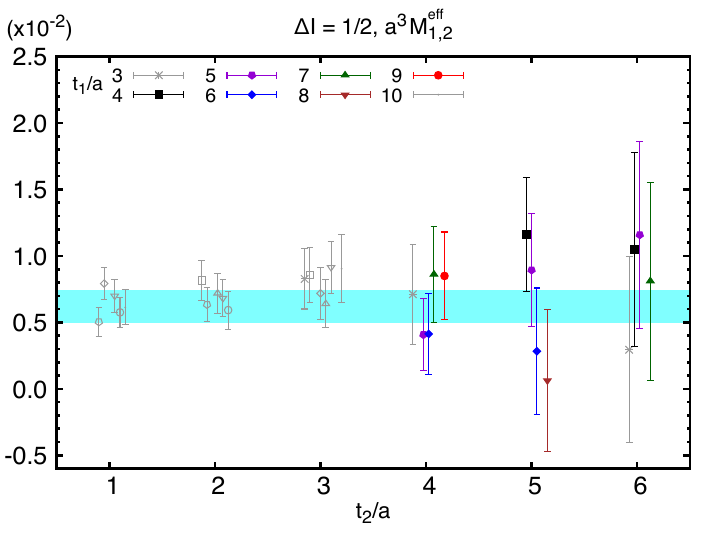}
\end{tabular}
\hfill
\begin{tabular}{c}
\includegraphics[width=80mm]{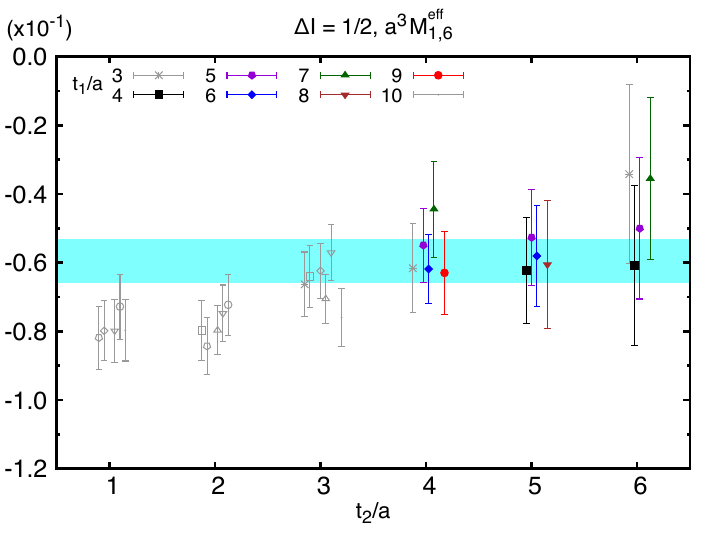}
\end{tabular}
\caption{Effective matrix elements of $I=0$ $K\to\pi\pi$ decay with near on-shell kinematics for $Q_2$~(left) and $Q_6$~(right).  The band represents the result for constant fit with the fit range $t_1\ge4a$ and $t_2\ge4a$ for both $Q_2$ and $Q_6$.  The data points in the fit range are plotted as filled colored points, while unfilled gray points are out of the fit range.  
}
\label{fig:efmlm0_n1_fit}
\end{center}
\end{figure}

\begin{figure}[tbp]
\begin{center}
\begin{tabular}{c}
\includegraphics[width=80mm]{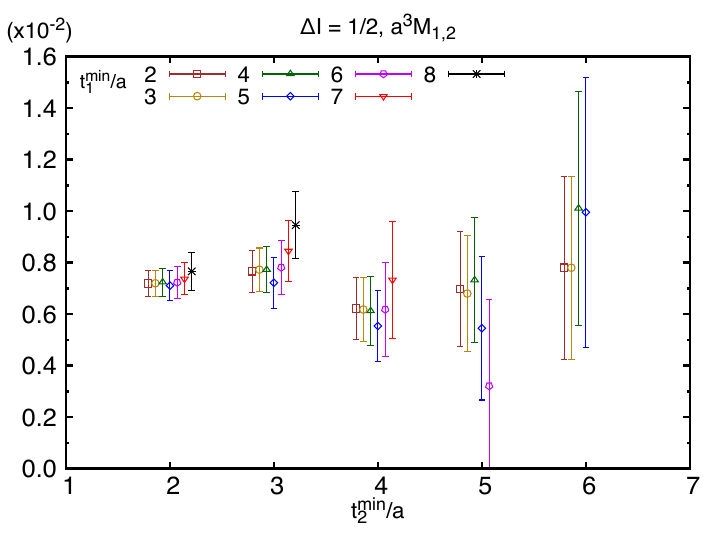}
\end{tabular}
\hfill
\begin{tabular}{c}
\includegraphics[width=80mm]{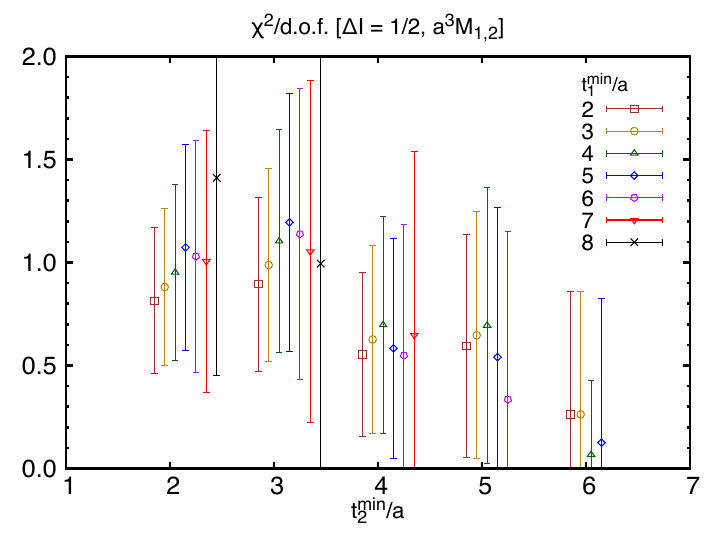}
\end{tabular}
\begin{tabular}{c}
\includegraphics[width=80mm]{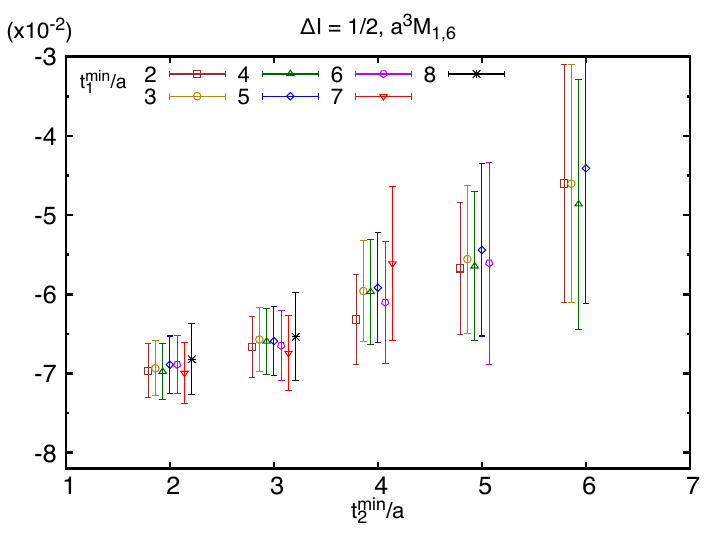}
\end{tabular}
\hfill
\begin{tabular}{c}
\includegraphics[width=80mm]{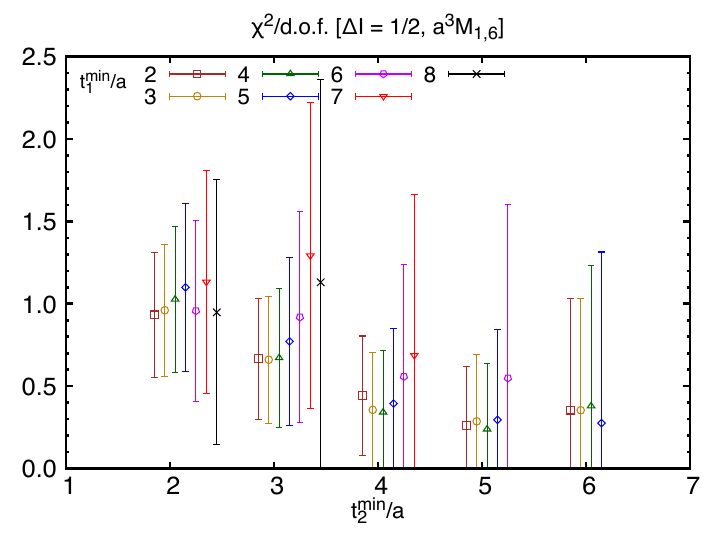}
\end{tabular}
\caption{Same as figure~\ref{fig:range_dps_I2_n0} but the results for the $\Delta I = 1/2$ matrix elements $M_{1,2}$ (upper) and $M_{1,6}$ (lower) with the first-excited two-pion state.}
\label{fig:range_dps_I0_n1}
\end{center}
\end{figure}

Figure~\ref{fig:efmlm0_n1_fit} shows the $\Delta I = 1/2$ effective $K\to\pi\pi$ matrix elements $M_{1,i}^{\rm eff}$ with the first-excited two-pion state and with the four-quark operators labeled by $i=2$ and $i=6$ along with the result for a correlated constant fit with the range $(t_1^{\rm min},t_2^{\rm min})=(4a,4a)$.
The fit results with other various fit ranges and the corresponding values of $\chi^2/d.o.f.$ are shown in Figure~\ref{fig:range_dps_I0_n1}.
These matrix elements are the main targets of this work as they correspond to $\Delta I = 1/2$ channel of near on-shell $K\to\pi\pi$ matrix elements, which are necessary to calculate the measure $\varepsilon'$ of direct $CP$ violation and require considerable effort to compute.
Similar to the case of the ground two-pion final state, the $\chi^2/d.o.f.$ is fairly stable for various fit ranges.
One noticeable difference from the ground-state case is that the contamination from kaon excited states is less significant so that we can choose a wider fit range with a smaller $t_1^{\rm min}$.  Again we use the same fit range for all four-quark operators to quote the final result.

\begin{figure}[tbp]
\begin{center}
\begin{tabular}{c}
\includegraphics[width=80mm]{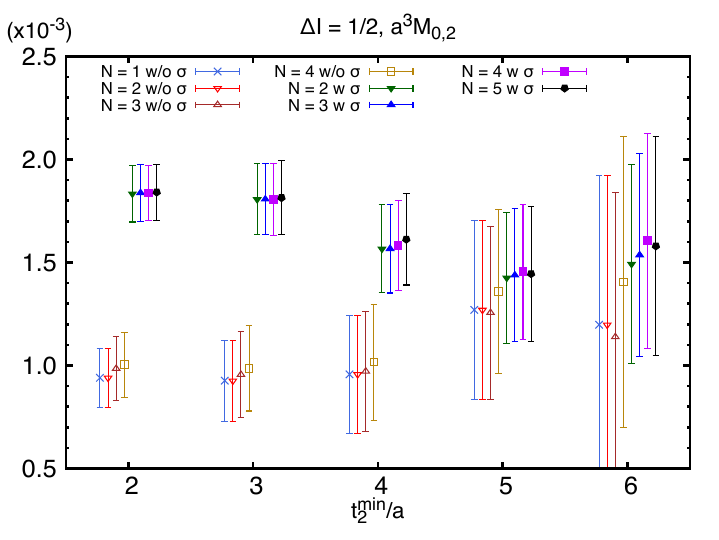}
\end{tabular}
\hfill
\begin{tabular}{c}
\includegraphics[width=80mm]{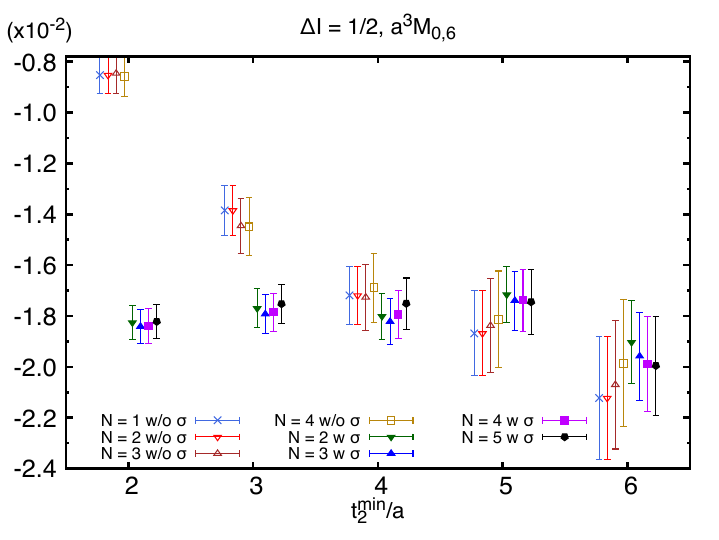}
\end{tabular}
\begin{tabular}{c}
\includegraphics[width=80mm]{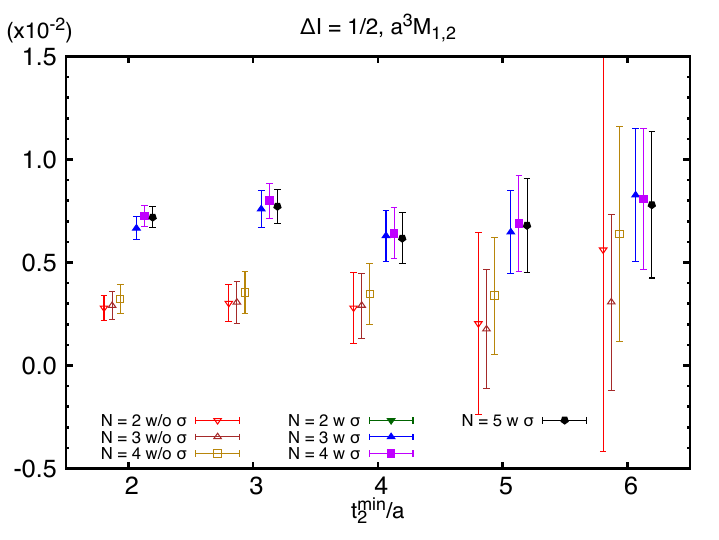}
\end{tabular}
\hfill
\begin{tabular}{c}
\includegraphics[width=80mm]{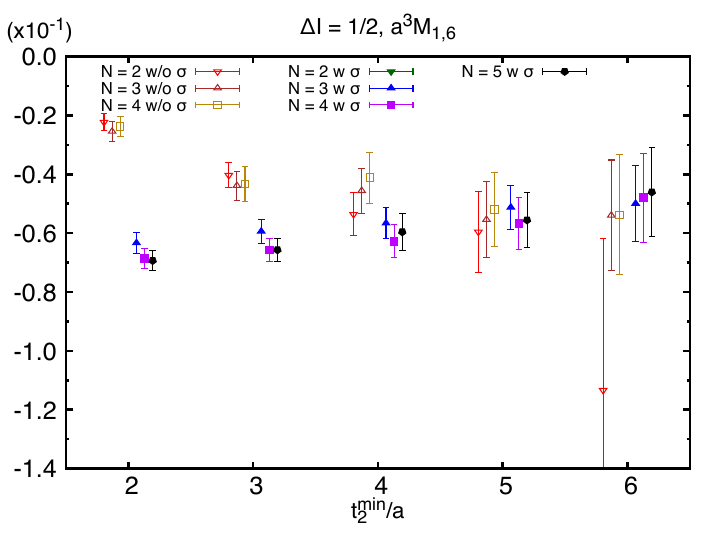}
\end{tabular}
\caption{Fit results for $\Delta I = 1/2$ matrix elements $M_{0,2}$ (upper/left) and $M_{0,6}$ (upper/right) at $t_1^{\rm min}=5a$, $M_{1,2}$ (lower/left) and $M_{1,6}$ (lower/right) at $t_1^{\rm min}=4a$ with various sets of two-pion operators used for the GEVP procedure.  $N$ stands for the number of two-pion operators in the set of operators and ``w $\sigma$'' indicates the set includes the $\sigma$-like operator, while the operator sets indicated by ``w/o $\sigma$'' do not.  Besides the $\sigma$-like operator the $\pi\pi(000)$ operator is always included, while the $\pi\pi(001)$, $\pi\pi(011)$ and $\pi\pi(111)$ are added to the operator set in this order until the number of operators reaches $N$.
}
\label{fig:gevp_dps_I0}
\end{center}
\end{figure}

Figure~\ref{fig:gevp_dps_I0} shows the fit results for $\Delta I = 1/2$ matrix elements obtained through the GEVP analyses with various sets of two-pion sink operators with and without the $\sigma$-like operator.  
The upper panels indicate that, while the results for the matrix elements with the two-pion ground state are mostly independent of the $\pi\pi$-like operators with a non-zero relative momentum, adding the $\sigma$-like operator makes a significant impact up to $t_2^{\rm min}=4a$ for $M_{0,2}$ and $t_2^{\rm min}=3a$ for $M_{0,6}$.  The benefit of including the $\sigma$-like operator is quite apparent in $M_{0,6}$, which shows a plateau beginning at $t_2^{\rm min}=2a$ with the $\sigma$-like operator, whereas the values differ by more than 50\% in this low-time region when the $\sigma$-like operator is excluded.
A similar trend is observed for the matrix elements with the first-excited two-pion state shown on the lower panels.  
The only difference in the trend between the ground- and first-excited two-pion final states is observed for ``$N=2$ w $\sigma$,'' the results with the two operators $\pi\pi(000)$ and $\sigma$, where the matrix elements with the first-excited state are too noisy to be in the lower panels.
This may result from insufficient coupling between the first excited state and these two operators.  We observe that the precision is significantly improved by including the $\pi\pi(001)$ operator.
The $\pi\pi(001)$ operator is the most na\"\i ve operator for creating a two-pion state with an energy close to the kaon mass. It plays a similar role to the `$\pi\pi(111)$' operator in the GPBC work, which comprised two single-pion operators with back-to-back momenta of $\pi/L$ in each spatial direction. The `$\pi\pi(111)$' operator was the only two-pion operator included in the original GPBC work, but was shown in Ref.~\cite{RBC:2020kdj} to be insufficient to reliably isolate the required two-pion state. Our PBC results appear to verify this behavior.
The lower panels indicate that, while the $\sigma$-like operator plays an important role in removing excited-state contamination, including the $\pi\pi(001)$ operator is also important to obtain a well-resolved signal for the first excited state.

The fit result for $M_{1,6}$, which provides the dominant contribution to ${\rm Im}(A_0)$ and therefore plays a crucial role in determining $\varepsilon'$, appears to depend on $t_2^{\rm min}$ still around our choice of $t_2^{\rm min}=4a$ as seen in Figure~\ref{fig:range_dps_I0_n1}.
The consistency between ``$N=4$ w $\sigma$'' and ``$N=5$ w $\sigma$'' seen in Figure~\ref{fig:gevp_dps_I0} may not necessarily rule out the possibility of further excited-state contamination because the fourth excited state that can in principle be extracted by including the fifth operator, $\pi\pi(111)$, is unresolved in our data and adding this operator might not play a role in isolating excited state effects. 
It is therefore valuable to demonstrate there is no evidence of further excited-state contamination for $t_2^{\rm min}=4a$.
We find the difference between the fit results for $M_{1,6}$ from ``$N=5$ w $\sigma$'' with $t_2^{\rm min}=4a$ and $t_2^{\rm min}=3a$ under jackknife is consistent with zero, 0.0056(63) in lattice units.
The $O(1\sigma)$ difference between ``$N=3$ w $\sigma$'' and ``$N=4$ w $\sigma$,'' noticeable in $M_{1,6}$ may indicate a removal of the contamination from the third excited state realized by additionally including the $\pi\pi(011)$ operator.
Because this difference is roughly equal to the statistical error in the region $t_2^{\rm min}=4a$, the remaining contamination from the fourth and higher excited states is expected to be smaller than the statistical error.
Our choice of $t_2^{\rm min}=4a$ for the $I=0$ first-excited two-pion final state roughly corresponds to 0.8~fm in physical units.
A similar trend in the $t^{\rm min}_2$-dependence of the fit result for the matrix element of $Q_6$ was observed at roughly equal values of $t_2^{\rm min}$ in physical units in our earlier work with GPBC~\cite{RBC:2020kdj}, where we concluded, after a serious investigation of systematic errors, it is not due to contamination from excited two-pion states but a simple statistical effects.

\begin{table*}[tbp]
\centering
\begin{tabular}{|c|ccc|}
\hline
$i$ & $a^3M_{0,i}$ & $a^3M_{1,i}$ & $a^3M_{2,i}$\\
\hline
1 & $0.00120(20)$[1.2] & $0.0024(13)$[1.2] & $-0.0017(29)$[0.7]\\
2 & $-0.00161(22)$[0.5] & $-0.0061(13)$[0.7] & $-0.0070(25)$[0.4]\\
3 & $0.00021(58)$[0.7] & $-0.0045(38)$[0.4] & $-0.0124(78)$[0.4]\\
4 & $-0.00268(62)$[0.4] & $-0.0118(40)$[0.6] & $-0.0157(67)$[0.3]\\
5 & $0.00582(52)$[1.1] & $0.0134(34)$[0.5] & $-0.0007(83)$[0.2]\\
6 & $0.0175(10)$[0.8] & $0.0597(66)$[0.3] & $0.045(14)$[0.7]\\
7 & $-0.02863(24)$[0.9] & $-0.0367(13)$[0.6] & $-0.0232(34)$[0.8]\\
8 & $-0.08944(50)$[1.6] & $-0.1115(28)$[0.6] & $-0.0565(93)$[1.9]\\
9 & $0.00158(22)$[0.4] & $0.0049(14)$[1.1] & $0.0057(24)$[0.8]\\
10 & $-0.00118(19)$[1.0] & $-0.0037(12)$[0.3] & $-0.0007(29)$[0.6]\\
\hline
\end{tabular}
\caption{Fit results for the $\Delta I = 1/2$ channel of the $K\to\pi\pi$ matrix elements with the three lowest energy two-pion states and with each four-quark operator.  For the ground state $(n=0)$ we choose the fit range $(t_1^{min},t_2^{min}) = (5a,4a)$ for each four-quark operator.  For the first-excited state $(n=1)$ we choose the fit range $(t_1^{min},t_2^{min}) = (4a,4a)$ for each four-quark operator.  For the second-excited state $(n=2)$ we choose the fit range $(t_1^{min},t_2^{min}) = (3a,4a)$ for each four-quark operator.  The first error is statistical and the following error represents the systematic error due to the potential contamination from kaon excited states.  The values of $\chi^2/d.o.f.$ are shown in the square brackets.}
\label{tab:fitres_ME_I0}
\end{table*}

Table~\ref{tab:fitres_ME_I0} summarizes the fit results for the $\Delta I = 1/2$ channel of $K\to\pi\pi$ matrix elements $M_{n,i}$ with $n=0,1,2$ and $i = 1$--10.
Compared with the $\Delta I = 3/2$ channel most of the fits are stable and $\chi^2/d.o.f.$ is fairly small despite the wider fit ranges.
It is interesting to observe the accuracy of $\Delta I=1/2$ matrix elements for the ground, first-excited and second-excited states. 
For the ground state, which corresponds to two pions at rest, the matrix elements appear the most precise among the three. 
For the first-excite state, which is the closest among the three to the physical kinematics, the matrix elements are well determined.
For the second-excited state, the matrix elements are also somewhat resolved for several four-quark operators.
These observations suggest that the states have been successfully disentangled by the GEVP.
It is thus remarkable that we have succeeded in extracting the matrix elements with excited two-pion final states with well-resolved signals.

\section{Renormalization of four-quark operators}
\label{sec:NPR}

Renormalization of composite operators is an essential step for the lattice calculation of weak processes to remove unphysical divergences, as is the case in general for quantum field theory, before we can remove the regulator, $i.e.$, take the continuum limit $a\to0$.  While the lattice operators are usually renormalized in a non-perturbative scheme, the corresponding Wilson coefficients are computed in the $\rm\overline{MS}$ scheme~\cite{Buras:1993dy,Buchalla:1995vs}.
In order to appropriately construct the weak Hamiltonian~\eqref{eq:Hw} and obtain the corresponding decay amplitudes, we need to convert the nonperturbatively renormalized matrix elements to the $\rm\overline{MS}$ scheme. 

As mentioned in Section~\ref{sec:4qOps} we use the chiral basis of the seven four-quark operators $Q_i'$~\cite{RBC:2001pmy}, which has two advantages in nonperturbative renormalization (NPR).  First, the linear independence of operators minimizes the number of renormalization conditions and gives well-defined inverse renormalization matrices, which are useful for step scaling~\cite{Arthur:2010ht}.
Second, each operator in this basis transforms as a specific representation of ${\rm SU}(3)_L\times {\rm SU}(3)_R$ chiral symmetry, while the operators $Q_i$ in the ten-operator basis in general combine multiple representations.  Out of the seven operators, one transforms as the (27,1) representation, four as (8,1) and two as (8,8).  The operators in the same representations still mix with each other but not with the other operators.  As a result, the corresponding renormalization matrices are block diagonal.

We employ a common momentum-space procedure using the regularization independent symmetric momentum (RI-SMOM) schemes, in which renormalization conditions are imposed nonperturbatively.
Since the perturbative matching between the RI-SMOM and $\overline{\rm MS}$ schemes is available to the one-loop level~\cite{Lehner:2011fz}, RI-SMOM gives convenient intermediate schemes to obtain the matrix elements in the $\overline{\rm MS}$ scheme.
We also perform step scaling as we previously observed a significant, three times improvement in the estimated systematic errors associated with the truncation of the perturbative matching to the $\rm\overline{MS}$ scheme when raising the renormalization scale from 1.53~GeV to 4~GeV through the step-scaling procedure~\cite{RBC:2020kdj}. This procedure is more important for the present calculation due to the much coarser, 1.0~GeV inverse lattice spacing versus 1.4~GeV in the previous calculation, restricting the range of usable energy scales even further.
Since the RI-SMOM renormalization conditions do not obey the equations of motion and require gauge fixing, we need to take into account the mixing with extra operators that vanish by the equations of motion or are gauge variant.  In this work we take the mixing with only quark bilinears into account since the mixing with irrelevant operators with higher dimensions such as $G_1=\bar s\gamma_\nu(1-\gamma_5)D_\mu G_{\mu\nu}d$, where $D_\mu$ is a covariant-derivative operator and $G_{\mu\nu}$ the gluon field strength and the space-time indices $\mu$ and $\nu$ are summed over, were found to be less significant than other systematic errors~\cite{RBC:2020kdj}.

In this section we give a summary for the renormalization procedure and then show the results for the renormalization and step-scaling matrices.  The more detailed procedure is presented in Refs.~\cite{McGlynn:2016ahb,RBC:2020kdj}.

\subsection{Determination of renormalization matrix}

Because of mixing with lower dimensional operators the four-quark operators are power divergent, and this needs to be subtracted before a multiplicative renormalization condition to remove the remaining logarithmic divergence can be imposed.  We consider three bilinear operators $S_1 = \bar s\gamma_5d$, $S_2 = \bar s\overrightarrow{\slashchar{D}}\gamma_5d$ and $S_3 = \bar s\overleftarrow{\slashchar{D}}\gamma_5d$, where the arrow indicates the direction of the discrete covariant derivative $\slashchar{D}$.  The subtracted four-quark operators are written as
\begin{equation}
    Q_i^{\prime\rm\,sub,lat} = Q_i^{\prime\rm\,lat} + \sum_{j\in\cal C} b_{i,j} S_j^{\rm lat},
\end{equation}
where we have added the superscript `lat' to indicate bare operators and the sum over $j$ runs over the set ${\cal C} = \{1,2,3,5,6,7,8\}$.
The subtraction coefficients $b_{i,j}$ are determined by the condition
\begin{equation}
P^{\beta\alpha}_{2,j}(p_1,p_2)\left\langle
s(-p_1)\bar d(p_2)Q_i^{\prime\rm\,sub,lat}(q)
\right\rangle^{\alpha\beta}_{\rm amp} = 0,
\label{eq:subt cond_NPR}
\end{equation}
where the indices $\alpha$ and $\beta$ are combined spin and color indices and the projection operators $P_{2,j}$ are given in Section~7.2.6 of Ref.~\cite{McGlynn:2016ahb}.
The conditions \eqref{eq:subt cond_NPR} to determine the subtraction coefficients depend on external momenta $p_1$, $p_2$ and their difference $q=p_1-p_2$.  We use these momenta for the multiplicative renormalization condition as well, and they determine the renormalization scale $\mu$.  In the case of RI-SMOM schemes they satisfy
\begin{equation}
p_1^2 = p_2^2 = q^2 \equiv \mu^2.
\end{equation}

After the subtractions we renormalize the four-quark operators
\begin{equation}
Q_i^{\prime\rm\,RI}(\mu) = \sum_{j\in\cal C} Z_{ij}^{\rm RI\leftarrow lat}(\mu)Q_j^{\prime\rm\,sub,lat},
\end{equation}
where the superscript `RI' is a generic expression of an RI-SMOM scheme.
The renormalized operators $Q_i^{\prime\rm\,RI}(\mu)$ are defined by the conditions
\begin{align}
&Z_q(\mu)^{-2}P_{4,m}^{\beta\alpha\delta\gamma}\left[\Gamma_{im}^{\rm RI}\right]^{\alpha\beta\gamma\delta}(p_1,p_2) = F_{im},\\
&\left[\Gamma_{im}^{\rm RI}\right]^{\alpha\beta\gamma\delta}(p_1,p_2) = \left\langle
E_m\sum_x\e^{2\img qx}Q_i^{\prime\rm\,RI}(x)
\right\rangle_{\rm amp}^{\alpha\beta\gamma\delta},
\end{align}
where $Z_q$ is the quark field renormalization factor, and $F_{im}$ is the free-field expression corresponding to the left hand side. 
$E_m$ denotes a product of external quark fields,
\begin{align}
E_{1,2,4,5} &= s(-p_1)\bar d(p_2)u(-p_1)\bar u(p_2),\\
E_{3,6,7} &= s(-p_1)\bar d(p_2)\sum_{q=u,d,s}q(-p_1)\bar q(p_2).
\end{align}
We consider two different schemes by employing different projection operators, which have the following spin structure:
\begin{equation*}
\begin{array}{rl}
\mbox{$\gamma_\mu$ scheme:} & P_{4,m} \sim \pm \gamma_\mu\otimes\gamma_5\gamma_\mu - \gamma_5\gamma_\mu\otimes\gamma_\mu,\\
\mbox{$\slashchar{q}$ scheme:} & P_{4,m} \sim \pm \slashchar{q}\otimes\gamma_5\slashchar{q} - \gamma_5\slashchar{q}\otimes\slashchar{q},
\end{array}
\end{equation*}
for the renormalization procedure with the parity-odd part of the four-quark operators.
The sign $\pm$ and color structure depend on $m$.  The explicit forms of these projection operators are given in Section~3.3.2 of Ref.~\cite{Zhang:2015hbq}.

To calculate $Z_q(\mu)$, we first compute $Z_A/Z_q$ from the RI-SMOM renormalization condition for the local axial current.  We can thus obtain $Z_q(\mu)$
using $Z_A$ computed from the ratio of the pion-to-vacuum matrix elements of the local and conserved currents.  While we can also use the SMOM renormalization $Z_V$ of the vector current to determine $Z_q$, we use $Z_A = 0.73457(11)$~\cite{Tu:2020vpn} in this work because of an advantage in the statistical precision. 

Here we can consider the $\gamma_\mu$ and $\slashchar{q}$ schemes for determination of $Z_q$ and  $Z_{ij}^{\rm RI}$.  We define the SMOM$(A,B)$ scheme as the scheme that employs the $A$ scheme for the projectors $P_4$ and the $B$ scheme for the projector to determine $Z_q$.
While there are four combinations, we only consider the SMOM$(\gamma_\mu,\gamma_\mu)$ and SMOM$(\slashchar{q},\slashchar{q}$) schemes because we have observed that the nonperturbative running in these schemes is better described by perturbation theory than in the other two schemes~\cite{Aoki:2010pe}.

On the $24^3$ ensemble, we calculate the vertex functions at the SU(3) symmetric valence quark mass $am_q=0.01$.  While renormalization schemes are usually defined in the chiral limit, the vertex functions were found to depend on the quark mass very little for SMOM $\Delta F=2$ operators~\cite{Boyle:2017skn}.  Therefore we calculate NPR at the single finite quark mass.
We choose two renormalization scales $\mu=\mu_1\approx1.3$~GeV and $\mu=\mu_2\approx1.5$~GeV.  Table~\ref{tab:Z24ID} shows a summary of these scales and the corresponding renormalization factors $Z_q$ for quark fields in the $\gamma_\mu$ and $\slashchar{q}$ schemes.  The higher scale $\mu_2$ corresponds to the intermediate scale chosen in Ref.~\cite{RBC:2020kdj} in physical units.

\begin{table}[tbp]
\centering
\begin{tabular}{|c|cc|}
\hline
 & $\mu_1$ & $\mu_2$ \\
lattice units & 1.2825 & 1.4810 \\
\,[GeV] &  1.312(3) & 1.515(3) \\
\hline
$p_1\times(\frac{L}{2\pi})$ & (2,2,4,0) & (0,4,4,0) \\
$p_2\times(\frac{L}{2\pi})$ & $(-2,4,2,0)$ & (4,4,0,0) \\
\hline
$Z_q^{\gamma_\mu}$ & 0.79232(12) & 0.79600(12) \\
$Z_q^{\slashchar{\scriptstyle q}}$ & 0.88660(16) & 0.88459(15) \\
\hline
\end{tabular}
\caption{Summary for chosen renormalization scales and corresponding renormalization factors of the quark field with both $\gamma_\mu$ and $\slashchar{q}$ schemes on the $24^3$ ensemble.}
\label{tab:Z24ID}
\end{table}
\begin{table*}[tp]
\centering
\begin{tabular}{|c|ccccccc|}
\hline
 & 1 & 2 & 3 & 5 & 6 & 7 & 8\\
\hline
1 & $0.52424(16)$ & $0$ & $0$ & $0$ & $0$ & $0$ & $0$\\
2 & $0$ & $0.3937(58)$ & $-0.3691(60)$ & $-0.0056(25)$ & $0.0042(13)$ & $0$ & $0$\\
3 & $0$ & $0.196(12)$ & $1.022(12)$ & $0.0024(48)$ & $-0.0055(27)$ & $0$ & $0$\\
5 & $0$ & $-0.018(35)$ & $-0.025(31)$ & $0.576(13)$ & $-0.1015(63)$ & $0$ & $0$\\
6 & $0$ & $-0.051(14)$ & $-0.089(16)$ & $-0.0727(43)$ & $0.4768(33)$ & $0$ & $0$\\
7 & $0$ & $0$ & $0$ & $0$ & $0$ & $0.56587(17)$ & $-0.104786(41)$\\
8 & $0$ & $0$ & $0$ & $0$ & $0$ & $-0.078889(57)$ & $0.46935(19)$\\
\hline
\end{tabular}
\caption{Renormalization matrix with the chiral basis on the $24^3$ ensemble at the renormalization scale $\mu_1$ in the SMOM$(\slashchar{q},\slashchar{q})$ scheme.}
\label{tab:NPR_1p3GeV_qslash_qslash_odd.bs1.L24}
\end{table*}

Table~\ref{tab:NPR_1p3GeV_qslash_qslash_odd.bs1.L24} shows the $7\times7$ renormalization matrix for the SMOM$(\slashchar{q},\slashchar{q})$ scheme at $\mu=\mu_1$.  
While the sea light and strange quark masses on the $24^3$ ensemble are 0.0017 and 0.085 in lattice units, respectively, we set the valence quark mass to 0.01 for both.  While the renormalization matrix should in principle be extrapolated to the chiral limit, we neglect the systematic error due to these finite masses because the renormalization scales are relatively large and the mass effect is expected to be small.
The results with different pairs of the scheme and scale are given in Table~\ref{tab:NPR_rest.L24} in Appendix~\ref{sec:supplemental}.

\subsection{Step scaling}

Since our lattice ensemble is coarse, the renormalization window problem needs to be addressed; the renormalization scales $\mu_1$ and $\mu_2$ used in the present work are not high enough to perform a precise perturbative matching but renormalizing at higher scales may lead to significant discretization errors.  To bypass this problem we employ the step-scaling technique~\cite{Arthur:2010ht} using a finer $32^3\times64$ lattice ensemble which we call the 32Ifine ensemble.  Details are given in Ref.~\cite{RBC:2014ntl}, but we note that the inverse lattice spacing and pion mass on this ensemble are $3.148(17)$~GeV and 371(5)~MeV, respectively.  We choose the valence quark mass $am_q = 0.0047$, which is the same as the sea light quark mass.  As noted above we expect the systematic error due to the finite mass is small since it was found to be the case for the $\Delta F = 2$ four-quark operators in the SMOM schemes~\cite{Boyle:2017skn}. 

On this fine lattice we calculate the step-scaling matrix
\begin{equation}
\Sigma^{\rm RI}(\mu',\mu) = Z^{\rm RI\leftarrow lat}(\mu'){Z^{\rm RI\leftarrow lat}(\mu)}^{-1},
\end{equation}
where the indices for the matrices are suppressed, and we suppose $\mu'>\mu$ and obtain the four-quark operators renormalized at a higher scale,
\begin{equation}
Q_i^{\prime\rm\,RI}(\mu') = \sum_{j\in\cal C} \Sigma^{\rm RI}_{ij}(\mu',\mu)Q_j^{\prime\rm\,RI}(\mu).
\label{eq:stepscaling}
\end{equation}
While the step-scaling matrix in principle needs to be extrapolated to the continuum limit, we use a single lattice ensemble. 
In Ref.~\cite{Boyle:2017skn} , bilinear and $VV+AA$ step scaling functions have displayed modest discretization effects at similar scales in lattice units.  We therefore anticipate these non-removed discretization errors are not dominant. 

\begin{table}[tbp]
\centering
\begin{tabular}{|c|ccc|}
\hline
 & $\mu_1$ & $\mu_2$ & $\mu_h$ \\
lattice units & 0.39270 & 0.48096 & 1.2725 \\
\,[GeV] &  1.236(7) & 1.514(8) & 4.006(22)\\
\hline
$p_1\times(\frac{L}{2\pi})$ & (1,1,1,1) & (1,1,2,0) & (4,4,3,1) \\
$p_2\times(\frac{L}{2\pi})$ & $(0,0,0,2)$ & (0,1,1,2) & (0,1,4,5) \\
\hline
$Z_q^{\gamma_\mu}/Z_A$ & 1.03455(36) & 1.03518(21) & 1.03258(3) \\
$Z_q^{\slashchar{\scriptstyle q}}/Z_A$ & 1.16074(96) & 1.14096(55) & 1.07070(7) \\
\hline
\end{tabular}
\caption{Summary for chosen renormalization scales and corresponding renormalization factors of the quark field with both $\gamma_\mu$ and $\slashchar{q}$ schemes divided by $Z_A$ on the 32Ifine ensemble.}
\label{tab:Z32Ifine}
\end{table}

Table~\ref{tab:Z32Ifine} shows a summary of the scales used for the calculation of the renormalization matrices on the 32Ifine lattice ensemble.  The corresponding $Z_q/Z_A$ is also shown for each scheme and scale.  The reason for showing $Z_q/Z_A$ instead of $Z_q$ shown in Table~\ref{tab:Z24ID} is that the purpose of using the 32Ifine ensemble is to calculate the step-scaling matrix, which is independent of $Z_A$, and that $Z_q/Z_A$ is calculated more directly without inputting $Z_A$ separately.
Note that although the fourth elements of the external momenta $p_1$ and $p_2$ are for the time direction and their actual unit is hence $2\pi/L_T=\pi/L$, they are shown in units of $2\pi/L$ in the table.
Note also that we identify the lower two scales as $\mu_1$ and $\mu_2$, although they are not identical to the ones listed in Table~\ref{tab:Z24ID} for the $24^3$ ensemble. 
Namely we intend to perform the step scaling \eqref{eq:stepscaling} with $\mu=\mu_1,\mu_2$ and $\mu'=\mu_h$ neglecting the impact of the difference in these scales between the two ensembles.  The difference in $\mu_1$ is larger by about 6\%, which would only cause a systematic error of $O(\frac{\alpha_s}{2\pi}\times6\%)$ on the renormalized matrix element after the step scaling because evolution from $\mu$ to $\mu'$ scales like $1+O(\frac{\alpha_s}{4\pi}\ln(\frac{\mu'}{\mu})^2)$. 

\begin{table*}[tp]
\centering
\begin{tabular}{|c|ccccccc|}
\hline
 & 1 & 2 & 3 & 5 & 6 & 7 & 8\\
\hline
1 & $0.87586(45)$ & $0$ & $0$ & $0$ & $0$ & $0$ & $0$\\
2 & $0$ & $1.33(12)$ & $0.09(12)$ & $-0.004(39)$ & $0.055(30)$ & $0$ & $0$\\
3 & $0$ & $-0.536(77)$ & $0.468(78)$ & $-0.000(26)$ & $-0.002(23)$ & $0$ & $0$\\
5 & $0$ & $0.03(42)$ & $-0.48(42)$ & $0.92(13)$ & $0.42(10)$ & $0$ & $0$\\
6 & $0$ & $0.15(23)$ & $0.07(25)$ & $0.039(69)$ & $1.921(61)$ & $0$ & $0$\\
7 & $0$ & $0$ & $0$ & $0$ & $0$ & $0.97336(57)$ & $0.2804(13)$\\
8 & $0$ & $0$ & $0$ & $0$ & $0$ & $0.1047(22)$ & $1.9670(61)$\\
\hline
\end{tabular}
\caption{Step-scaling matrix from $\mu_1$ to $\mu_h$ in the SMOM$(\slashchar{q},\slashchar{q})$ scheme calculated on the 32Ifine ensemble.
}
\label{tab:NPR_1p3GeV_ss_qslash_qslash_odd.bs1}
\end{table*}

\begin{table*}[tp]
\centering
\begin{tabular}{|c|ccccccc|}
\hline
 & 1 & 2 & 3 & 5 & 6 & 7 & 8\\
\hline
1 & $0.45916(28)$ & $0$ & $0$ & $0$ & $0$ & $0$ & $0$\\
2 & $0$ & $0.539(68)$ & $-0.409(89)$ & $-0.014(23)$ & $0.032(15)$ & $0$ & $0$\\
3 & $0$ & $-0.119(45)$ & $0.676(56)$ & $0.004(15)$ & $-0.006(12)$ & $0$ & $0$\\
5 & $0$ & $-0.12(24)$ & $-0.56(30)$ & $0.498(79)$ & $0.110(52)$ & $0$ & $0$\\
6 & $0$ & $-0.03(14)$ & $-0.16(19)$ & $-0.118(41)$ & $0.912(32)$ & $0$ & $0$\\
7 & $0$ & $0$ & $0$ & $0$ & $0$ & $0.52867(38)$ & $0.02963(61)$\\
8 & $0$ & $0$ & $0$ & $0$ & $0$ & $-0.0959(15)$ & $0.9123(30)$\\
\hline
\end{tabular}
\caption{Renormalization matrix with the chiral basis on the $24^3$ ensemble at the renormalization scale $\mu_h$ in the SMOM$(\slashchar{q},\slashchar{q})$ scheme calculated by applying the step-scaling matrix in Table~\ref{tab:NPR_1p3GeV_ss_qslash_qslash_odd.bs1} to the renormalization matrix at lower scale $\mu_1$ in Table~\ref{tab:NPR_1p3GeV_qslash_qslash_odd.bs1.L24}.
}
\label{tab:NPR_1p3GeV_qslash_qslash_odd_4GeV.bs1.L24}
\end{table*}

Table~\ref{tab:NPR_1p3GeV_ss_qslash_qslash_odd.bs1} shows the result for the step-scaling matrix from $\mu_1$ to $\mu_h$ in the SMOM($\slashchar{q},\slashchar{q}$) scheme calculated on the 32Ifine lattice.  We also show the corresponding renormalization matrix $\Sigma^{{\rm SMOM}(\slashchar{\scriptstyle q},\slashchar{\scriptstyle q})}(\mu_h,\mu_1)Z^{{\rm SMOM}(\slashchar{\scriptstyle q},\slashchar{\scriptstyle q})\rm\leftarrow lat}(\mu_1)$ after the step scaling in Table~\ref{tab:NPR_1p3GeV_qslash_qslash_odd_4GeV.bs1.L24}.
The results with different pairs of the scheme and lower scale are shown in Table~\ref{tab:ss_rest} for the step-scaling matrix and in Table~\ref{tab:ssZ_rest} for the renormalization matrix multiplied with the step-scaling matrix.

\subsection{Perturbative matching}

While the nonperturbative renormalization procedure described above removes the ultraviolet divergences of four-quark operators, and we can take the continuum limit of matrix elements of the renormalized four-quark operators, it is also necessary to convert them into the $\rm\overline{MS}$ scheme in order to appropriately construct the weak Hamiltonian and obtain the $K\to\pi\pi$ amplitudes because the corresponding Wilson coefficients are calculated in the $\rm\overline{MS}$ scheme.

We perform the perturbative matching from RI-SMOM schemes to the $\rm\overline{MS}$ scheme
\begin{equation}
Q_i^{\prime\,\rm\overline{MS}}(\mu) = \sum_{j\in\cal C} R_{ij}^{\rm\overline{MS}\leftarrow RI}(\mu) Q_j^{\prime\,\rm RI}(\mu),
\end{equation}
where we suppose $\mu$ to be the high scale $\mu_h$ after the step scaling.

We use the one-loop matching matrices between the $\overline{\rm MS}$ and RI-SMOM schemes given in Ref.~\cite{Lehner:2011fz}.
We use the strong coupling constant
\begin{equation}
\alpha_s^{N_f=3}(\mu_h\approx4.0\mbox{~GeV}) = 0.2167(31),
\label{eq:alpha_s}
\end{equation}
which is determined as follows. We first input the Particle Data Group (PDG)  value $\alpha_s^{N_f=5}(M_Z)=0.1179(9)$~\cite{ParticleDataGroup:2022pth} given at the $Z$ pole in the five flavor theory and perform its scale evolution with the four-loop $\beta$ function~\cite{vanRitbergen:1997va,Chetyrkin:1997sg}, changing the number $N_f$ of active quark flavors from 5 to 4 at the bottom threshold. From this procedure, we obtain the four-flavor $\Lambda_{\rm QCD}^{N_f=4}$ parameter.  Using this parameter as input, we use two-loop scale evolution to the charm threshold with $N_f=4$ and, again, two-loop evolution with $N_f=3$ to obtain $\alpha_s^{N_f=3}$ given in Eq.~\eqref{eq:alpha_s} which is consistent with the order of the perturbative calculation of the Wilson coefficients.

\begin{table*}[tp]
\centering
\begin{tabular}{|c|ccccccc|}
\hline
 & 1 & 2 & 3 & 5 & 6 & 7 & 8\\
\hline
1 & $1.00365$ & 0 & 0 & 0 & 0 & 0 & 0\\
2 & 0 & $0.99817$ & $0.00548$ & 0 & 0 & 0 & 0\\
3 & 0 & $0.00165$ & $0.98923$ & $0.00128$ & $-0.00383$ & 0 & 0\\
5 & 0 & 0 & 0 & $1.00074$ & $-0.00223$ & 0 & 0\\
6 & 0 & $-0.02874$ & $-0.06706$ & $-0.01825$ & $1.04878$ & 0 & 0\\
7 & 0 & 0 & 0 & 0 & 0 & $1.00074$ & $-0.00223$\\
8 & 0 & 0 & 0 & 0 & 0 & $-0.02783$ & $1.07752$\\
\hline
\hline
1 & $0.99216$ & 0 & 0 & 0 & 0 & 0 & 0\\
2 & 0 & $1.04531$ & $0.08480$ & 0 & 0 & 0 & 0\\
3 & 0 & $-0.07768$ & $0.85013$ & $0.00128$ & $-0.00383$ & 0 & 0\\
5 & 0 & 0 & 0 & $1.00074$ & $-0.00223$ & 0 & 0\\
6 & 0 & $-0.02874$ & $-0.06706$ & $-0.00100$ & $0.99705$ & 0 & 0\\
7 & 0 & 0 & 0 & 0 & 0 & $1.00074$ & $-0.00223$\\
8 & 0 & 0 & 0 & 0 & 0 & $-0.01058$ & $1.02579$\\
\hline
\end{tabular}
\caption{Perturbative matching matrix from the SMOM$(\gamma_\mu,\gamma_\mu)$ (upper) and SMOM($\slashchar{q},\slashchar{q}$) (lower) schemes to the $\overline{\rm MS}$ scheme in the chiral basis calculated at one-loop level.  The error from the uncertainty of $\alpha_s$ in Eq.~\eqref{eq:alpha_s} is omitted because the truncation uncertainty, whose effect on the $K\to\pi\pi$ amplitudes is considered in Section~\ref{sec:AI}, is expected to be more significant.}
\label{tab:PTmatchingR}
\end{table*}

Table~\ref{tab:PTmatchingR} shows the perturbative matching matrices from the SMOM$(\gamma_\mu,\gamma_\mu)$ and SMOM($\slashchar{q},\slashchar{q}$) schemes to the $\overline{\rm MS}$ scheme. 

\subsection{Systematic uncertainty}
\label{sec:systErrNPT}

These renormalization processes can create systematic errors because of the mixing with excluded operators, the discretization error on the renormalization condition and  the truncation of perturbative matching.  
It was found in the previous work~\cite{RBC:2020kdj} that these sources of systematic errors are less significant than some major ones such as Wilson coefficients.  
However it should be noted that the investigation on the systematic errors was done on a finer lattice with the inverse lattice spacing of $a_{\rm32ID}^{-1}\approx1.38$~GeV, where the subscript `32ID' denotes the ensemble used in Ref.~\cite{RBC:2020kdj}, and that one of the intermediate scales we employ in this work is the same as the one used in the previous work in physical units, $\mu_2\approx1.48a^{-1}\approx1.11a^{-1}_{\rm 32ID}$.  This is because we need to match the momenta $p_1$ and $p_2$ in the renormalization conditions to ensure the renormalized matrix elements on both ensembles are on the same scaling trajectory for the continuum extrapolation.  While we do not take the continuum limit at this stage, it is important to verify whether $\mu_2$ is a good choice for the continuum limit, since we plan to take it in the near future.
In summary the renormalization conditions in this work are imposed at relatively high scales in lattice units and therefore it is important to revisit the investigation of the systematic error from NPR.

\begin{table*}[tp]
\centering
\begin{tabular}{|c|ccccccc|}
\hline
 & 1 & 2 & 3 & 5 & 6 & 7 & 8\\
\hline
1 & $1.11947(41)$ & $0$ & $0$ & $0$ & $0$ & $0$ & $0$\\
2 & $0$ & $1.05(12)$ & $-0.11(11)$ & $0.046(54)$ & $0.019(36)$ & $0$ & $0$\\
3 & $0$ & $-0.040(57)$ & $1.138(55)$ & $0.031(27)$ & $-0.000(19)$ & $0$ & $0$\\
5 & $0$ & $-0.15(35)$ & $-0.11(34)$ & $1.20(16)$ & $0.12(12)$ & $0$ & $0$\\
6 & $0$ & $0.07(15)$ & $0.05(15)$ & $-0.001(59)$ & $1.291(47)$ & $0$ & $0$\\
7 & $0$ & $0$ & $0$ & $0$ & $0$ & $1.11215(69)$ & $0.0694(11)$\\
8 & $0$ & $0$ & $0$ & $0$ & $0$ & $-0.0058(11)$ & $1.2851(34)$\\
\hline
\hline
1 & $0.89666(48)$ & $0$ & $0$ & $0$ & $0$ & $0$ & $0$\\
2 & $0$ & $1.05(24)$ & $-0.29(25)$ & $0.001(77)$ & $0.036(52)$ & $0$ & $0$\\
3 & $0$ & $0.03(12)$ & $0.82(12)$ & $0.020(40)$ & $-0.002(27)$ & $0$ & $0$\\
5 & $0$ & $0.16(72)$ & $-0.84(74)$ & $0.97(22)$ & $0.13(15)$ & $0$ & $0$\\
6 & $0$ & $0.07(30)$ & $-0.25(31)$ & $0.024(91)$ & $1.034(58)$ & $0$ & $0$\\
7 & $0$ & $0$ & $0$ & $0$ & $0$ & $0.89855(64)$ & $0.0534(11)$\\
8 & $0$ & $0$ & $0$ & $0$ & $0$ & $-0.0029(19)$ & $1.0228(37)$\\
\hline
\end{tabular}
\caption{Matrix $I^{\rm RI}(\mu_2,\mu_1)$ for both the SMOM($\gamma_\mu,\gamma_\mu$) (upper) and SMOM($\slashchar{q},\slashchar{q}$) (lower) schemes.  
These matrices should be the unit matrix if no systematic errors exist.}
\label{tab:ZinvZ}
\end{table*}

Since the systematic error considered in this context is a discretization effect in the renormalization conditions, it is reasonable to isolate the various components from continuum perturbation theory.
Thus we analyze the matrix
\begin{equation}
I^{\rm RI}(\mu_2,\mu_1)
= Z^{\rm RI\leftarrow lat}(\mu_2)^{-1}\Sigma^{\rm RI}(\mu_2,\mu_1)Z^{\rm RI\leftarrow lat}(\mu_1),
\end{equation}
which is computed purely on the lattice.
Here the Z matrices are computed on the $24^3$ lattice and the step-scaling matrix $\Sigma$ on the 32Ifine lattice.
Table~\ref{tab:ZinvZ} shows the results for $I^{\rm RI}(\mu_2,\mu_1)$ for the SMOM($\gamma_\mu,\gamma_\mu$) and SMOM($\slashchar{q},\slashchar{q}$) schemes.
The difference between this matrix and the $7\times7$ unit matrix should represent the systematic error in the nonperturbative renormalization.
While most of the off-diagonal elements are consistent with zero, many of the diagonal elements significantly deviate from one.  Especially $I_{66}^{\rm SMOM(\gamma_\mu,\gamma_\mu)}$ and $I_{88}^{\rm SMOM(\gamma_\mu,\gamma_\mu)}$ show almost 30\% deviations from unity.  If the systematic errors on the Z matrix at $\mu_1$ and $\mu_2$ are similar, the corresponding renormalization matrix would have roughly a 15\% systematic error.  However we expect the systematic error at $\mu_2$ could be significantly larger than at $\mu_1\approx1.28a^{-1}$ because a similar investigation made in Section~VII.G in Ref.~\cite{RBC:2020kdj} indicates a much smaller systematic error at $\mu_2\approx1.11a_{\rm32ID}^{-1}$.
We discuss the actual impact of this difference on the $K\to\pi\pi$ amplitudes and quote the corresponding systematic error in the following section.

\section{$K\to\pi\pi$ amplitudes and $\varepsilon'$}
\label{sec:AandEpslon'}

The $K\to\pi\pi$ amplitudes $A_I = \langle (\pi\pi)_I|H_W|K\rangle$ with isospin-$I$ two-pion final state and the weak Hamiltonian given in Eq.~\eqref{eq:Hw} can be calculated by
\begin{equation}
A_I = \sum_{i=1}^{10} C_i^{\overline{\rm MS}}(\mu_h) M_{I,i}^{\overline{\rm MS}}(\mu_h),
\label{eq:def_AI}
\end{equation}
where we define 
\begin{align}
C_i^{\overline{\rm MS}}(\mu_h) &= \frac{G_F}{\sqrt2}V_{us}^*V_{ud}[z_i^{\overline{\rm MS}}(\mu_h)+\tau y_i^{\overline{\rm MS}}(\mu_h)]
\label{eq:Wcoef_C}
\end{align}
with the Wilson coefficients $z_i^{\overline{\rm MS}}(\mu_h)$ and $y_i^{\overline{\rm MS}}(\mu_h)$ in the ${\overline{\rm MS}}$ scheme at the scale $\mu_h$ in the three-flavor theory~\cite{Buras:1993dy,Buchalla:1995vs}.  The $\overline{\rm MS}$ matrix elements $M_{I,i}^{\overline{\rm MS}}(\mu_h)$ with the ten-operator basis can be expressed in terms of the chiral-basis RI matrix elements $M_{I,k}^{\prime\,\rm RI}(\mu_h)$ and the perturbative matching matrix $R_{jk}^{\rm\overline{MS}\leftarrow RI}(\mu_h)$,
\begin{equation}
M_{I,i}^{\overline{\rm MS}}(\mu_h)
= \sum_{j,k\in{\cal C}}\left(T_{ij}+\Delta T_{ij}^{\rm\overline{MS}}\right)
R_{jk}^{\rm\overline{MS}\leftarrow RI}(\mu_h)
M_{I,k}^{\prime\,\rm RI}(\mu_h),
\label{eq:MERItoMEMS}
\end{equation}
where the $10\times7$ matrices $T$ and $\Delta T^{\rm\overline{MS}}$ are the leading order (LO) and next-to-leading order (NLO) contributions to the conversion matrix from the chiral seven-operator basis to the physical ten-operator basis in the $\rm\overline{MS}$ scheme and were given in Eqs.~(59) and (65) of Ref.~\cite{Lehner:2011fz}, respectively.  We use the strong coupling constant given in Eq.~\eqref{eq:alpha_s} to evaluate $\Delta T^{\rm\overline{MS}}$. The indices $j$ and $k$ run over the set ${\cal C} = \{1,2,3,5,6,7,8\}$.
Although the $\Delta I = 3/2$ ($I=2$) channel is not contributed from the four QCD penguin operators, and the chiral seven-operator basis is not required, 
we put zero in the corresponding irrelevant matrix elements and keep the formula consistent with the $\Delta I = 1/2$ ($I=0$) channel.

Note that the above formulae are valid for the on-shell matrix elements, while none of the two-pion states on our lattice ensemble has the same energy as the kaon mass.
Thus our lattice results for matrix elements need to be interpolated to the physical kinematics.
We spend the first two subsections for the calculation of the renormalized on-shell matrix elements using the results obtained in former sections.
Then we quote our results for the on-shell $K\to\pi\pi$ amplitude and test the $\Delta I = 1/2$ rule after briefly providing the numerical values of Wilson coefficients $z_i^{\overline{\rm MS}}(\mu_h)$ and $y_i^{\overline{\rm MS}}(\mu_h)$.  Lastly we quote our result for Re($\varepsilon'/\varepsilon$).

\subsection{Matrix elements in the chiral basis with unphysical kinematics}
\label{sec:MEChBasis}

We now explicitly introduce the isospin index $I$ of the final two-pion state to express the matrix elements $M_{I,n,i}^{\rm lat}$ on the lattice, although it was omitted in Section~\ref{sec:3pt}.  Here we also introduce the superscript `lat' indicating the unrenormalized matrix elements.  In this subsection we convert these matrix elements obtained in Section~\ref{sec:3pt} to those in the chiral seven-operator basis, $M_{I,n,j}^{\prime\,\rm lat}$.

We follow the procedure applied in Ref.~\cite{RBC:2020kdj} to optimize the statistical precision of the chiral-basis matrix elements.  
While the matrix elements in the ten-operator basis are uniquely expressed in terms of those in the chiral basis,
\begin{equation}
M_{I,n,i}^{\rm lat} = \sum_{j\in{\cal C}}T_{ij}M_{I,n,j}^{\prime\,\rm lat}
\label{eq:ME10_ME7}
\end{equation}
with the known matrix $T$, the inverse problem to determine $M_{I,n,j}^{\prime\,\rm lat}$ is ill-defined.
In other words, there are infinite number of combinations of the matrix elements in the ten-operator basis to express those in the chiral basis.  Because of the violation of the Fierz symmetry due to the stochastic A2A method, the statistical precision of the chiral-basis matrix elements depends on the combinations of the ten matrix elements and can be optimized by a correlated $\chi^2$ fit to Eq.~\eqref{eq:ME10_ME7}.
We perform the fit for each isospin channel and each two-pion final state labeled by $n$ individually.
For the $\Delta I = 1/2$ channel, we use the ten values of matrix elements, $i=1,2,\ldots,10$, for the fit to determine the seven values of $M_{0,n,j}^{\prime\,\rm lat}$ for $j\in{\cal C}$.
For the $\Delta I = 3/2$ channel there is no contribution from the QCD penguin operators.  Out of the remaining six matrix elements with the relevant operators, $M_{(I=)2,n,9}^{\rm lat}$ and $M_{2,n,10}^{\rm lat}$ are obtained from the identical contractions as $M_{2,n,1}^{\rm lat}$ and $M_{2,n,2}^{\rm lat}$, respectively, as seen in Appendix~\ref{sec:contractions}, and these pairs of matrix elements are therefore 100\% correlated.  We hence use only four matrix elements $M_{2,n,1}^{\rm lat}$, $M_{2,n,2}^{\rm lat}$, $M_{2,n,7}^{\rm lat}$ and $M_{2,n,8}^{\rm lat}$ for the fit to determine the three linearly independent matrix elements $M_{2,n,j}^{\prime\,\rm lat}$ for $j=1,7,8$.

\begin{table}[tbp]
\centering
\begin{tabular}{|cc|ccc|}
\hline
$I$ & $i$ & $a^3M_{I,0,i}^{\prime\,\rm lat}$ & $a^3M_{I,1,i}^{\prime\,\rm lat}$ & $a^3M_{I,2,i}^{\prime\,\rm lat}$ \\
\hline
 & 1 & $0.001444(46)$ & $0.00504(49)$ & $0.0105(13)$\\
2 & 7 & $0.015392(99)$ & $0.01218(19)$ & $0.01319(61)$\\
 & 8 & $0.04979(30)$ & $0.04378(41)$ & $0.0480(10)$\\
\hline
 & 1 & $-0.00002(42)$ & $0.0029(26)$ & $0.0030(59)$\\
 & 2 & $-0.00120(16)$ & $-0.0028(11)$ & $-0.0004(24)$\\
 & 3 & $0.00164(19)$ & $0.0057(12)$ & $0.0059(19)$\\
0 & 5 & $-0.00589(52)$ & $-0.0136(34)$ & $-0.0019(80)$\\
 & 6 & $-0.0175(10)$ & $-0.0596(66)$ & $-0.049(14)$\\
 & 7 & $0.02861(23)$ & $0.0368(13)$ & $0.0232(34)$\\
 & 8 & $0.08940(50)$ & $0.1112(28)$ & $0.0555(91)$\\
\hline
\end{tabular}
\caption{Unrenormalized matrix elements $a^3M_{I,n,i}^{\prime\,\rm lat}$ with the three lowest two-pion states ($n=0,1,2$) and the four-quark operators in the chiral basis.}
\label{tab:MEchiral}
\end{table}

The results are summarized in Table~\ref{tab:MEchiral} for the both $I=2$ and 0 two-pion final states of the three lowest energy levels $n=0,1,2$.  

\subsection{Renormalized matrix elements with physical kinematics}
\label{sec:onshellME}

We renormalize these chiral-basis matrix elements nonperturbatively to the RI-SMOM schemes and perform a step scaling,
\begin{equation}
M_{I,n,j}^{\prime\,\rm RI}(\mu_h)
= \sum_{k,l\in{\cal C}}\Sigma^{\rm RI}_{jk}(\mu_h,\mu)Z_{kl}^{\rm RI\leftarrow lat}(\mu)
F_{I,n}M_{I,n,l}^{\prime\,\rm lat}.
\label{eq:RIMEs}
\end{equation}
In Section~\ref{sec:NPR} we obtained the results for the renormalization matrix $Z_{kl}^{\rm RI\leftarrow lat}(\mu)$ and the step-scaling matrix $\Sigma^{\rm RI}_{jk}(\mu_h,\mu)$ for $\mu = \mu_1,\mu_2$ and for two RI schemes, SMOM($\gamma_\mu,\gamma_\mu$) and SMOM($\slashchar q,\slashchar q$).  We also perform a finite-volume correction using the LL factors $F_{I,n}$ obtained in Section~\ref{sec:2pt}.  Here the isospin and state indices $I$ and $n$, respectively, which are now explicitly added to the LL factor, are not summed over.

Now we interpolate these results to the physical kinematics and determine the on-shell matrix elements $M_{I,j}^{\prime\,\rm RI}(\mu_h)$, which are expressed without a state index. 
For the $\Delta I = 3/2$ channel the behavior of the $K\to\pi\pi$ matrix elements in the unphysical-kinematics region encompassing the $E_{\pi\pi}$-dependence was investigated using ChPT in Ref.~\cite{Lin:2002nq}, which indicates the matrix elements in the rest frame receive the same order of contributions from $E_{\pi\pi}$-independent, $E_{\pi\pi}$ and $E_{\pi\pi}^2$ terms in units of $m_K$.  Because of many unknown low-energy constants, the coefficients for these terms are not known.  
Ideally, we would perform this interpolation with at least three data points in order to account for these different contributions.
However we expect involving the second excited two-pion state with the energy 700~MeV for $I=0$ and 800~MeV for $I=2$ in the interpolation would unnecessarily increase the systematic error mainly because this energy level is well beyond the range that any ChPT-inspired interpolation function might be expected to be reliable.  Also this energy region is above the four-pion threshold where the LL factor is not a strictly accurate factor to normalize the finite-volume two-pion state to the infinite-volume one.  For these reasons we only use the off-shell matrix elements with the ground ($M_{2,0,j}^{\prime\,\rm RI}(\mu_h)$) and first-excited ($M_{2,1,j}^{\prime\,\rm RI}(\mu_h)$) two-pion final states to determine the matrix elements at the on-shell point.  
While the energy of the first-excited state is also above the four-pion threshold for this isospin channel, some earlier works suggest that the systematic error on the L\"uscher formalism for scattering phase shifts, which is closely related to the LL factor, is not significant when the energy is slightly above the four-pion threshold~\cite{NPLQCD:2011htk,Fischer:2020jzp,RBC:2023xqv,Bruno:2023pde}.  

While further comments on the $\Delta I = 1/2$ channel are given in the next paragraph, we perform two interpolations with the following functions for both $\Delta I=3/2$ and $\Delta I = 1/2$ channels
\begin{align}
M_{I,n,j}^{\prime\,\rm RI}(\mu_h) &= M_{I,j}^{\prime\,\rm RI,lin}(\mu_h) + c_{I,j}^{\rm RI,lin}\frac{E_{I,n} - m_K}{m_K},
\label{eq:linint}\\
M_{I,n,j}^{\prime\,\rm RI}(\mu_h) &= M_{I,j}^{\prime\,\rm RI,quad}(\mu_h) + c_{I,j}^{\rm RI,quad}\frac{E_{I,n}^2 - m_K^2}{m_K^2},
\label{eq:quadint}
\end{align}
where we input the kaon mass $m_K$ and isospin-$I$ two-pion energies $E_{I,n}$ obtained in Section~\ref{sec:2pt} and the off-shell matrix elements $M_{I,n,j}^{\prime\,\rm RI}(\mu_h)$ obtained in Eq.~\eqref{eq:RIMEs} to determine the parameters $M_{I,j}^{\prime\,\rm RI,lin/quad}(\mu_h)$ and $c_{I,j}^{\rm lin/quad}$.
We use $M^{\prime\,\rm RI,lin}_{I,j}(\mu_h)$ as our primary result for $M^{\prime\,\rm RI}_{I,j}(\mu_h)$ for calculation of the $\rm\overline{MS}$ matrix elements $M_{I,i}^{\rm\overline{MS}}(\mu_h)$ in Eq.~\eqref{eq:MERItoMEMS}, while $M^{\prime\,\rm RI,quad}_{I,j}(\mu_h)$ is used to estimate the systematic error from the interpolation (see Appendix~\ref{sec:syserr_onshell_limit}).
We quote the systematic error due to the assumption of the somewhat incomplete functions~\eqref{eq:linint} and \eqref{eq:quadint} by taking the difference between $M_{I,j}^{\prime\,\rm RI,lin}(\mu_h)$ and $M_{I,j}^{\prime\,\rm RI,quad}(\mu_h)$. 

For the $\Delta I = 1/2$ channel, since the mixing contribution from lower dimensional operators that vanish by the equations of motion is subtracted in a different way for determining the unrenormalized matrix elements in Section~\ref{sec:Meff} and for calculating the renormalization matrix in Section~\ref{sec:NPR}, multiplying the renormalization matrix with the matrix elements may not eliminate the logarithmic divergence.  This remaining divergence depends on the two-pion energy and vanishes at the on-shell point.
Therefore the on-shell $K\to\pi\pi$ amplitude is still physical as long as the interpolation to the on-shell point is successfully done.  We perform the interpolations based on the same assumptions as for the $\Delta I = 3/2$ channel using Eqs~\eqref{eq:linint} and \eqref{eq:quadint}.  

\begin{figure}[tp]
\begin{center}
\begin{tabular}{c}
\includegraphics[width=80mm]{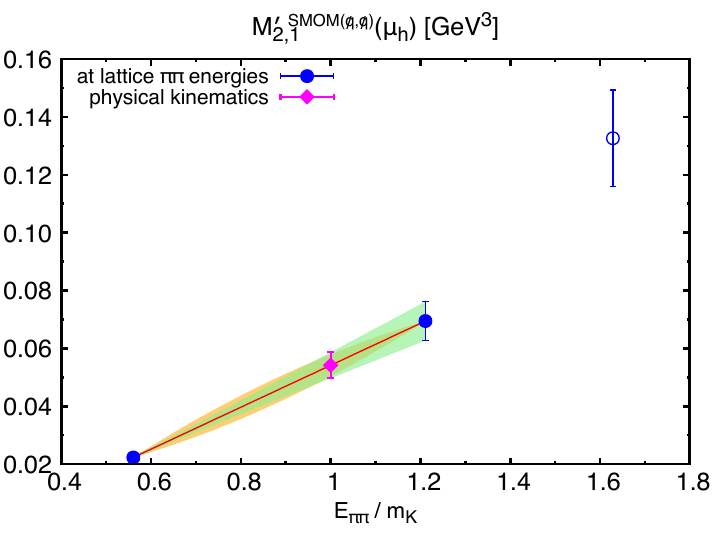}
\end{tabular}
\hfill
\begin{tabular}{c}
\includegraphics[width=80mm]{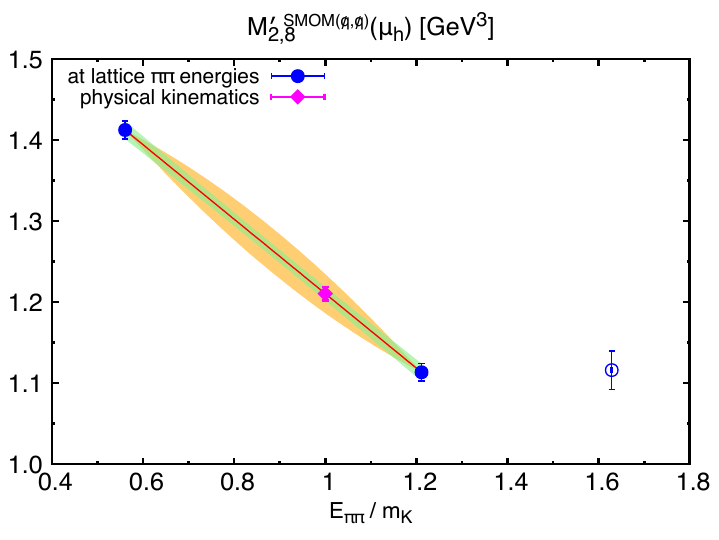}
\end{tabular}
\begin{tabular}{c}
\includegraphics[width=80mm]{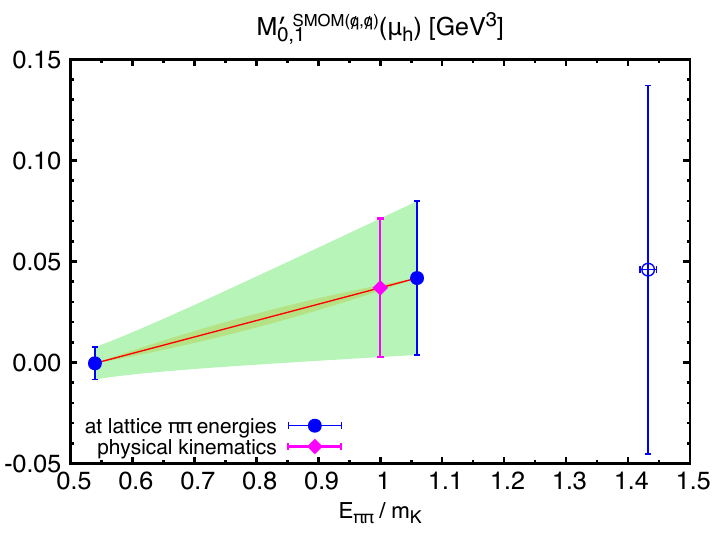}
\end{tabular}
\hfill
\begin{tabular}{c}
\includegraphics[width=80mm]{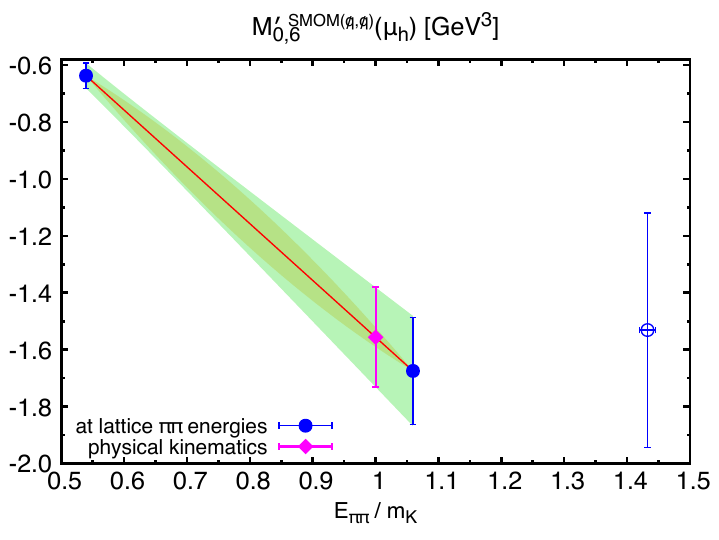}
\end{tabular}
\caption{
Examples of interpolation results for the SMOM($\slashchar q,\slashchar q$) matrix elements to the physical kinematics.  The renormalization scale is nonperturbatively raised from $\mu_1$ to $\mu_h$ by a step scaling.  Matrix elements of $Q_1'$ (upper/left) and $Q_8'$ (upper/right) are shown for the $\Delta I = 3/2$ channel, while those of $Q_1'$ (lower/left) and $Q_6'$ (lower/right) are shown for the $\Delta I = 1/2$ channel.  
We use the matrix elements with the ground and first-excited two-pion states (filled circles) for the interpolation, while those with the second excited two-pion state (unfilled circle) are excluded because of the expected inaccuracy of the LL factor at the energy level far above the four-pion threshold.
The solid red line, green band and diamond represent the result for the linear interpolation in Eq.~\eqref{eq:linint}, while the systematic uncertainty from this assumption is estimated as the difference from the quadratic interpolation in Eq.~\eqref{eq:quadint} and expressed by the yellow band, which is totally hidden behind the green band of statistical error for the $\Delta I = 1/2$ channel. 
}
\label{fig:interpolateME}
\end{center}
\end{figure}

\begin{table}[tbp]
\centering
\footnotesize
\begin{tabular}{|c|cccc|cccc|}
\hline
RI & \multicolumn{4}{|c|}{SMOM$(\gamma_\mu,\gamma_\mu)$} & \multicolumn{4}{c|}{SMOM$(\slashchar{q},\slashchar{q})$} \\
$\mu$  & \multicolumn{2}{|c}{$\mu_1$} & \multicolumn{2}{c|}{$\mu_2$} & \multicolumn{2}{c}{$\mu_1$} & \multicolumn{2}{c|}{$\mu_2$} \\
 $X$ & lin & quad & lin & quad & lin & quad & lin & quad \\
\hline
$M_{2,1}^{\prime\,{\rm RI},X}$ & 0.0541(45) & 0.0503(39) & 0.0484(40) & 0.0450(35) & 0.0554(46) & 0.0515(40) & 0.0617(51) & 0.0574(45)\\
$M_{2,7}^{\prime\,{\rm RI},X}$ & 0.2748(26) & 0.2824(24) & 0.2033(21) & 0.2093(19) & 0.2676(26) & 0.2751(24) & 0.2470(26) & 0.2543(24)\\
$M_{2,8}^{\prime\,{\rm RI},X}$ & 1.2104(85) & 1.2345(82) & 0.9445(64) & 0.9633(62) & 1.2942(95) & 1.3199(92) & 1.2689(88) & 1.2941(85)\\
$c_{2,1}^{{\rm RI},X}$ & 0.072(11) & 0.0409(60) & 0.0647(95) & 0.0365(54) & 0.074(11) & 0.0418(61) & 0.083(12) & 0.0466(68)\\
$c_{2,7}^{{\rm RI},X}$ & -0.1455(63) & -0.0821(36) & -0.1140(52) & -0.0644(29) & -0.1425(62) & -0.0804(35) & -0.1390(63) & -0.0785(36)\\
$c_{2,8}^{{\rm RI},X}$ & -0.460(22) & -0.260(13) & -0.358(18) & -0.2023(99) & -0.490(24) & -0.277(14) & -0.480(24) & -0.271(13)\\
\hline
$M_{0,1}^{\prime\,{\rm RI},X}$ & 0.036(34) & 0.035(32) & 0.032(30) & 0.031(29) & 0.037(34) & 0.036(33) & 0.041(38) & 0.040(37)\\
$M_{0,2}^{\prime\,{\rm RI},X}$ & -0.107(25) & -0.105(24) & -0.069(18) & -0.067(18) & -0.162(36) & -0.159(35) & -0.116(26) & -0.114(25)\\
$M_{0,3}^{\prime\,{\rm RI},X}$ & 0.109(23) & 0.106(23) & 0.093(20) & 0.091(19) & 0.131(30) & 0.128(29) & 0.141(28) & 0.138(28)\\
$M_{0,5}^{\prime\,{\rm RI},X}$ & -0.337(74) & -0.331(72) & -0.205(52) & -0.202(51) & -0.48(11) & -0.47(11) & -0.311(78) & -0.306(76)\\
$M_{0,6}^{\prime\,{\rm RI},X}$ & -1.44(16) & -1.40(16) & -1.12(13) & -1.10(12) & -1.56(17) & -1.52(17) & -1.51(17) & -1.48(16)\\
$M_{0,7}^{\prime\,{\rm RI},X}$ & 0.731(21) & 0.732(20) & 0.553(17) & 0.553(17) & 0.714(21) & 0.714(20) & 0.673(21) & 0.673(21)\\
$M_{0,8}^{\prime\,{\rm RI},X}$ & 2.883(63) & 2.888(60) & 2.249(49) & 2.253(47) & 3.080(67) & 3.086(65) & 3.019(66) & 3.024(63)\\
$c_{0,1}^{{\rm RI},X}$ & 0.079(63) & 0.050(39) & 0.071(56) & 0.044(35) & 0.081(64) & 0.051(40) & 0.090(72) & 0.057(45)\\
$c_{0,2}^{{\rm RI},X}$ & -0.118(41) & -0.074(26) & -0.073(32) & -0.045(20) & -0.191(56) & -0.120(35) & -0.133(43) & -0.083(27)\\
$c_{0,3}^{{\rm RI},X}$ & 0.136(42) & 0.085(26) & 0.115(36) & 0.072(22) & 0.165(49) & 0.103(31) & 0.175(50) & 0.109(31)\\
$c_{0,5}^{{\rm RI},X}$ & -0.35(12) & -0.217(78) & -0.191(90) & -0.120(56) & -0.53(17) & -0.33(10) & -0.32(12) & -0.200(78)\\
$c_{0,6}^{{\rm RI},X}$ & -1.83(30) & -1.15(19) & -1.44(23) & -0.90(15) & -1.99(32) & -1.25(20) & -1.94(31) & -1.22(19)\\
$c_{0,7}^{{\rm RI},X}$ & -0.037(46) & -0.023(29) & -0.022(37) & -0.014(23) & -0.036(46) & -0.022(29) & -0.027(46) & -0.017(29)\\
$c_{0,8}^{{\rm RI},X}$ & -0.32(16) & -0.200(97) & -0.25(12) & -0.156(76) & -0.34(17) & -0.21(10) & -0.34(16) & -0.21(10)\\
\hline
\end{tabular}
\caption{Results for interpolations of infinite-volume matrix elements in the chiral basis to the physical kinematics.  Results renormalized in the multiple renormalization schemes, SMOM$(\gamma_\mu,\gamma_\mu)$ and SMOM$(\slashchar q,\slashchar q)$, after a step scaling from the low scales $\mu=\mu_1\approx1.3$~GeV and $\mu_2\approx1.5$~GeV to the high scale $\mu_h\approx4.0$~GeV are shown in units of GeV$^3$.  The superscript $X$ specifies the interpolation procedure, Eq.~\eqref{eq:linint} for $X=\rm lin$ and Eq.~\eqref{eq:quadint} for $X = \rm quad$. Errors are statistical, only.}
\label{tab:interpolation summary}
\end{table}

Figure~\ref{fig:interpolateME} shows some results for the linear interpolation \eqref{eq:linint} of the matrix elements in the SMOM($\slashchar q,\slashchar q$) scheme at $\mu_h$ to the physical kinematics.  As noted above the only ground and first-excited two-pion final states are used for the interpolation.  The systematic error due to the interpolation is estimated as the difference from the quadratic interpolation \eqref{eq:quadint} and expressed by the yellow band.
For the $\Delta I = 1/2$ channel since the energy $E_{0,1}$ of the first-excited two-pion state is close to $m_K$ and the disconnected diagram increases the statistical error, the systematic error due to the interpolation in this isospin channel is small compared to the statistical error (green band).
Table~\ref{tab:interpolation summary} summarizes the results for all interpolation parameters in Eqs~\eqref{eq:linint} and \eqref{eq:quadint} for the SMOM($\slashchar q,\slashchar q$) and the SMOM($\gamma_\mu,\gamma_\mu$) schemes with step scaling performed from $\mu=\mu_1,\mu_2$ to $\mu_h$ before the interpolations.

\begin{table*}[tbp]
\centering
\begin{tabular}{|cc|ccc|ccc|}
\hline
& & \multicolumn{3}{|c|}{via SMOM$(\gamma_\mu,\gamma_\mu)$} & \multicolumn{3}{c|}{via SMOM$(\slashchar{q},\slashchar{q})$} \\
 $I$ & $i$ & $\mu_1$ & $\mu_2$ & GPBC ($\mu_2$) & $\mu_1$ & $\mu_2$ & GPBC ($\mu_2$) \\
\hline
\multirow{3}{*}{2}
& 2 & $0.01087(90)$ & $0.00971(81)$ & \ldots & $0.01098(91)$ & $0.0123(10)$ & \ldots\\
& 7 & $0.2723(26)$ & $0.2014(21)$ & \ldots & $0.2649(26)$ & $0.2443(26)$ & \ldots\\
& 8 & $1.2966(91)$ & $1.0121(69)$ & \ldots & $1.3247(97)$ & $1.2991(90)$ & \ldots\\
\hline
\multirow{10}{*}{0}
& 1 & $-0.099(26)$ & $-0.062(20)$ & $-0.093(18)$ & $-0.151(39)$ & $-0.101(28)$ & $-0.107(22)$\\
& 2 & $0.120(24)$ & $0.102(20)$ & $0.143(14)$ & $0.136(26)$ & $0.142(26)$ & $0.147(15)$\\
& 3 & $-0.094(73)$ & $-0.012(52)$ & $-0.053(44)$ & $-0.22(12)$ & $-0.060(77)$ & $-0.086(61)$\\
& 4 & $0.147(69)$ & $0.169(52)$ & $0.200(40)$ & $0.09(10)$ & $0.205(73)$ & $0.185(53)$\\
& 5 & $-0.334(74)$ & $-0.203(52)$ & $-0.311(48)$ & $-0.47(11)$ & $-0.308(78)$ & $-0.348(62)$\\
& 6 & $-1.50(17)$ & $-1.18(13)$ & $-1.272(86)$ & $-1.56(17)$ & $-1.51(17)$ & $-1.308(90)$\\
& 7 & $0.726(21)$ & $0.548(17)$ & $0.784(23)$ & $0.707(21)$ & $0.666(21)$ & $0.769(23)$\\
& 8 & $3.086(67)$ & $2.408(52)$ & $3.308(63)$ & $3.152(69)$ & $3.089(67)$ & $3.389(64)$\\
& 9 & $-0.102(24)$ & $-0.086(21)$ & $-0.114(19)$ & $-0.118(27)$ & $-0.122(27)$ & $-0.117(20)$\\
& 10 & $0.117(26)$ & $0.078(20)$ & $0.123(19)$  & $0.169(38)$ & $0.122(28)$ & $0.137(22)$\\
\hline
\end{tabular}
\caption{$K\to\pi\pi$ matrix elements $M_{I,i}^{\rm\overline{MS}}(\mu_h)$ in the $\overline{\rm MS}$ scheme at $\mu_h\approx4.01$~GeV with the physical kinematics.  Results obtained through multiple intermediate renormalization schemes (SMOM$(\gamma_\mu,\gamma_\mu)$ and SMOM$(\slashchar q,\slashchar q)$) and scales ($\mu_1$ and $\mu_2$) are shown in units of GeV$^3$.  For the $\Delta I = 1/2$ channel we also show the results from the GPBC calculation, which were obtained through an intermediate renormalization scale near $\mu_2$ and summarized in Table~XIV of Ref.~\cite{RBC:2020kdj}.  Since $\Delta T^{\rm\overline{MS}}_{ij}$ in Eq.~\eqref{eq:MERItoMEMS} is zero at NLO level for $i=1,2,9,10$, the relation $M_{2,1}^{\rm\overline{MS}}(\mu_h) = M_{2,2}^{\rm\overline{MS}}(\mu_h) = \frac{2}{3}M_{2,9}^{\rm\overline{MS}}(\mu_h) = \frac{2}{3}M_{2,10}^{\rm\overline{MS}}(\mu_h)$ still obeys at this order of perturbation theory in the $\rm\overline{MS}$ scheme and hence the results for $i=1,9,10$ in the $\Delta I = 3/2$ channel are dropped.  The QCD penguin ($i=3,4,5,6$) contributions to the $\Delta I = 3/2$ channel are zero also in the $\rm\overline{MS}$ scheme and dropped. Errors are statistical, only.}
\label{tab:MSbarMEs}
\end{table*}

\begin{table}[tp]
\centering
\begin{tabular}{|c|ccc|}
\hline
& 2 & 7 & 8\\
\hline
2 & $8.31 \times10^{-7}$ & $-1.75 \times10^{-7}$ & $1.20 \times10^{-6}$\\
7 & $-1.75 \times10^{-7}$ & $6.56 \times10^{-6}$ & $1.54 \times10^{-5}$\\
8 & $1.20 \times10^{-6}$ & $1.54 \times10^{-5}$ & $9.37 \times10^{-5}$\\
\hline
\end{tabular}
\caption{Covariance matrix between the $\Delta I = 3/2$ matrix elements in the $\rm\overline{MS}$ scheme at $\mu_h$ obtained through the SMOM$(\slashchar q,\slashchar q)$ scheme and the low scale $\mu_1$ given in Table~\ref{tab:MSbarMEs}.  Results are shown in units of GeV$^6$.}
\label{tab:cov_mtx3_1p3GeV_qslash_qslash_odd}
\end{table}

\begin{table}[tp]
\centering
\scriptsize
\begin{tabular}{|c|cccccccccc|}
\hline
& 1 & 2 & 3 & 4 & 5 & 6 & 7 & 8 & 9 & 10\\
\hline
1 & $0.00148$ & {\tiny $3.72 \times10^{-5}$} & $0.00423$ & $0.00276$ & $0.00364$ & $0.00136$ & {\tiny $5.99 \times10^{-5}$} & $-0.000478$ & $0.000110$ & $-0.00134$\\
2 & {\tiny $3.72 \times10^{-5}$} & $0.000700$ & $0.00125$ & $0.00187$ & $0.000807$ & $0.00130$ & {\tiny $4.20 \times10^{-5}$} & $0.000288$ & $-0.000568$ & {\tiny $9.50 \times10^{-5}$}\\
3 & $0.00423$ & $0.00125$ & $0.0149$ & $0.0118$ & $0.0121$ & $0.00612$ & {\tiny -$2.48 \times10^{-5}$} & $-0.00118$ & $-0.00112$ & $-0.00411$\\
4 & $0.00276$ & $0.00187$ & $0.0118$ & $0.0108$ & $0.00909$ & $0.00553$ & {\tiny -$2.77 \times10^{-5}$} & $-0.000342$ & $-0.00176$ & $-0.00265$\\
5 & $0.00364$ & $0.000807$ & $0.0121$ & $0.00909$ & $0.0124$ & $0.00881$ & $-0.000151$ & $-0.00177$ & $-0.000586$ & $-0.00342$\\
6 & $0.00136$ & $0.00130$ & $0.00612$ & $0.00553$ & $0.00881$ & $0.0301$ & $-0.000893$ & $-0.00470$ & $-0.00102$ & $-0.00108$\\
7 & {\tiny $5.99 \times10^{-5}$} & {\tiny $4.20 \times10^{-5}$} & {\tiny -$2.48 \times10^{-5}$} & {\tiny -$2.77 \times10^{-5}$} & $-0.000151$ & $-0.000893$ & $0.000431$ & $0.000930$ & $0.000102$ & {\tiny $8.44 \times10^{-5}$}\\
8 & $-0.000478$ & $0.000288$ & $-0.00118$ & $-0.000342$ & $-0.00177$ & $-0.00470$ & $0.000930$ & $0.00473$ & $-0.000127$ & $0.000639$\\
9 & $0.000110$ & $-0.000568$ & $-0.00112$ & $-0.00176$ & $-0.000586$ & $-0.00102$ & $0.000102$ & $-0.000127$ & $0.000725$ & {\tiny $4.73 \times10^{-5}$}\\
10 & $-0.00134$ & {\tiny $9.50 \times10^{-5}$} & $-0.00411$ & $-0.00265$ & $-0.00342$ & $-0.00108$ & {\tiny $8.44 \times10^{-5}$} & $0.000639$ & {\tiny $4.73 \times10^{-5}$} & $0.00148$\\
\hline
\end{tabular}
\caption{Covariance matrix between the $\Delta I = 1/2$ matrix elements in the $\rm\overline{MS}$ scheme at $\mu_h$ obtained through the SMOM$(\slashchar q,\slashchar q)$ scheme and the low scale $\mu_1$ given in Table~\ref{tab:MSbarMEs}.  Results are shown in units of GeV$^6$.}
\label{tab:cov_mtx_1p3GeV_qslash_qslash_odd}
\end{table}

Table~\ref{tab:MSbarMEs} shows the $\rm\overline{MS}$ matrix elements $M_{I,i}^{\rm\overline{MS}}(\mu_h)$ obtained through various intermediate schemes and scales.  The result should in principle be independent of the intermediate scheme and scale and the difference seen for different intermediate schemes or scales is because of the systematic error discussed in Section~\ref{sec:systErrNPT}.  We see significant difference in well-resolved matrix elements obtained through the SMOM($\gamma_\mu,\gamma_\mu$) scheme at $\mu_2$ from those through different scheme or scale.  
We attribute this deviation to finite-lattice spacing errors in Section~\ref{sec:systErrNPT}.
For the $\Delta I = 1/2$ channel, we also show the earlier results obtained from the GPBC calculation~\cite{RBC:2020kdj} carried out on a smaller lattice spacing $a^{-1}_{\rm32ID}\approx1.38$~GeV for comparison.  Ignoring the results through the SMOM$(\gamma_\mu,\gamma_\mu)$ scheme at $\mu_2$, which may have significant systematic error as mentioned aobove, most of the results are consistent between PBC and GPBC but somewhat significant difference is observed in the matrix element of $Q_6$.  While the difference could be due to the finite lattice spacing error, the statistical error is not small enough to conclude at this stage.
We choose the matrix elements obtained through the SMOM($\slashchar q,\slashchar q$) scheme at $\mu_1$ to calculate the central value of the amplitudes.  The corresponding covariance matrix is tabulated in Tables~\ref{tab:cov_mtx3_1p3GeV_qslash_qslash_odd} and \ref{tab:cov_mtx_1p3GeV_qslash_qslash_odd} for the $\Delta I = 3/2$ and $\Delta I = 1/2$ channels, respectively.

\subsection{Wilson coefficients}
\label{sec:WCs}

\begin{table}[tbp]
\centering
\begin{tabular}{ccc}
\hline
$i$ & $z_i^{\rm\overline{MS}}(\mu_h)$ & $y_i^{\rm\overline{MS}}(\mu_h)$ \\
\hline
1 & $-0.201$ & 0 \\
2 & $1.091$ & 0 \\
3 & $-0.00536$ & $0.0193$ \\
4 & $0.0251$  & $-0.0568$ \\
5 & $-0.00619$ & $0.01337$ \\
6 & $0.0178$  & $-0.0569$ \\
7 & $0.000134$ & $-0.000260$ \\
8 & $-0.000122$  & $0.000535$ \\
9 & $0.0000537$ & $-0.00965$ \\
10 & $0.0000945$ & $0.00191$ \\
\hline
\end{tabular}
\caption{Wilson coefficients $z_i$ and $y_i$ renormalized in the $\rm\overline{MS}$ used in this work.
}
\label{tab:Wcoefs}
\end{table}

We follow the procedure explained in Ref.~\cite{RBC:2020kdj} for computing the Wilson coefficients using the NLO perturbative expansion given in Ref.~\cite{Buras:1993dy,Buchalla:1995vs}.  The Wilson coefficients depend on many Standard-Model parameters that accommodate the effects of higher-energy physics and were summarized in Table~XI of Ref.~\cite{RBC:2020kdj}.  Since some of the values of these parameters have been updated \cite{ParticleDataGroup:2022pth}, we update the Wilson coefficients accordingly employing the latest values of these parameters.  Table~\ref{tab:Wcoefs} summarizes the Wilson coefficients.  Here we use the value of $\alpha_s$ given in Eq.~\eqref{eq:alpha_s}.

Some of the CKM matrix elements have also been updated \cite{ParticleDataGroup:2022pth}, and we therefore use their updated values and the parameter $\tau = -V_{ts}^*V_{td}/V_{us}^*V_{ud}$ in Eq.~\eqref{eq:Wcoef_C}.  The new value of $\tau = 0.001473(67) - 0.000640(33)\img$ is determined by using the Wolfenstein parametrization expanded to the eighth order with the parameters given in Ref.~\cite{ParticleDataGroup:2022pth}. 

\subsection{$K\to\pi\pi$ amplitudes}
\label{sec:AI}

We use Eq.~\eqref{eq:def_AI} to determine the $K\to\pi\pi$ amplitudes $A_I$ with the Wilson coefficients and the $\rm\overline{MS}$ matrix elements given in the previous subsections.  
It is known that the matrix elements of different four-quark operators contribute in different ways to Re($A_I$) and Im($A_I$).  In this subsection we discuss these different compositions as well as the final results for each Re($A_I$) and Im($A_I$).

\begin{table*}[tbp]
\centering
\begin{tabular}{|c|cc|cc|}
\hline
& \multicolumn{2}{|c|}{via SMOM$(\gamma_\mu,\gamma_\mu)$} & \multicolumn{2}{c|}{via SMOM$(\slashchar{q},\slashchar{q})$} \\
$i$ & $\mu_1$ & $\mu_2$ & $\mu_1$ & $\mu_2$ \\
\hline
1 & $-0.393(33)$ & $-0.351(29)$ & $-0.397(33)$ & $-0.443(37)$\\
2 & $2.14(18)$ & $1.91(16)$ & $2.16(18)$ & $2.41(20)$\\
7 & $0.006549(61)$ & $0.004843(50)$ & $0.006372(62)$ & $0.005876(61)$\\
8 & $-0.02822(20)$ & $-0.02202(15)$ & $-0.02883(21)$ & $-0.02827(20)$\\
9 & $0.0001160(96)$ & $0.0001036(86)$ & $0.0001172(97)$ & $0.000131(11)$\\
10 & $0.000286(24)$ & $0.000255(21)$ & $0.000289(24)$ & $0.000322(27)$\\
\hline
Total & $1.72(14)$ & $1.54(13)$ & $1.74(15)$ & $1.94(16)$\\
\hline
\end{tabular}
\caption{Contribution from each $\rm\overline{MS}$ matrix element to ${\rm Re}(A_2)$.  Results with multiple intermediate renormalization schemes (SMOM$(\gamma_\mu,\gamma_\mu)$ and SMOM$(\slashchar{q},\slashchar{q})$) and scales ($\mu_1$ and $\mu_2$) are shown in units of $10^{-8}$~GeV.  Errors are statistical, only.}
\label{tab:breakdown_ReA2_MSbarME}
\end{table*}

\begin{table*}[tbp]
\centering
\begin{tabular}{|c|cc|cc|}
\hline
& \multicolumn{2}{|c|}{via SMOM$(\gamma_\mu,\gamma_\mu)$} & \multicolumn{2}{c|}{via SMOM$(\slashchar{q},\slashchar{q})$} \\
$i$ & $\mu_1$ & $\mu_2$ & $\mu_1$ & $\mu_2$ \\
\hline
1 & $0$ & $0$ & $0$ & $0$\\
2 & $0$ & $0$ & $0$ & $0$\\
7 & $0.8162(77)$ & $0.6036(62)$ & $0.7941(77)$ & $0.7324(76)$\\
8 & $-7.997(56)$ & $-6.242(42)$ & $-8.170(60)$ & $-8.012(55)$\\
9 & $1.81(15)$ & $1.62(13)$ & $1.83(15)$ & $2.04(17)$\\
10 & $-0.359(30)$ & $-0.320(27)$ & $-0.362(30)$ & $-0.404(34)$\\
\hline
Total & $-5.73(12)$ & $-4.34(11)$ & $-5.91(13)$ & $-5.64(14)$\\
\hline
\end{tabular}
\caption{Same as Tab~\ref{tab:breakdown_ReA2_MSbarME} but the breakdown of ${\rm Im}(A_2)$ shown in units of $10^{-13}$~GeV.}
\label{tab:breakdown_ImA2_MSbarME}
\end{table*}

The contributions from each $\rm\overline{MS}$ matrix element at $\mu_h$ to $A_2$ (the summands on the right hand side of Eq.~\eqref{eq:def_AI} for $I=2$) and their total value are summarized in Table~\ref{tab:breakdown_ReA2_MSbarME} for the real part and in Table~\ref{tab:breakdown_ImA2_MSbarME} for the imaginary part.  The results obtained through multiple intermediate schemes and scales are presented.  Re($A_2$) receives the dominant contribution from $Q_2$ with 20\% cancellation from the subdominant contribution of $Q_1$.  Im($A_2$) is contributed mainly by $Q_8$ with 30\% cancellation from $Q_7$ and $Q_9$.

\begin{table*}[tbp]
\centering
\begin{tabular}{|c|cc|cc|}
\hline
& \multicolumn{2}{|c|}{via SMOM$(\gamma_\mu,\gamma_\mu)$} & \multicolumn{2}{c|}{via SMOM$(\slashchar{q},\slashchar{q})$} \\
$i$ & $\mu_1$ & $\mu_2$ & $\mu_1$ & $\mu_2$ \\
\hline
1 & $0.358(95)$ & $0.223(72)$ & $0.55(14)$ & $0.37(10)$\\
2 & $2.35(47)$ & $2.01(40)$ & $2.68(52)$ & $2.79(51)$\\
3 & $0.0090(70)$ & $0.0012(50)$ & $0.021(12)$ & $0.0058(73)$\\
4 & $0.066(31)$ & $0.076(23)$ & $0.042(47)$ & $0.092(33)$\\
5 & $0.0371(82)$ & $0.0225(58)$ & $0.052(12)$ & $0.0342(87)$\\
6 & $-0.480(54)$ & $-0.376(42)$ & $-0.496(55)$ & $-0.483(53)$\\
7 & $0.001745(51)$ & $0.001319(42)$ & $0.001701(50)$ & $0.001603(51)$\\
8 & $-0.00672(15)$ & $-0.00524(11)$ & $-0.00686(15)$ & $-0.00672(15)$\\
9 & $-0.000072(17)$ & $-0.000061(15)$ & $-0.000084(19)$ & $-0.000087(19)$\\
10 & $0.000205(45)$ & $0.000136(34)$ & $0.000296(67)$ & $0.000213(49)$\\
\hline
Total & $2.34(54)$ & $1.95(45)$ & $2.84(57)$ & $2.80(58)$\\
\hline
\end{tabular}
\caption{Same as Table~\ref{tab:breakdown_ReA2_MSbarME} but the breakdown of Re($A_0$) shown in units of $10^{-7}$~GeV.}
\label{tab:breakdown_ReA0_MSbarME}
\end{table*}

\begin{table*}[tbp]
\centering
\begin{tabular}{|c|cc|cc|}
\hline
& \multicolumn{2}{|c|}{via SMOM$(\gamma_\mu,\gamma_\mu)$} & \multicolumn{2}{c|}{via SMOM$(\slashchar{q},\slashchar{q})$} \\
$i$ & $\mu_1$ & $\mu_2$ & $\mu_1$ & $\mu_2$ \\
\hline
1 & $0$ & $0$ & $0$ & $0$\\
2 & $0$ & $0$ & $0$ & $0$\\
3 & $0.21(16)$ & $0.03(12)$ & $0.48(27)$ & $0.13(17)$\\
4 & $0.96(45)$ & $1.11(34)$ & $0.61(68)$ & $1.34(48)$\\
5 & $0.51(11)$ & $0.313(80)$ & $0.73(17)$ & $0.47(12)$\\
6 & $-9.9(1.1)$ & $-7.73(86)$ & $-10.2(1.1)$ & $-9.9(1.1)$\\
7 & $0.02175(63)$ & $0.01644(52)$ & $0.02120(62)$ & $0.01998(64)$\\
8 & $-0.1903(41)$ & $-0.1485(32)$ & $-0.1944(42)$ & $-0.1905(41)$\\
9 & $-0.113(27)$ & $-0.096(23)$ & $-0.131(30)$ & $-0.135(30)$\\
10 & $-0.0258(57)$ & $-0.0171(43)$ & $-0.0372(85)$ & $-0.0267(61)$\\
\hline
Total & $-8.5(1.2)$ & $-6.53(94)$ & $-8.7(1.2)$ & $-8.3(1.2)$\\
\hline
\end{tabular}
\caption{Same as Table~\ref{tab:breakdown_ReA2_MSbarME} but the breakdown of Im($A_0$) shown in units of $10^{-11}$~GeV.}
\label{tab:breakdown_ImA0_MSbarME}
\end{table*}

The same breakdowns for the $I=0$ two-pion final state are tabulated in Tables~\ref{tab:breakdown_ReA0_MSbarME} and \ref{tab:breakdown_ImA0_MSbarME} for Re($A_0$) and Im($A_0$), respectively.  
Re($A_0$) receives the dominant contribution from the matrix element of $Q_2$. 
It is interesting to observe that $Q_2$, which accounts for most of Re($A_0$), is the original operator for weak interaction, $i.e.$  $s \to u W^- $ followed by $W^- \to  \bar u d$. 
Thus, what was required to quantitatively understand Re($A_0$) was a proper non-perturbative framework for calculating the $K\to\pi\pi$ matrix element of $Q_2$.
As in the previous lattice calculation~\cite{RBC:2020kdj} we again find the QCD penguin operator $Q_6$~\cite{Shifman:1975tn} only plays a subdominant ($\sim20\%$) role in Re$(A_0)$, which is significantly canceled by another subdominant contribution of $Q_1$.  On the other hand, Im($A_0$) is dominated by the matrix element of $Q_6$.  The contributions from the other operators shift the result by 15\%.  This observation about Im($A_0$) is also consistent with our previous GPBC calculation~\cite{RBC:2020kdj}.

There is a significant dependence on the intermediate renormalization procedure as observed in the NPR study (Section~\ref{sec:systErrNPT}) and in the renormalized matrix elements (Section~\ref{sec:onshellME}).
Such a large difference was not observed in the GPBC work~\cite{RBC:2020kdj} on a finer lattice of $a_{\rm 32ID}^{-1} \approx 1.38$~GeV and is likely due to finite lattice spacing error as it appears more significant for $\mu_2\approx1.5$~GeV.
The difference is taken into account as a part of systematic uncertainty discussed below.

\begin{table}[tbp]
\centering
\begin{tabular}{lcrrrr}
\hline
Error source && Re$(A_2)$ & Im$(A_2)$ & Re$(A_0)$ & Im$(A_0)$ \\
\hline
NPR & & 6\% & 13\% & 16\% & 13\%\\
On-shell limit & & 7\% & 3\% & 1.8\% & 3\%\\
LL factor$^*$ & & 1.5\% & 1.5\% & 1.5\% & 1.5\%\\
Finite-volume corrections$^*$ & & 7\% & 7\% & 7\% & 7\%\\
Missing $G_1$ operator$^*$ & & 3\% & 3\% & 3\% & 3\%\\
Finite lattice spacing$^\dag$ & & 22\% & 22\% & 22\% & 22\%\\
Wilson coefficients$^*$ (HE) & & 12\% & 12\% & 12\% & 12\%\\
Parametric errors$^{*\,{\rm for}\,I=0}$ (HE) & & $< 1\%$ & 7\% & $<1\%$ & 6\%\\
\hline
Total & & 28\% & 30\% & 31\% & 30\%\\
\hline
\end{tabular}
\caption{Systematic error breakdowns for the $K\to\pi\pi$ amplitudes.  The errors with the symbol `$*$' are inherited from Ref.~\cite{RBC:2020kdj}.  The finite lattice spacing error with the symbol `$\dag$' is also estimated based on Ref.~\cite{RBC:2020kdj} with an assumption of $O(a^2)$ scaling.  The error sources with `(HE)' are associated with high-energy particle properties and independent of lattice calculations.
}
\label{tab:syserrAI}
\end{table}

We choose the results obtained through the SMOM$(\slashchar{q},\slashchar{q})$ intermediate scheme and scale $\mu_1$ as the central value of our final results, which read
\begin{align}
{\rm Re}(A_2) &= 1.74(15)(48)\times10^{-8}~{\rm GeV},
\label{eq:res_ReA2}\\
{\rm Im}(A_2) &= -5.91(13)(1.75)\times10^{-13}~{\rm GeV},
\label{eq:res_ImA2}\\
{\rm Re}(A_0) &= 2.84(57)(87)\times10^{-7}~{\rm GeV},
\label{eq:res_ReA0}\\
{\rm Im}(A_0) &= -8.7(1.2)(2.6)\times10^{-11}~{\rm GeV},
\label{eq:res_ImA0}
\end{align}
and are consistent with our earlier works, Ref.~\cite{Blum:2015ywa} for $A_2$ and Ref.~\cite{RBC:2020kdj} for $A_0$.
The first error is statistical, while the second error is the combined systematic error, whose breakdown is summarized in Table~\ref{tab:syserrAI}.  
The NPR error is estimated as half of the maximum difference from the other sets of intermediate scheme and scale.  The reason for halving is because the intermediate scale $\mu_2$, which always results in the largest deviation from the central value and hence determines the NPR error, may not be sufficiently small on this coarse lattice as discussed in Section~\ref{sec:systErrNPT}.  
The error from the on-shell limit is discussed in detail in Appendix~\ref{sec:syserr_onshell_limit} but basically estimated by propagating the difference between $M_{I,j}^{\prime\,{\rm SMOM}({\scriptsize \slashchar q},{\scriptsize \slashchar q}),\rm lin}(\mu_h)$ and $M_{I,j}^{\prime\,{\rm SMOM}({\scriptsize \slashchar q},{\scriptsize \slashchar q}),\rm quad}(\mu_h)$ as an error on $M_{I,j}^{\prime\,{\rm SMOM}({\scriptsize \slashchar q},{\scriptsize \slashchar q})}(\mu_h)$ in Eq.~\eqref{eq:MERItoMEMS}.
The other errors are quoted based on Ref.~\cite{RBC:2020kdj} as there is no reason to expect significant difference from our earlier estimations.  The finite lattice spacing error is extended with an assumption of the $O(a^2)$ scaling and the other errors with `$*$' are the same as in Ref.~\cite{RBC:2020kdj}.
The error from the Wilson coefficients is an estimate dominated mostly by the matching between the three- and four-flavor theories, that needs to be done below the charm threshold.  
We could in principle bypass the matching for the $\Delta I = 3/2$ channel at one-loop level because it is to accommodate the effects of matrix elements of four-quark operators including the charm and anti-charm quarks, which are all $\Delta I = 1/2$.  In this paper we still assign a 12\% Wilson-coefficient error to $\Delta I = 3/2$ channel since we have not computed the renormalization matrices in the four-flavor theory that could be used with the more accurate Wilson coefficients for this isospin channel and instead use the same renormalization procedure and Wilson coefficients as for the $\Delta I = 1/2$ channel.  The last error, which is from finite lattice spacing, is estimated with the assumption of $O(a^2)$ scaling with the corresponding error estimated in Ref.~\cite{RBC:2020kdj}.  This estimate is based on our earlier results for the $\Delta I = 3/2$ channel. 

\begin{table}[tp]
\centering
\begin{tabular}{ccc}
\hline
 & Re$(A_2)\,[10^{-8}$~GeV] & Im$(A_2)\,[10^{-13}$~GeV] \\
\hline
This work & $1.74(15)(30)$ & $-5.91(13)(1.18)$ \\
$a^{-1}_{\rm32ID}\approx1.38$~GeV & $1.381(46)(135)$ & $-6.93(49)(76)$ \\
Continuum limit & $1.501(39)(140)$ & $-8.05(23)(97)$ \\
\hline
\end{tabular}
\caption{Comparison of $A_2$ with our earlier results obtained with the anti-periodic down quark on a finer lattice $a^{-1}_{\rm32ID}\approx1.38$~GeV  \cite{Blum:2012uk} and the continuum limit with finer $48^3$ and $64^3$ Iwasaki gauge ensembles \cite{Blum:2015ywa}.
The imaginary part of earlier results are modified according to the significant (15\%) change in the $\tau$ parameter.  The first error is statistical and the second one combines the listed systematic errors in quadrature excluding the finite lattice spacing error.}
\label{tab:comparisonA2}
\end{table}
\begin{table}[tp]
\centering
\begin{tabular}{ccc}
\hline
 & Re$(A_0)\,[10^{-7}$~GeV] & Im$(A_0)\,[10^{-11}$~GeV] \\
\hline
This work & $2.84(57)(62)$ & $-8.7(1.2)(1.8)$ \\
GPBC, $a^{-1}_{\rm32ID}\approx1.38$~GeV & $2.99(32)(47)$ & $-6.98(62)(1.17)$ \\
\hline
\end{tabular}
\caption{Comparison of $A_0$ with our earlier work \cite{RBC:2020kdj} on a finer lattice $a^{-1}_{\rm32ID}\approx1.38$~GeV with GPBC.  The first error is statistical and the second one combines the listed systematic errors in quadrature excluding the finite lattice spacing error.}
\label{tab:comparisonA0}
\end{table}

We also considered another estimation of the systematic error from the finite lattice spacing by comparing the results from this work with our earlier results.  Table~\ref{tab:comparisonA2} shows a comparison of $A_2$ from this work and earlier works on the $32^3\times64$ lattice with the Iwasaki $+$ DSDR gauge action~\cite{Blum:2012uk} and on the $48^3\times64$ and $64^3\times128$ lattices with the Iwasaki gauge action~\cite{Blum:2015ywa}.  The chiral extrapolation of the results obtained with non-unitary light quark masses was performed in the former reference, while the physical pion and kaon masses are realized, and the continuum limit was taken in the latter work.  The values of Im($A_2$) from earlier works are modified according to the significantly updated value of the $\tau$ parameter.  Table~\ref{tab:comparisonA0} shows a comparison of $A_0$ between this work and our earlier work on the $32^3\times64$ lattice with GPBC at the physical pion and kaon masses~\cite{RBC:2020kdj}.  The first error on each value in these tables is statistical and the second error combines the listed systematic errors in quadrature except for the discretization error.  The tables indicate that these errors are not small enough to estimate the finite lattice spacing error from the comparisons.  Therefore we stick with the estimation of the finite lattice spacing error given in the previous paragraph, though it is estimated from the results for the $\Delta I = 3/2$ channel and may not be an ideal estimation for the $\Delta I = 1/2$ channel.  Unmasking the discretization error more accurately and taking the continuum limit of the $\Delta I = 1/2$ channel are important goals in the near future.

With the quoted values of the $K\to\pi\pi$ amplitude we can test the $\Delta I = 1/2$ rule, which is an experimental fact that kaons are roughly 450 times more likely to decay into the $I=0$ channel of two pions than the $I=2$ counterpart, $\mbox{Re}(A_0)/\mbox{Re}(A_2) = 22.45(6)$.  While leading-order perturbative QCD can explain the difference in the amplitudes only up to the factor of 2~\cite{Gaillard:1974nj,Altarelli:1974exa,Gilman:1978wm,Gilman:1979bc}, some theoretical indications of this fact were seen when we performed lattice calculations of the $\Delta I = 3/2$ channel~\cite{Blum:2012uk,Blum:2015ywa,RBC:2012ynq}, where we observed a sizable cancellation between the two Wick contractions of the $K\to\pi\pi$ three-point function with the dominant $(27,1)$ operator.  In our most recent work on the $\Delta I = 1/2$ channel with GPBC~\cite{RBC:2020kdj}, we were able to numerically verify it by combining the result for Re$(A_0)$ from the GPBC work with the continuum limit of Re$(A_2)$ from Ref.~\cite{Blum:2015ywa}, $\mbox{Re}(A_0)/\mbox{Re}(A_2) = 19.9(2.3)(4.4)$, with the statistical and systematic errors, respectively.  Here we quote the corresponding value calculated on the $24^3$ ensemble
\begin{equation}
    \mbox{Re}(A_0)/\mbox{Re}(A_2) = 16.3(3.7)(6.7),
\end{equation}
where the errors are, again, statistical and systematic, respectively.  

\subsection{Determination of $\varepsilon'$}

Now we quote our numerical result for the measure of direct $CP$ violation, Re$(\varepsilon'/\varepsilon)$.  Writing $\varepsilon = |\varepsilon|\e^{\img\phi_\varepsilon}$, the phase of $\varepsilon'/\varepsilon$ defined in Eq.~\eqref{eq:epsilonPrime} is found to be $\delta_2-\delta_0+\pi/2-\phi_\varepsilon$, where $\pi/2$ comes from the overall factor $\img$ on the right hand side.  Using the value $\phi_\varepsilon = 43.52(2)^\circ$ \cite{ParticleDataGroup:2022pth} and the prediction from the dispersion theory with inputs from experiment and ChPT \cite{Colangelo:2001df} $\delta_2-\delta_0=-44.8(3.0)^\circ$ at 497.6~MeV, the experimental value of the $K^0$ mass, we can verify the phase of $\varepsilon'/\varepsilon$ is approximately zero.  Lattice predictions for the phase shifts are also consistent with the dispersive approach \cite{RBC:2021acc,RBC:2023xqv}.  The potential difference of the phase from zero does not affect the following results that have much larger statistical and systematic errors.
Therefore we simply drop the phase term for the real part of $\varepsilon'/\varepsilon$ and find 
\begin{equation}
\mbox{Re}\left(\frac{\varepsilon'}{\varepsilon}\right)
= \frac{\omega}{\sqrt2|\varepsilon|}
\left[
\frac{\mbox{Im}(A_2)}{\mbox{Re}(A_2)}-\frac{\mbox{Im}(A_0)}{\mbox{Re}(A_0)}
\right].
\end{equation}

It is illustrative to analyze the two terms on the right hand side separately.  The first term of the $\Delta I = 3/2$ channel is called electroweak penguin (EWP) term Re$(\varepsilon'/\varepsilon)_{\rm EWP}$, while the second term of $\Delta I = 1/2$ is the QCD penguin (QCDP) term Re$(\varepsilon'/\varepsilon)_{\rm QCDP}$.  
Using the results for $A_I$ discussed in the previous subsection we obtain
\begin{align}
\mbox{Re}\left(\frac{\varepsilon'}{\varepsilon}\right)_{\rm EWP} &= -4.80(49)(1.95)\times10^{-4},\\
\mbox{Re}\left(\frac{\varepsilon'}{\varepsilon}\right)_{\rm QCDP} &= 43.4(12.9)(18.5)\times10^{-4},\\
\mbox{Re}\left(\frac{\varepsilon'}{\varepsilon}\right) &= 38.6(13.0)(18.6)\times10^{-4},
\end{align}
where the errors are statistical and systematic, respectively.  The results indicate that the QCDP part is the dominant source of both the statistical and systematic errors.

We could reduce the errors if we input experimental values of the real parts, Re$(A_2)_{\rm exp} = 1.479(4)\times10^{-8}$~GeV and Re$(A_0)_{\rm exp} = 3.3201(18)\times10^{-7}$~GeV, which are much more precise than those determined on the lattice.  
To give our best prediction for the $\varepsilon'/\varepsilon$ we also use the continuum limit of Im($A_2$) determined in Ref.~\cite{Blum:2015ywa}.  Since Im($A_2$) depends on the $\tau$ parameter that has been significantly updated, we use the modified value shown in Table~\ref{tab:comparisonA2}.  We only use Im$(A_0)$ from this work.  The EWP and QCDP contributions to Re($\varepsilon'/\varepsilon$) obtained with this approach read
\begin{align}
\mbox{Re}\left(\frac{\varepsilon'}{\varepsilon}\right)_{\rm EWP} &= -7.69(22)(92)\times10^{-4},\\
\mbox{Re}\left(\frac{\varepsilon'}{\varepsilon}\right)_{\rm QCDP} &= 37.1(5.2)(11.1)\times10^{-4},
\end{align}
where the errors are statistical and systematic, respectively.  In is interesting to note that the contribution from QCD penguin operators to Re($\varepsilon'/\varepsilon$) is sizable compared to that from the electroweak penguin operators.  We obtain the ratio 
${\rm Re}(\varepsilon'/\varepsilon)_{\rm QCDP}/{\rm Re}(\varepsilon'/\varepsilon)_{\rm EWP} = -4.83(69)(1.55)$, which is consistent with the one from the GPBC calculation~\cite{RBC:2020kdj}, $-3.73(56)(89)$.

Before quoting our final result for Re($\varepsilon'/\varepsilon$) we note that this quantity could receive significant corrections due to electromagnetic and isospin-breaking effects as discussed in Refs.~\cite{Cirigliano:2019cpi,RBC:2020kdj}.  
We quote the same size of this systematic error as the third error as in Ref.~\cite{RBC:2020kdj} and obtain the final result
\begin{equation}
\mbox{Re}\left(\frac{\varepsilon'}{\varepsilon}\right) = 29.4(5.2)(11.1)(5.0)\times10^{-4}.
\end{equation}
While the errors are somewhat large, this result is compatible with the experimental value, Re($\varepsilon'/\varepsilon)_{\rm exp} = 16.6(2.3)$~\cite{ParticleDataGroup:2022pth}.

\section{Conclusion}
\label{sec:conclusion}

We have presented a new calculation of $\Delta I = 1/2$ and $\Delta I = 3/2$ processes in $K\to\pi\pi$ decays using periodic boundary conditions, essentially extracting the physical amplitudes from the dominant  excited state with near physical kinematics, $E_{\pi\pi}=m_K$.  The calculation was done on a $24^3\times64$ domain wall fermion ensemble with $a^{-1}=1.023$~GeV at the physical pion and kaon masses.
The GEVP method was used to compute the matrix elements and amplitudes with multiple two-pion final states in finite volume.  After applying the Lellouch-L\"uscher formalism to normalize these two-pion states to their infinite volume counterparts, the amplitudes were interpolated to physical kinematics while the energy of the $I=0$ first excited two-pion state is only 5\% off from $m_K$.

\begin{figure}[tbp]
\begin{center}
\begin{tabular}{c}
\includegraphics[width=100mm]{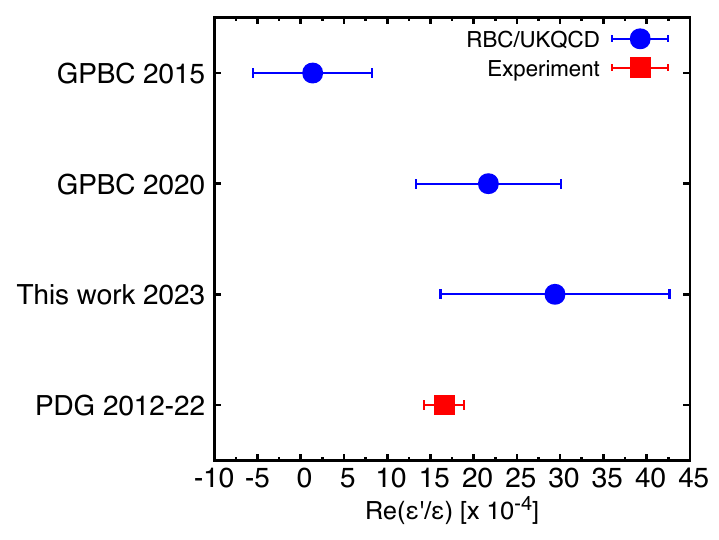}
\end{tabular}
\caption{Result for Re($\varepsilon'/\varepsilon$) comparing with our earlier works~\cite{Bai:2015nea,RBC:2020kdj} and the world average of experimental values~\cite{ParticleDataGroup:2022pth}.  All quoted errors are combined in quadrature.}
\label{fig:history_epsilonp}
\end{center}
\end{figure}

We obtained a value for the well known measure of direct $CP$ violation, Re$(\varepsilon'/\varepsilon)=29.4(5.2)(11.1)(5.0)\times10^{-4}$, where the first error is statistical, the second is all systematics in an isospin symmetric world, and the third combines uncertainties in estimates of electromagnetic and isospin-breaking corrections in the real world.  The result is consistent with our previous GPBC result, Re$(\varepsilon'/\varepsilon)_{2020}=21.7(2.6)(6.2)(5.0)\times10^{-4}$~\cite{RBC:2020kdj}.
In Figure~\ref{fig:history_epsilonp} we summarize all our lattice results for Re$(\varepsilon'/\varepsilon)$ obtained earlier with GPBC~\cite{Bai:2015nea,RBC:2020kdj} and in this work comparing with the world average of experimental results~\cite{NA48:2002tmj,Worcester:2009qt,KTeV:2010sng,ParticleDataGroup:2022pth}.
Note that our recent analyses~\cite{RBC:2020kdj} with multiple two-pion operators and larger statistics revealed that the excited-state systematic error was substantially underestimated in our original, 2015 work~\cite{Bai:2015nea} that included only a single operator, hence the discrepancy observable in the figure. The inclusion of the additional operators, particularly the $\sigma$-like operator, dramatically reduced this error.

While the statistical and systematic errors in the present work are larger because of fewer measurements on a coarser lattice, the results are promising enough to motivate us to perform calculations with more configurations and on a finer lattice ensemble that are on-going.  Therefore we expect more precise results will be obtained in the near future.
The on-going calculations are on the $32^3\times64$ Iwasaki $+$ DSDR ensemble, which has the same lattice spacing as on the GPBC ensemble, as well as to increase the statistics on the $24^3\times64$ ensemble employed here.  We will attempt to take the continuum limit with these two ensembles.  This may be challenging due to possible discretization effects on the $a^{-1}=1.023$ GeV ensemble, both for NPR factors and matrix elements, beyond the leading $O(a^2)$ scaling errors.  The validity of taking the continuum limit with these ensembles needs to be carefully investigated.

There is another important systematic error due to the perturbative truncation of the Wilson coefficients in the three-flavor theory.  This is not improved by step scaling because the matching of the Wilson coefficients in the four-flavor theory to those in the three-flavor theory needs to be done below the charm threshold.  This error gives a 12\% systematic uncertainty in the $K\to\pi\pi$ amplitudes, which is as large as the finite lattice spacing error on the $32^3$ lattice.
While the error will eventually be reduced when we have enough computational resources to calculate the matrix elements on fine enough lattices where we can introduce the charm quark and the corresponding four-quark operators explicitly, it is still desired to fix this defect sooner.
Perturbative matching of the Wilson coefficients to NNLO~\cite{Buras:1999st,Gorbahn:2004my,Brod:2010mj,Cerda-Sevilla:2016yzo,Cerda-Sevilla:2018hjk} and an alternative nonperturbative matching~\cite{Tomii:2020smd} are being studied.

The single largest systematic error on $\varepsilon'/\varepsilon$ is from the electromagnetic and isospin-breaking corrections.  Despite $O(1\%)$ effects on many quantities, the small value of $A_2$ relative to $A_0$ could enhance the impact on $\varepsilon'/\varepsilon$ to $O(20\%)$.  In order to introduce electromagnetic effects in the lattice calculation there is a significant challenge posed by the extension of the formalism by L\"uscher~\cite{Luscher:1990ux} and Lellouch-L\"uscher~\cite{Lellouch:2000pv} to two-hadron systems in finite volumes due to the long-distance character of electromagnetism.  
One such approach under development by our collaboration~\cite{Christ:2021guf} uses the Coulomb gauge to break the electromagnetic contributions into those of a static Coulomb potential and of transverse radiation. The truncation of the Coulomb potential allows the former to be treated on the lattice, and the effects of the truncation can be reintroduced through conventional quantum mechanics on the basis that they describe long-distance physics that is not capable of resolving the structure of the interaction. There has been progress in the former, but challenges remain to be overcome with the transverse radiation contributions before the approach can be applied to the $K\to\pi\pi$ decay calculation.
Although a complete formalism for introducing these effects has not been constructed, we expect that the incorporation of elctromagnetism and strong isospin-breaking in the method with PBC may be easier compared with GPBC where the electric charge is not conserved.

\acknowledgments

We thank the U.S. Department of Energy (DOE) for partial support under awards 
DE-FG02-92ER40716 (TB, DH), DE-SC0010339 (TB, LJ, MT), DE-SC0012704 (PAB, TI, CJ, AS) and DE-SC001704 (AS). LJ and MT were also supported in part by the U.S. DOE under Early Career Award No. DE-SC0021147.  The work of CK was supported by the U.S. DOE Exascale Computing Project.  
TI is also supported by the Department of Energy, Laboratory Directed Research and Development (LDRD No. 23-051) of BNL and RIKEN BNL Research Center.
The work of DH was supported by the Swiss National Science Foundation (SNSF) through grant No. 200020\_208222.
We thank the collaborators in the RBC and UKQCD collaborations, especially N.H. Christ, for fruitful discussions.
The software used for this work is based on 
\href{https://github.com/RBC-UKQCD/CPS}{CPS},
\href{https://github.com/paboyle/Grid}{Grid}, and \href{https://github.com/aportelli/hadrons}{Hadrons}.  
Computations were carried out with USQCD resources funded by the US DOE at BNL and JLab.

\appendix

\section{State normalization and amplitude definition}
\label{sec:normalization}

In this appendix we summarize the conventions on state normalization and the definition of the $K\to\pi\pi$ amplitudes.  While there have been different expressions in the overall factor for the definition of the amplitudes in the series of RBC/UKQCD papers~\cite{RBC:2001pmy,RBC:2012ynq,Blum:2015ywa,RBC:2020kdj}, this appendix clarifies the interpretation of these differences.
To give a brief clarification we remark here that the differences were due to the difference in the definition of the LL factor and hence in the normalization of the infinite-volume two-pion states as one can see below and that the amplitudes $A_I$ themselves have been consistent through the series.

We normalize hadronic states in finite volume to unity.  The two-pion states are normalized by
\begin{equation}
_L\langle (\pi\pi)_I,n|{\pi\pi}_{I'},n'\rangle_L = \delta_{I,I'}\delta_{n,n'},
\end{equation}
where and $n$ and $n'$ stand for state indices. 
This convention is consistent with the insertion of the complete set in Eq.~\eqref{eq:state_expansion} for the two-pion states.  We have added the subscript $L$ to express finite-volume states.  The kaon states $|K,n\rangle$ are normalized to unity in an equivalent manner.

The infinite-volume states are normalized relativistically,
\begin{align}
_\infty\langle K(\vec p) | K(\vec p')\rangle_\infty
&= 2E_K(\vec p)(2\pi)^3\delta^3(\vec p-\vec p'),
\end{align}
for the kaon state.  Here we have added the subscript $\infty$ to express hadronic states in infinite volume.
For two-pion states we also employ the relativistic normalization
\begin{align}
_\infty\langle \pi^+(\vec p_1) \pi^-(\vec p_2) | \pi^+(\vec p_1') \pi^-(\vec p_2'))\rangle_\infty
&= 2E_\pi(\vec p_1)\cdot2E_\pi(\vec p_2)(2\pi)^6\delta^3(\vec p_1-\vec p_1')\delta^3(\vec p_2-\vec p_2'),
\label{eq:infpippim}\\
_\infty\langle \pi^0(\vec p_1) \pi^0(\vec p_2) | \pi^0(\vec p_1') \pi^0(\vec p_2'))\rangle_\infty
&= \frac{1}{2}\Big(2E_\pi(\vec p_1)\cdot2E_\pi(\vec p_2)(2\pi)^6\delta^3(\vec p_1-\vec p_1')\delta^3(\vec p_2-\vec p_2')\notag\\
&\ \ \ \ \ \ \ + (\vec p_1' \leftrightarrow \vec p_2')\Big).
\end{align}
The isospin-definite two-pion states are then given by
\begin{align}
|(\pi\pi)_{I=2}, \vec p_1, \vec p_2\rangle_\infty
&= \frac{1}{\sqrt3}|\pi^+(\vec p_1) \pi^-(\vec p_2)\rangle_\infty + \sqrt{\frac{2}{3}}|\pi^0(\vec p_1) \pi^0(\vec p_2)\rangle_\infty,
\label{eq:infpipiI2}
\\
|(\pi\pi)_{I=0}, \vec p_1, \vec p_2\rangle_\infty
&= \sqrt{\frac{2}{3}}|\pi^+(\vec p_1) \pi^-(\vec p_2)\rangle_\infty - \frac{1}{\sqrt3}|\pi^0(\vec p_1) \pi^0(\vec p_2)\rangle_\infty.
\label{eq:infpipiI0}
\end{align}

The LL factor~\eqref{eq:LLfactor} compensates the normalization difference between finite and infinite volumes discussed above as well as the effects of two-pion interaction in finite volume.  If the on-shell kinematics is realized by a finite-volume state labeled by $n_{\rm phys}$, the $K\to\pi\pi$ amplitudes are given by
\begin{equation}
A_{I} = F_{I,n_{\rm phys}L} \langle(\pi\pi)_I,n_{\rm phys}| H_W | K^0, 0\rangle_L.
\end{equation}

\begin{table}[tp]
\centering
\begin{tabular}{ccc}
\hline
Year \& Ref. & Amplitude definition & LL factor \\
\hline
2003 \cite{RBC:2001pmy} & \eqref{eq:wosqrt2} & not used \\
2011 \cite{Blum:2011pu} & \eqref{eq:wosqrt2} & \eqref{eq:LL4pi} \\
2012 \cite{Blum:2011ng} & not specified & not specified \\
2012 \cite{Blum:2012uk} & \eqref{eq:wsqrt2} & \eqref{eq:LL8pi} \\
2015 \cite{Blum:2015ywa} & \eqref{eq:wsqrt2} & \eqref{eq:LL8pi} \\
2015 \cite{Bai:2015nea} & not specified & not specified \\
2020 \cite{RBC:2020kdj} & \eqref{eq:wosqrt2} & \eqref{eq:LL4pi} \\
This paper & \eqref{eq:wosqrt2} & \eqref{eq:LL4pi} \\
\hline
\end{tabular}
\caption{Expressions for definition of $A_I$ in RBC/UKQCD papers.}
\label{tab:AI_def_history}
\end{table}

For the rest of the appendix all states are in infinite volume and correspond to the physical kinematics of $K\to\pi\pi$.  Therefore we drop the subscripts $L$ and $\infty$ and the momentum arguments. 
In the series of $K\to\pi\pi$ paper by the RBC and UKQCD collaborations there have been two expressions in the definition for the amplitudes,
\begin{equation}
A_I = \langle(\pi\pi)_I,({\rm A})|H_W|K^0\rangle
\label{eq:wosqrt2}
\end{equation}
and
\begin{equation}
\sqrt2A_I = \langle(\pi\pi)_I,({\rm B})|H_W|K^0\rangle.
\label{eq:wsqrt2}
\end{equation}
Here (A) or (B) was not explicitly shown in the earlier papers but denotes a specific normalization of the two-pion state.  
Some of the earlier papers also multiplied $A_I$ with the phase factor $\e^{\img\delta_I}$ on the left hand side related to the definition of the two-pion state.  This phase factor does not affect the magnitude of the amplitudes and it should be noted that the imaginary parts of $A_I$ in our convention have been associated with only $CP$-violating effects that appear in the CKM matrix elements (more specifically, the $\tau$ parameter).
The difference in the normalization between these two two-pion states is obviously by the factor of $\sqrt2$ and originates from the difference in the overall factor of the LL factor.  In this paper we define the LL factor by
\begin{equation}
F({\rm A})^2 = \frac{4\pi m_KE_{\pi\pi}^2}{k^3}
\left(k\frac{\td\delta_0}{\td k} + q\frac{\td\phi}{\td q}\right),
\label{eq:LL4pi}
\end{equation}
while some of the earlier papers defined
\begin{equation}
F({\rm B})^2 = \frac{8\pi m_KE_{\pi\pi}^2}{k^3}
\left(k\frac{\td\delta_0}{\td k} + q\frac{\td\phi}{\td q}\right).
\label{eq:LL8pi}
\end{equation}
Again we distinguish these two different definitions by (A) and (B).
Table~\ref{tab:AI_def_history} summarizes the expressions of the amplitude definition in the series of RBC/UKQCD papers.  One can easily recognize that the definition of $A_I$ has been consistent despite the two different expressions.
Note that expression (A) corresponds to our convention described in Eqs.~\eqref{eq:infpippim}--\eqref{eq:infpipiI0} and we omit (A) in what follows.

It is also valuable to relate these amplitudes with the decay rates, which is independent of convention.  We consider the amplitudes for decay of $K^+$ into $\pi^+\pi^0$ ($A_{+0}$) and those of $K_S$ into $\pi^+\pi^-$ ($A_{+-}$) and $\pi^0\pi^0$ ($A_{00}$). 
Applying Fermi's golden rule yields their relations with the decay rates
\begin{align}
\Gamma_{+0} &= \frac{1}{8\pi}\frac{p_{+0}}{m_{K^+}^2}|A_{+0}|^2,
\label{eq:Gammap0}
\\
\Gamma_{+-} &= \frac{1}{8\pi}\frac{\sqrt{\frac{m_{K^0}^2}{4}-m_{\pi^+}^2}}{m_{K^0}^2}|A_{+-}|^2,
\label{eq:Gammapm}
\\
\Gamma_{00} &= \frac{1}{8\pi}\frac{\sqrt{\frac{m_{K^0}^2}{4}-m_{\pi^0}^2}}{m_{K^0}^2}|A_{00}|^2.
\label{eq:Gamma00}
\end{align}
Here $p_{+0}$ is given in Eq.~(27) of Ref.~\cite{Blum:2011pu} and $m_X$ stands for the mass of meson $X$ ($=K^+, K^0, \pi^+$ and $\pi^0$).
Since the latter two amplitudes are a linear combination of $A_0$ and $A_2$, the phase factors $\e^{\img\delta_I}$ should be taken into account separately.
The relations between these amplitudes and the isospin-definite amplitudes, $A_0$ and $A_2$, are given by
\begin{align}
A_{+0} &= \sqrt{\frac{3}{2}}A_2\e^{\img\delta_2},
\label{eq:Ap0_A2}\\
A_{+-} &= \frac{2}{\sqrt3}A_0\e^{\img\delta_0}+\sqrt{\frac{2}{3}}A_2\e^{\img\delta_2},
\label{eq:Apm_A0A2}\\
A_{00} &= -\sqrt{\frac{2}{3}}A_0\e^{\img\delta_0}+\frac{2}{\sqrt3}A_2\e^{\img\delta_2},
\label{eq:A00_A0A2}
\end{align}
where we have used the Wigner-Eckart theorem for $A_{+0}$.  The extra factor of $\sqrt2$ on the right hand sides of Eqs.~\eqref{eq:Apm_A0A2} and \eqref{eq:A00_A0A2} despite the relations of two-pion states in Eqs.~\eqref{eq:infpipiI2} and \eqref{eq:infpipiI0} is because of the difference in the initial kaon states, $|K_S\rangle = \frac{1}{\sqrt2}(|K^0\rangle+|\overline K^0\rangle)$ for $A_{+-}$ and $A_{00}$ versus $|K^0\rangle$ for $A_2$ and $A_0$.  Assuming ${\rm Re}(A_I)\gg{\rm Im}(A_I)$, we can calculate Re$(A_I)\approx |A_I|$ from experimental values of decay rates as follows.  Calculation of $|A_2|$ is trivial from Eqs.~\eqref{eq:Gammap0} and \eqref{eq:Ap0_A2}.
Summing up the absolute squares of Eqs.~\eqref{eq:Apm_A0A2} and \eqref{eq:A00_A0A2} gives
\begin{equation}
|A_{+-}|^2 + |A_{00}|^2 = 2|A_0|^2 + 2|A_2|^2,
\label{eq:ApmA00toA0A2}
\end{equation}
which enables us to calculate $|A_0|\approx{\rm Re}(A_0)$ as the other terms in this equation are related to the decay rates as explained above.  In the actual world with electromagnetism and without exact isospin symmetry, however, Eqs~\eqref{eq:Ap0_A2}--\eqref{eq:A00_A0A2} do not have to hold consistently and the definitions of $A_0$ and $A_2$ are ambiguous. 
In the series of the RBC/UKQCD papers on $K\to\pi\pi$, we have been using the value of Re$(A_0)$ defined as $|A_0|$ in Eq.~\eqref{eq:ApmA00toA0A2} but without $|A_2|^2$ ($\ll|A_0|^2$), while Re$(A_2)$ has been defined as $|A_2|$ obtained from Eq.~\eqref{eq:Ap0_A2}.

\section{Wick contractions}
\label{sec:contractions}

Throughout the section the indices of Greek characters are summed over the color space, while the trace `Tr' and matrix operation of quark propagators $S_{l,s}^{\alpha\beta}$ are implemented within the spin space.  The four-quark operators $Q_1$ and $Q_2$ are replaced with $\widetilde Q_1 = \bar s_\alpha\gamma_\mu(1-\gamma_5)d_\alpha\cdot \bar u_\beta \gamma_\mu(1-\gamma_5)u_\beta$ and $\widetilde Q_2 = \bar s_\alpha \gamma_\mu(1-\gamma_5)u_\beta\cdot\bar u_\beta \gamma_\mu(1-\gamma_5)d_\alpha$, respectively, using the Fierz symmetry.  They are identical to the original definitions on the lattice and give the same numerical results, although the explicit forms of the contractions given below are different.  
As noted in Section~\ref{sec:4qOps} we only take the parity-odd part of the four-quark operators because the parity-even part increases the statistical error but is zero in the infinite statistics.

\subsection{$K\to\pi\pi$}

All $K\to\pi\pi$ three-point functions are a linear combination of the following contractions:
\begin{align}
&\mbox{type1}_{\rm1\Tr,cdiag}^{\rm LR/LL}(x_K,x_Q,x_1,x_2)
\notag\\*
=& \pm\Tr[
S_l^{\alpha\beta}(x_Q,x_1) \gamma_5
S_l^{\beta\gamma}(x_1,x_Q)\gamma_\mu\gamma_5
S_l^{\gamma\delta}(x_Q,x_2)\gamma_5
S_l^{\delta\epsilon}(x_2,x_K)\gamma_5
S_s^{\epsilon\alpha}(x_K,x_Q)\gamma_\mu
]
\notag\\
& \pm\Tr[
S_l^{\alpha\beta}(x_Q,x_2) \gamma_5
S_l^{\beta\gamma}(x_2,x_Q)\gamma_\mu\gamma_5
S_l^{\gamma\delta}(x_Q,x_1)\gamma_5
S_l^{\delta\epsilon}(x_1,x_K)\gamma_5
S_s^{\epsilon\alpha}(x_K,x_Q)\gamma_\mu
]
\notag\\
& -\Tr[
S_l^{\alpha\beta}(x_Q,x_1) \gamma_5
S_l^{\beta\gamma}(x_1,x_Q)\gamma_\mu
S_l^{\gamma\delta}(x_Q,x_2)\gamma_5
S_l^{\delta\epsilon}(x_2,x_K)\gamma_5
S_s^{\epsilon\alpha}(x_K,x_Q)\gamma_\mu\gamma_5
]
\notag\\
& -\Tr[
S_l^{\alpha\beta}(x_Q,x_2) \gamma_5
S_l^{\beta\gamma}(x_2,x_Q)\gamma_\mu
S_l^{\gamma\delta}(x_Q,x_1)\gamma_5
S_l^{\delta\epsilon}(x_1,x_K)\gamma_5
S_s^{\epsilon\alpha}(x_K,x_Q)\gamma_\mu\gamma_5
],
\label{eq:1sttype1}
\\
&\mbox{type1}_{\rm2\Tr,cdiag}^{\rm LR/LL}(x_K,x_Q,x_1,x_2)
\notag\\*
=& \pm\Tr[
S_l^{\alpha\beta}(x_Q,x_1) \gamma_5
S_l^{\beta\alpha}(x_1,x_Q)\gamma_\mu\gamma_5
]\cdot\Tr[
S_l^{\gamma\delta}(x_Q,x_2)\gamma_5
S_l^{\delta\epsilon}(x_2,x_K)\gamma_5
S_s^{\epsilon\gamma}(x_K,x_Q)\gamma_\mu
]
\notag\\
& \pm\Tr[
S_l^{\alpha\beta}(x_Q,x_2) \gamma_5
S_l^{\beta\alpha}(x_2,x_Q)\gamma_\mu\gamma_5
]\cdot\Tr[
S_l^{\gamma\delta}(x_Q,x_1)\gamma_5
S_l^{\delta\epsilon}(x_1,x_K)\gamma_5
S_s^{\epsilon\gamma}(x_K,x_Q)\gamma_\mu
]
\notag\\
& -\Tr[
S_l^{\alpha\beta}(x_Q,x_1) \gamma_5
S_l^{\beta\alpha}(x_1,x_Q)\gamma_\mu
]\cdot\Tr[
S_l^{\gamma\delta}(x_Q,x_2)\gamma_5
S_l^{\delta\epsilon}(x_2,x_K)\gamma_5
S_s^{\epsilon\gamma}(x_K,x_Q)\gamma_\mu\gamma_5
]
\notag\\
& -\Tr[
S_l^{\alpha\beta}(x_Q,x_2) \gamma_5
S_l^{\beta\alpha}(x_2,x_Q)\gamma_\mu
]\cdot\Tr[
S_l^{\gamma\delta}(x_Q,x_1)\gamma_5
S_l^{\delta\epsilon}(x_1,x_K)\gamma_5
S_s^{\epsilon\gamma}(x_K,x_Q)\gamma_\mu\gamma_5
],
\\
&\mbox{type1}_{\rm1\Tr,cmix}^{\rm LR/LL}(x_K,x_Q,x_1,x_2)
\notag\\*
=& \pm\Tr[
S_l^{\alpha\beta}(x_Q,x_1) \gamma_5
S_l^{\beta\alpha}(x_1,x_Q)\gamma_\mu\gamma_5
S_l^{\gamma\delta}(x_Q,x_2)\gamma_5
S_l^{\delta\epsilon}(x_2,x_K)\gamma_5
S_s^{\epsilon\gamma}(x_K,x_Q)\gamma_\mu
]
\notag\\
& \pm\Tr[
S_l^{\alpha\beta}(x_Q,x_2) \gamma_5
S_l^{\beta\alpha}(x_2,x_Q)\gamma_\mu\gamma_5
S_l^{\gamma\delta}(x_Q,x_1)\gamma_5
S_l^{\delta\epsilon}(x_1,x_K)\gamma_5
S_s^{\epsilon\gamma}(x_K,x_Q)\gamma_\mu
]
\notag\\
& -\Tr[
S_l^{\alpha\beta}(x_Q,x_1) \gamma_5
S_l^{\beta\alpha}(x_1,x_Q)\gamma_\mu
S_l^{\gamma\delta}(x_Q,x_2)\gamma_5
S_l^{\delta\epsilon}(x_2,x_K)\gamma_5
S_s^{\epsilon\gamma}(x_K,x_Q)\gamma_\mu\gamma_5
]
\notag\\
& -\Tr[
S_l^{\alpha\beta}(x_Q,x_2) \gamma_5
S_l^{\beta\alpha}(x_2,x_Q)\gamma_\mu
S_l^{\gamma\delta}(x_Q,x_1)\gamma_5
S_l^{\delta\epsilon}(x_1,x_K)\gamma_5
S_s^{\epsilon\gamma}(x_K,x_Q)\gamma_\mu\gamma_5
],
\\
&\mbox{type1}_{\rm2\Tr,cmix}^{\rm LR/LL}(x_K,x_Q,x_1,x_2)
\notag\\*
=& \pm\Tr[
S_l^{\alpha\beta}(x_Q,x_1) \gamma_5
S_l^{\beta\gamma}(x_1,x_Q)\gamma_\mu\gamma_5
]\cdot\Tr[
S_l^{\gamma\delta}(x_Q,x_2)\gamma_5
S_l^{\delta\epsilon}(x_2,x_K)\gamma_5
S_s^{\epsilon\alpha}(x_K,x_Q)\gamma_\mu
]
\notag\\
& \pm\Tr[
S_l^{\alpha\beta}(x_Q,x_2) \gamma_5
S_l^{\beta\gamma}(x_2,x_Q)\gamma_\mu\gamma_5
]\cdot\Tr[
S_l^{\gamma\delta}(x_Q,x_1)\gamma_5
S_l^{\delta\epsilon}(x_1,x_K)\gamma_5
S_s^{\epsilon\alpha}(x_K,x_Q)\gamma_\mu
]
\notag\\
& -\Tr[
S_l^{\alpha\beta}(x_Q,x_1) \gamma_5
S_l^{\beta\gamma}(x_1,x_Q)\gamma_\mu
]\cdot\Tr[
S_l^{\gamma\delta}(x_Q,x_2)\gamma_5
S_l^{\delta\epsilon}(x_2,x_K)\gamma_5
S_s^{\epsilon\alpha}(x_K,x_Q)\gamma_\mu\gamma_5
]
\notag\\
& -\Tr[
S_l^{\alpha\beta}(x_Q,x_2) \gamma_5
S_l^{\beta\gamma}(x_2,x_Q)\gamma_\mu
]\cdot\Tr[
S_l^{\gamma\delta}(x_Q,x_1)\gamma_5
S_l^{\delta\epsilon}(x_1,x_K)\gamma_5
S_s^{\epsilon\alpha}(x_K,x_Q)\gamma_\mu\gamma_5
],
\\
&\mbox{type2}_{\rm1\Tr,cdiag}^{\rm LR/LL}(x_K,x_Q,x_1,x_2)
\notag\\*
=& \pm\Tr[
S_l^{\alpha\beta}(x_Q,x_1) \gamma_5
S_l^{\beta\gamma}(x_1,x_2)\gamma_5
S_l^{\gamma\delta}(x_2,x_Q)\gamma_\mu\gamma_5
S_l^{\delta\epsilon}(x_Q,x_K)\gamma_5
S_s^{\epsilon\alpha}(x_K,x_Q)\gamma_\mu
]
\notag\\
& \pm\Tr[
S_l^{\alpha\beta}(x_Q,x_2) \gamma_5
S_l^{\beta\gamma}(x_2,x_1)\gamma_5
S_l^{\gamma\delta}(x_1,x_Q)\gamma_\mu\gamma_5
S_l^{\delta\epsilon}(x_Q,x_K)\gamma_5
S_s^{\epsilon\alpha}(x_K,x_Q)\gamma_\mu
]
\notag\\
& -\Tr[
S_l^{\alpha\beta}(x_Q,x_1) \gamma_5
S_l^{\beta\gamma}(x_1,x_2)\gamma_5
S_l^{\gamma\delta}(x_2,x_Q)\gamma_\mu
S_l^{\delta\epsilon}(x_Q,x_K)\gamma_5
S_s^{\epsilon\alpha}(x_K,x_Q)\gamma_\mu\gamma_5
]
\notag\\
& -\Tr[
S_l^{\alpha\beta}(x_Q,x_2) \gamma_5
S_l^{\beta\gamma}(x_2,x_1)\gamma_5
S_l^{\gamma\delta}(x_1,x_Q)\gamma_\mu
S_l^{\delta\epsilon}(x_Q,x_K)\gamma_5
S_s^{\epsilon\alpha}(x_K,x_Q)\gamma_\mu\gamma_5
],
\\
&\mbox{type2}_{\rm2\Tr,cdiag}^{\rm LR/LL}(x_K,x_Q,x_1,x_2)
\notag\\*
=& \pm\Tr[
S_l^{\alpha\beta}(x_Q,x_1) \gamma_5
S_l^{\beta\gamma}(x_1,x_2)\gamma_5
S_l^{\gamma\alpha}(x_2,x_Q)\gamma_\mu\gamma_5
]\cdot\Tr[
S_l^{\delta\epsilon}(x_Q,x_K)\gamma_5
S_s^{\epsilon\delta}(x_K,x_Q)\gamma_\mu
]
\notag\\
& \pm\Tr[
S_l^{\alpha\beta}(x_Q,x_2) \gamma_5
S_l^{\beta\gamma}(x_2,x_1)\gamma_5
S_l^{\gamma\alpha}(x_1,x_Q)\gamma_\mu\gamma_5
]\cdot\Tr[
S_l^{\delta\epsilon}(x_Q,x_K)\gamma_5
S_s^{\epsilon\delta}(x_K,x_Q)\gamma_\mu
]
\notag\\
& -\Tr[
S_l^{\alpha\beta}(x_Q,x_1) \gamma_5
S_l^{\beta\gamma}(x_1,x_2)\gamma_5
S_l^{\gamma\alpha}(x_2,x_Q)\gamma_\mu
]\cdot\Tr[
S_l^{\delta\epsilon}(x_Q,x_K)\gamma_5
S_s^{\epsilon\delta}(x_K,x_Q)\gamma_\mu\gamma_5
]
\notag\\
& -\Tr[
S_l^{\alpha\beta}(x_Q,x_2) \gamma_5
S_l^{\beta\gamma}(x_2,x_1)\gamma_5
S_l^{\gamma\alpha}(x_1,x_Q)\gamma_\mu
]\cdot\Tr[
S_l^{\delta\epsilon}(x_Q,x_K)\gamma_5
S_s^{\epsilon\delta}(x_K,x_Q)\gamma_\mu\gamma_5
],
\\
&\mbox{type2}_{\rm1\Tr,cmix}^{\rm LR/LL}(x_K,x_Q,x_1,x_2)
\notag\\*
=& \pm\Tr[
S_l^{\alpha\beta}(x_Q,x_1) \gamma_5
S_l^{\beta\gamma}(x_1,x_2)\gamma_5
S_l^{\gamma\alpha}(x_2,x_Q)\gamma_\mu\gamma_5
S_l^{\delta\epsilon}(x_Q,x_K)\gamma_5
S_s^{\epsilon\delta}(x_K,x_Q)\gamma_\mu
]
\notag\\
& \pm\Tr[
S_l^{\alpha\beta}(x_Q,x_2) \gamma_5
S_l^{\beta\gamma}(x_2,x_1)\gamma_5
S_l^{\gamma\alpha}(x_1,x_Q)\gamma_\mu\gamma_5
S_l^{\delta\epsilon}(x_Q,x_K)\gamma_5
S_s^{\epsilon\delta}(x_K,x_Q)\gamma_\mu
]
\notag\\
& -\Tr[
S_l^{\alpha\beta}(x_Q,x_1) \gamma_5
S_l^{\beta\gamma}(x_1,x_2)\gamma_5
S_l^{\gamma\alpha}(x_2,x_Q)\gamma_\mu
S_l^{\delta\epsilon}(x_Q,x_K)\gamma_5
S_s^{\epsilon\delta}(x_K,x_Q)\gamma_\mu\gamma_5
]
\notag\\
& -\Tr[
S_l^{\alpha\beta}(x_Q,x_2) \gamma_5
S_l^{\beta\gamma}(x_2,x_1)\gamma_5
S_l^{\gamma\alpha}(x_1,x_Q)\gamma_\mu
S_l^{\delta\epsilon}(x_Q,x_K)\gamma_5
S_s^{\epsilon\delta}(x_K,x_Q)\gamma_\mu\gamma_5
],
\\
&\mbox{type2}_{\rm2\Tr,cmix}^{\rm LR/LL}(x_K,x_Q,x_1,x_2)
\notag\\*
=& \pm\Tr[
S_l^{\alpha\beta}(x_Q,x_1) \gamma_5
S_l^{\beta\gamma}(x_1,x_2)\gamma_5
S_l^{\gamma\delta}(x_2,x_Q)\gamma_\mu\gamma_5
]\cdot\Tr[
S_l^{\delta\epsilon}(x_Q,x_K)\gamma_5
S_s^{\epsilon\alpha}(x_K,x_Q)\gamma_\mu
]
\notag\\
& \pm\Tr[
S_l^{\alpha\beta}(x_Q,x_2) \gamma_5
S_l^{\beta\gamma}(x_2,x_1)\gamma_5
S_l^{\gamma\delta}(x_1,x_Q)\gamma_\mu\gamma_5
]\cdot\Tr[
S_l^{\delta\epsilon}(x_Q,x_K)\gamma_5
S_s^{\epsilon\alpha}(x_K,x_Q)\gamma_\mu
]
\notag\\
& -\Tr[
S_l^{\alpha\beta}(x_Q,x_1) \gamma_5
S_l^{\beta\gamma}(x_1,x_2)\gamma_5
S_l^{\gamma\delta}(x_2,x_Q)\gamma_\mu
]\cdot\Tr[
S_l^{\delta\epsilon}(x_Q,x_K)\gamma_5
S_s^{\epsilon\alpha}(x_K,x_Q)\gamma_\mu\gamma_5
]
\notag\\
& -\Tr[
S_l^{\alpha\beta}(x_Q,x_2) \gamma_5
S_l^{\beta\gamma}(x_2,x_1)\gamma_5
S_l^{\gamma\delta}(x_1,x_Q)\gamma_\mu
]\cdot\Tr[
S_l^{\delta\epsilon}(x_Q,x_K)\gamma_5
S_s^{\epsilon\alpha}(x_K,x_Q)\gamma_\mu\gamma_5
],
\\
&\mbox{type3}_{\rm1\Tr,cdiag}^{{\rm LR/LL},l\rm\mathchar`-loop}(x_K,x_Q,x_1,x_2)
\notag\\*
=& \pm\Tr[
S_l^{\alpha\beta}(x_Q,x_Q)\gamma_\mu\gamma_5
S_l^{\beta\gamma}(x_Q,x_1)\gamma_5
S_l^{\gamma\delta}(x_1,x_2)\gamma_5
S_l^{\delta\epsilon}(x_2,x_K)\gamma_5
S_s^{\epsilon\alpha}(x_K,x_Q)\gamma_\mu
]
\notag\\
& \pm\Tr[
S_l^{\alpha\beta}(x_Q,x_Q)\gamma_\mu\gamma_5
S_l^{\beta\gamma}(x_Q,x_2)\gamma_5
S_l^{\gamma\delta}(x_2,x_1)\gamma_5
S_l^{\delta\epsilon}(x_1,x_K)\gamma_5
S_s^{\epsilon\alpha}(x_K,x_Q)\gamma_\mu
]
\notag\\
& -\Tr[
S_l^{\alpha\beta}(x_Q,x_Q)\gamma_\mu
S_l^{\beta\gamma}(x_Q,x_1)\gamma_5
S_l^{\gamma\delta}(x_1,x_2)\gamma_5
S_l^{\delta\epsilon}(x_2,x_K)\gamma_5
S_s^{\epsilon\alpha}(x_K,x_Q)\gamma_\mu\gamma_5
]
\notag\\
& -\Tr[
S_l^{\alpha\beta}(x_Q,x_Q)\gamma_\mu
S_l^{\beta\gamma}(x_Q,x_2)\gamma_5
S_l^{\gamma\delta}(x_2,x_1)\gamma_5
S_l^{\delta\epsilon}(x_1,x_K)\gamma_5
S_s^{\epsilon\alpha}(x_K,x_Q)\gamma_\mu\gamma_5
],
\\
&\mbox{type3}_{\rm1\Tr,cdiag}^{{\rm LR/LL},s\rm\mathchar`-loop}(x_K,x_Q,x_1,x_2)
\notag\\*
=& -\Tr[
S_s^{\alpha\beta}(x_Q,x_Q)\gamma_\mu\gamma_5
S_l^{\beta\gamma}(x_Q,x_1)\gamma_5
S_l^{\gamma\delta}(x_1,x_2)\gamma_5
S_l^{\delta\epsilon}(x_2,x_K)\gamma_5
S_s^{\epsilon\alpha}(x_K,x_Q)\gamma_\mu
]
\notag\\
& -\Tr[
S_s^{\alpha\beta}(x_Q,x_Q)\gamma_\mu\gamma_5
S_l^{\beta\gamma}(x_Q,x_2)\gamma_5
S_l^{\gamma\delta}(x_2,x_1)\gamma_5
S_l^{\delta\epsilon}(x_1,x_K)\gamma_5
S_s^{\epsilon\alpha}(x_K,x_Q)\gamma_\mu
]
\notag\\
& \pm\Tr[
S_s^{\alpha\beta}(x_Q,x_Q)\gamma_\mu
S_l^{\beta\gamma}(x_Q,x_1)\gamma_5
S_l^{\gamma\delta}(x_1,x_2)\gamma_5
S_l^{\delta\epsilon}(x_2,x_K)\gamma_5
S_s^{\epsilon\alpha}(x_K,x_Q)\gamma_\mu\gamma_5
]
\notag\\
& \pm\Tr[
S_s^{\alpha\beta}(x_Q,x_Q)\gamma_\mu
S_l^{\beta\gamma}(x_Q,x_2)\gamma_5
S_l^{\gamma\delta}(x_2,x_1)\gamma_5
S_l^{\delta\epsilon}(x_1,x_K)\gamma_5
S_s^{\epsilon\alpha}(x_K,x_Q)\gamma_\mu\gamma_5
],
\\
&\mbox{type3}_{\rm2\Tr,cdiag}^{{\rm LR/LL},q\rm\mathchar`-loop}(x_K,x_Q,x_1,x_2)
\notag\\*
=& \pm\Tr[
S_q^{\alpha\alpha}(x_Q,x_Q)\gamma_\mu\gamma_5
]\cdot\Tr[
S_l^{\beta\gamma}(x_Q,x_1)\gamma_5
S_l^{\gamma\delta}(x_1,x_2)\gamma_5
S_l^{\delta\epsilon}(x_2,x_K)\gamma_5
S_s^{\epsilon\beta}(x_K,x_Q)\gamma_\mu
]
\notag\\
& \pm\Tr[
S_q^{\alpha\alpha}(x_Q,x_Q)\gamma_\mu\gamma_5
]\cdot\Tr[
S_l^{\beta\gamma}(x_Q,x_2)\gamma_5
S_l^{\gamma\delta}(x_2,x_1)\gamma_5
S_l^{\delta\epsilon}(x_1,x_K)\gamma_5
S_s^{\epsilon\beta}(x_K,x_Q)\gamma_\mu
]
\notag\\
& -\Tr[
S_q^{\alpha\alpha}(x_Q,x_Q)\gamma_\mu
]\cdot\Tr[
S_l^{\beta\gamma}(x_Q,x_1)\gamma_5
S_l^{\gamma\delta}(x_1,x_2)\gamma_5
S_l^{\delta\epsilon}(x_2,x_K)\gamma_5
S_s^{\epsilon\beta}(x_K,x_Q)\gamma_\mu\gamma_5
]
\notag\\
& -\Tr[
S_q^{\alpha\alpha}(x_Q,x_Q)\gamma_\mu
]\cdot\Tr[
S_l^{\beta\gamma}(x_Q,x_2)\gamma_5
S_l^{\gamma\delta}(x_2,x_1)\gamma_5
S_l^{\delta\epsilon}(x_1,x_K)\gamma_5
S_s^{\epsilon\beta}(x_K,x_Q)\gamma_\mu\gamma_5
],
\\
&\mbox{type3}_{\rm1\Tr,cmix}^{{\rm LR/LL},l\rm\mathchar`-loop}(x_K,x_Q,x_1,x_2)
\notag\\*
=& \pm\Tr[
S_l^{\alpha\alpha}(x_Q,x_Q)\gamma_\mu\gamma_5
S_l^{\beta\gamma}(x_Q,x_1)\gamma_5
S_l^{\gamma\delta}(x_1,x_2)\gamma_5
S_l^{\delta\epsilon}(x_2,x_K)\gamma_5
S_s^{\epsilon\beta}(x_K,x_Q)\gamma_\mu
]
\notag\\
& \pm\Tr[
S_l^{\alpha\alpha}(x_Q,x_Q)\gamma_\mu\gamma_5
S_l^{\beta\gamma}(x_Q,x_2)\gamma_5
S_l^{\gamma\delta}(x_2,x_1)\gamma_5
S_l^{\delta\epsilon}(x_1,x_K)\gamma_5
S_s^{\epsilon\beta}(x_K,x_Q)\gamma_\mu
]
\notag\\
& -\Tr[
S_l^{\alpha\alpha}(x_Q,x_Q)\gamma_\mu
S_l^{\beta\gamma}(x_Q,x_1)\gamma_5
S_l^{\gamma\delta}(x_1,x_2)\gamma_5
S_l^{\delta\epsilon}(x_2,x_K)\gamma_5
S_s^{\epsilon\beta}(x_K,x_Q)\gamma_\mu\gamma_5
]
\notag\\
& -\Tr[
S_l^{\alpha\alpha}(x_Q,x_Q)\gamma_\mu
S_l^{\beta\gamma}(x_Q,x_2)\gamma_5
S_l^{\gamma\delta}(x_2,x_1)\gamma_5
S_l^{\delta\epsilon}(x_1,x_K)\gamma_5
S_s^{\epsilon\beta}(x_K,x_Q)\gamma_\mu\gamma_5
],
\\
&\mbox{type3}_{\rm1\Tr,cmix}^{{\rm LR/LL},s\rm\mathchar`-loop}(x_K,x_Q,x_1,x_2)
\notag\\*
=& -\Tr[
S_s^{\alpha\alpha}(x_Q,x_Q)\gamma_\mu\gamma_5
S_l^{\beta\gamma}(x_Q,x_1)\gamma_5
S_l^{\gamma\delta}(x_1,x_2)\gamma_5
S_l^{\delta\epsilon}(x_2,x_K)\gamma_5
S_s^{\epsilon\beta}(x_K,x_Q)\gamma_\mu
]
\notag\\
& -\Tr[
S_s^{\alpha\alpha}(x_Q,x_Q)\gamma_\mu\gamma_5
S_l^{\beta\gamma}(x_Q,x_2)\gamma_5
S_l^{\gamma\delta}(x_2,x_1)\gamma_5
S_l^{\delta\epsilon}(x_1,x_K)\gamma_5
S_s^{\epsilon\beta}(x_K,x_Q)\gamma_\mu
]
\notag\\
& \pm\Tr[
S_s^{\alpha\alpha}(x_Q,x_Q)\gamma_\mu
S_l^{\beta\gamma}(x_Q,x_1)\gamma_5
S_l^{\gamma\delta}(x_1,x_2)\gamma_5
S_l^{\delta\epsilon}(x_2,x_K)\gamma_5
S_s^{\epsilon\beta}(x_K,x_Q)\gamma_\mu\gamma_5
]
\notag\\
& \pm\Tr[
S_s^{\alpha\alpha}(x_Q,x_Q)\gamma_\mu
S_l^{\beta\gamma}(x_Q,x_2)\gamma_5
S_l^{\gamma\delta}(x_2,x_1)\gamma_5
S_l^{\delta\epsilon}(x_1,x_K)\gamma_5
S_s^{\epsilon\beta}(x_K,x_Q)\gamma_\mu\gamma_5
],
\\
&\mbox{type3}_{\rm2\Tr,cmix}^{{\rm LR/LL},q\rm\mathchar`-loop}(x_K,x_Q,x_1,x_2)
\notag\\*
=& \pm\Tr[
S_q^{\alpha\beta}(x_Q,x_Q)\gamma_\mu\gamma_5
]\cdot\Tr[
S_l^{\beta\gamma}(x_Q,x_1)\gamma_5
S_l^{\gamma\delta}(x_1,x_2)\gamma_5
S_l^{\delta\epsilon}(x_2,x_K)\gamma_5
S_s^{\epsilon\alpha}(x_K,x_Q)\gamma_\mu
]
\notag\\
& \pm\Tr[
S_q^{\alpha\beta}(x_Q,x_Q)\gamma_\mu\gamma_5
]\cdot\Tr[
S_l^{\beta\gamma}(x_Q,x_2)\gamma_5
S_l^{\gamma\delta}(x_2,x_1)\gamma_5
S_l^{\delta\epsilon}(x_1,x_K)\gamma_5
S_s^{\epsilon\alpha}(x_K,x_Q)\gamma_\mu
]
\notag\\
& -\Tr[
S_q^{\alpha\beta}(x_Q,x_Q)\gamma_\mu
]\cdot\Tr[
S_l^{\beta\gamma}(x_Q,x_1)\gamma_5
S_l^{\gamma\delta}(x_1,x_2)\gamma_5
S_l^{\delta\epsilon}(x_2,x_K)\gamma_5
S_s^{\epsilon\alpha}(x_K,x_Q)\gamma_\mu\gamma_5
]
\notag\\
& -\Tr[
S_q^{\alpha\beta}(x_Q,x_Q)\gamma_\mu
]\cdot\Tr[
S_l^{\beta\gamma}(x_Q,x_2)\gamma_5
S_l^{\gamma\delta}(x_2,x_1)\gamma_5
S_l^{\delta\epsilon}(x_1,x_K)\gamma_5
S_s^{\epsilon\alpha}(x_K,x_Q)\gamma_\mu\gamma_5
],
\label{eq:lasttype3}
\\
&\mbox{type4}_{\rm2\Tr,cdiag}^{{\rm LR/LL},l\rm\mathchar`-loop}(x_K,x_Q,x_1,x_2)
\notag\\*
=& \pm\Tr[
S_l^{\alpha\beta}(x_Q,x_Q)\gamma_\mu\gamma_5
S_l^{\beta\gamma}(x_Q,x_K)\gamma_5
S_s^{\gamma\alpha}(x_K,x_Q)\gamma_\mu
]\cdot\Tr[
S_l^{\delta\epsilon}(x_1,x_2)\gamma_5
S_l^{\epsilon\delta}(x_2,x_1)\gamma_5
]
\notag\\*
&-\Tr[
S_l^{\alpha\beta}(x_Q,x_Q)\gamma_\mu
S_l^{\beta\gamma}(x_Q,x_K)\gamma_5
S_s^{\gamma\alpha}(x_K,x_Q)\gamma_\mu\gamma_5
]\cdot\Tr[
S_l^{\delta\epsilon}(x_1,x_2)\gamma_5
S_l^{\epsilon\delta}(x_2,x_1)\gamma_5
],
\\
&\mbox{type4}_{\rm2\Tr,cdiag}^{{\rm LR/LL},s\rm\mathchar`-loop}(x_K,x_Q,x_1,x_2)
\notag\\*
=& -\Tr[
S_s^{\alpha\beta}(x_Q,x_Q)\gamma_\mu\gamma_5
S_l^{\beta\gamma}(x_Q,x_K)\gamma_5
S_s^{\gamma\alpha}(x_K,x_Q)\gamma_\mu
]\cdot\Tr[
S_l^{\delta\epsilon}(x_1,x_2)\gamma_5
S_l^{\epsilon\delta}(x_2,x_1)\gamma_5
]
\notag\\*
&\pm\Tr[
S_s^{\alpha\beta}(x_Q,x_Q)\gamma_\mu
S_l^{\beta\gamma}(x_Q,x_K)\gamma_5
S_s^{\gamma\alpha}(x_K,x_Q)\gamma_\mu\gamma_5
]\cdot\Tr[
S_l^{\delta\epsilon}(x_1,x_2)\gamma_5
S_l^{\epsilon\delta}(x_2,x_1)\gamma_5
],
\\
&\mbox{type4}_{\rm3\Tr,cdiag}^{{\rm LR/LL},q\rm\mathchar`-loop}(x_K,x_Q,x_1,x_2)
\notag\\*
=& \pm\Tr[
S_q^{\alpha\alpha}(x_Q,x_Q)\gamma_\mu\gamma_5
]\cdot\Tr[
S_l^{\beta\gamma}(x_Q,x_K)\gamma_5
S_s^{\gamma\beta}(x_K,x_Q)\gamma_\mu
]\cdot\Tr[
S_l^{\delta\epsilon}(x_1,x_2)\gamma_5
S_l^{\epsilon\delta}(x_2,x_1)\gamma_5
]
\notag\\*
&-\Tr[
S_q^{\alpha\alpha}(x_Q,x_Q)\gamma_\mu
]\cdot\Tr[
S_l^{\beta\gamma}(x_Q,x_K)\gamma_5
S_s^{\gamma\beta}(x_K,x_Q)\gamma_\mu\gamma_5
]\cdot\Tr[
S_l^{\delta\epsilon}(x_1,x_2)\gamma_5
S_l^{\epsilon\delta}(x_2,x_1)\gamma_5
],
\\
&\mbox{type4}_{\rm2\Tr,cmix}^{{\rm LR/LL},l\rm\mathchar`-loop}(x_K,x_Q,x_1,x_2)
\notag\\*
=& \pm\Tr[
S_l^{\alpha\alpha}(x_Q,x_Q)\gamma_\mu\gamma_5
S_l^{\beta\gamma}(x_Q,x_K)\gamma_5
S_s^{\gamma\beta}(x_K,x_Q)\gamma_\mu
]\cdot\Tr[
S_l^{\delta\epsilon}(x_1,x_2)\gamma_5
S_l^{\epsilon\delta}(x_2,x_1)\gamma_5
]
\notag\\*
&-\Tr[
S_l^{\alpha\alpha}(x_Q,x_Q)\gamma_\mu
S_l^{\beta\gamma}(x_Q,x_K)\gamma_5
S_s^{\gamma\beta}(x_K,x_Q)\gamma_\mu\gamma_5
]\cdot\Tr[
S_l^{\delta\epsilon}(x_1,x_2)\gamma_5
S_l^{\epsilon\delta}(x_2,x_1)\gamma_5
],
\\
&\mbox{type4}_{\rm2\Tr,cmix}^{{\rm LR/LL},s\rm\mathchar`-loop}(x_K,x_Q,x_1,x_2)
\notag\\*
=& -\Tr[
S_s^{\alpha\alpha}(x_Q,x_Q)\gamma_\mu\gamma_5
S_l^{\beta\gamma}(x_Q,x_K)\gamma_5
S_s^{\gamma\beta}(x_K,x_Q)\gamma_\mu
]\cdot\Tr[
S_l^{\delta\epsilon}(x_1,x_2)\gamma_5
S_l^{\epsilon\delta}(x_2,x_1)\gamma_5
]
\notag\\*
&\pm\Tr[
S_s^{\alpha\alpha}(x_Q,x_Q)\gamma_\mu
S_l^{\beta\gamma}(x_Q,x_K)\gamma_5
S_s^{\gamma\beta}(x_K,x_Q)\gamma_\mu\gamma_5
]\cdot\Tr[
S_l^{\delta\epsilon}(x_1,x_2)\gamma_5
S_l^{\epsilon\delta}(x_2,x_1)\gamma_5
],
\\
&\mbox{type4}_{\rm3\Tr,cmix}^{{\rm LR/LL},q\rm\mathchar`-loop}(x_K,x_Q,x_1,x_2)
\notag\\*
=& \pm\Tr[
S_q^{\alpha\beta}(x_Q,x_Q)\gamma_\mu\gamma_5
]\cdot\Tr[
S_l^{\beta\gamma}(x_Q,x_K)\gamma_5
S_s^{\gamma\alpha}(x_K,x_Q)\gamma_\mu
]\cdot\Tr[
S_l^{\delta\epsilon}(x_1,x_2)\gamma_5
S_l^{\epsilon\delta}(x_2,x_1)\gamma_5
]
\notag\\*
&-\Tr[
S_q^{\alpha\beta}(x_Q,x_Q)\gamma_\mu
]\cdot\Tr[
S_l^{\beta\gamma}(x_Q,x_K)\gamma_5
S_s^{\gamma\alpha}(x_K,x_Q)\gamma_\mu\gamma_5
]\cdot\Tr[
S_l^{\delta\epsilon}(x_1,x_2)\gamma_5
S_l^{\epsilon\delta}(x_2,x_1)\gamma_5
],
\end{align}
where the sign `$\pm$' is determined by the chirality LR/LL of the four-quark operator.  All color indices $\alpha$--$\epsilon$ are implicitly summed over.  A generic symbol `$q$' is employed for some of {\it type3} and {\it type4} diagrams to indicate either the light ($l$) or strange ($s$) quark.  $S_q$ denotes the propagator of the quark $q$.  The arguments $x_K,x_Q$ are the locations of the kaon ($O^K$), four-quark ($Q_i$) operators, respectively.  Two arguments $x_1$ and $x_2$ are used to indicate the location of the two individual pion operators of a $\pi\pi$-like operator $\pi\pi^I$ with a specific isospin $I$.  Omitting these position arguments, the $K\to\pi\pi$ three-point functions $C_{i}^{K\to\pi\pi|_I}=\langle\pi\pi^I(x_1,x_2)Q_i(x_Q)O^K(x_K)^\dag\rangle$ read
\begin{align}
-\sqrt{6}\img\, C_{1}^{K\to\pi\pi|_{I=2}}
=& +\,{\rm type1}_{\rm2\Tr,cdiag}^{\rm LL}
-\,{\rm type1}_{\rm1\Tr,cdiag}^{\rm LL}
,\\
-\sqrt{6}\img\, C_{2}^{K\to\pi\pi|_{I=2}}
=& +\,{\rm type1}_{\rm2\Tr,cmix}^{\rm LL}
-\,{\rm type1}_{\rm1\Tr,cmix}^{\rm LL}
,\\
-\sqrt{6}\img\, C_{7}^{K\to\pi\pi|_{I=2}}
=& +\frac{3}{2}\,{\rm type1}_{\rm2\Tr,cdiag}^{\rm LR}
-\frac{3}{2}\,{\rm type1}_{\rm1\Tr,cdiag}^{\rm LR}
,\\
-\sqrt{6}\img\, C_{8}^{K\to\pi\pi|_{I=2}}
=& +\frac{3}{2}\,{\rm type1}_{\rm2\Tr,cmix}^{\rm LR}
-\frac{3}{2}\,{\rm type1}_{\rm1\Tr,cmix}^{\rm LR}
,\\
C_{9}^{K\to\pi\pi|_{I=2}}
=& \frac{3}{2}\,C_{1}^{K\to\pi\pi|_{I=2}}
,\\
C_{10}^{K\to\pi\pi|_{I=2}}
=& \frac{3}{2}\,C_{2}^{K\to\pi\pi|_{I=2}}
\end{align}
for the $\Delta I = 3/2$ channel and
\begin{align}
-\sqrt{3}\img\, C_{1}^{K\to\pi\pi|_{I=0}}
=& -\frac{1}{2}\,{\rm type1}_{\rm2\Tr,cdiag}^{\rm LL}
-\,{\rm type1}_{\rm1\Tr,cdiag}^{\rm LL}
+\frac{3}{2}\,{\rm type2}_{\rm2\Tr,cdiag}^{\rm LL}
\notag\\
& +\frac{3}{2}\,{\rm type3}_{\rm2\Tr,cdiag}^{{\rm LL},l\rm\mathchar`-loop}
-3\,{\rm type4}_{\rm3\Tr,cdiag}^{{\rm LL},l\rm\mathchar`-loop}
,\\
-\sqrt{3}\img\, C_{2}^{K\to\pi\pi|_{I=0}}
=& -\frac{1}{2}\,{\rm type1}_{\rm2\Tr,cmix}^{\rm LL}
-\,{\rm type1}_{\rm1\Tr,cmix}^{\rm LL}
+\frac{3}{2}\,{\rm type2}_{\rm2\Tr,cmix}^{\rm LL}
\notag\\
& +\frac{3}{2}\,{\rm type3}_{\rm2\Tr,cmix}^{{\rm LL},l\rm\mathchar`-loop}
-3\,{\rm type4}_{\rm3\Tr,cmix}^{{\rm LL},l\rm\mathchar`-loop}
,\\
-\sqrt{3}\img\, C_{3}^{K\to\pi\pi|_{I=0}}
=& -\frac{3}{2}\,{\rm type1}_{\rm1\Tr,cdiag}^{\rm LL}
+3\,{\rm type2}_{\rm2\Tr,cdiag}^{\rm LL}
-\frac{3}{2}\,{\rm type2}_{\rm1\Tr,cdiag}^{\rm LL}
+3\,{\rm type3}_{\rm2\Tr,cdiag}^{{\rm LL},l\rm\mathchar`-loop}
\notag\\
& -\frac{3}{2}\,{\rm type3}_{\rm1\Tr,cdiag}^{{\rm LL},l\rm\mathchar`-loop}
+\frac{3}{2}\,{\rm type3}_{\rm2\Tr,cdiag}^{{\rm LL},s\rm\mathchar`-loop}
-\frac{3}{2}\,{\rm type3}_{\rm1\Tr,cdiag}^{{\rm LL},s\rm\mathchar`-loop}
-6\,{\rm type4}_{\rm3\Tr,cdiag}^{{\rm LL},l\rm\mathchar`-loop}
\notag\\
& +3\,{\rm type4}_{\rm2\Tr,cdiag}^{{\rm LL},l\rm\mathchar`-loop}
-3\,{\rm type4}_{\rm3\Tr,cdiag}^{{\rm LL},s\rm\mathchar`-loop}
+3\,{\rm type4}_{\rm2\Tr,cdiag}^{{\rm LL},s\rm\mathchar`-loop}
,\\
-\sqrt{3}\img\, C_{4}^{K\to\pi\pi|_{I=0}}
=& -\frac{3}{2}\,{\rm type1}_{\rm1\Tr,cmix}^{\rm LL}
+3\,{\rm type2}_{\rm2\Tr,cmix}^{\rm LL}
-\frac{3}{2}\,{\rm type2}_{\rm1\Tr,cmix}^{\rm LL}
+3\,{\rm type3}_{\rm2\Tr,cmix}^{{\rm LL},l\rm\mathchar`-loop}
\notag\\
& -\frac{3}{2}\,{\rm type3}_{\rm1\Tr,cmix}^{{\rm LL},l\rm\mathchar`-loop}
+\frac{3}{2}\,{\rm type3}_{\rm2\Tr,cmix}^{{\rm LL},s\rm\mathchar`-loop}
-\frac{3}{2}\,{\rm type3}_{\rm1\Tr,cmix}^{{\rm LL},s\rm\mathchar`-loop}
-6\,{\rm type4}_{\rm3\Tr,cmix}^{{\rm LL},l\rm\mathchar`-loop}
\notag\\
& +3\,{\rm type4}_{\rm2\Tr,cmix}^{{\rm LL},l\rm\mathchar`-loop}
-3\,{\rm type4}_{\rm3\Tr,cmix}^{{\rm LL},s\rm\mathchar`-loop}
+3\,{\rm type4}_{\rm2\Tr,cmix}^{{\rm LL},s\rm\mathchar`-loop}
,\\
-\sqrt{3}\img\, C_{5}^{K\to\pi\pi|_{I=0}}
=& -\frac{3}{2}\,{\rm type1}_{\rm1\Tr,cdiag}^{\rm LR}
+3\,{\rm type2}_{\rm2\Tr,cdiag}^{\rm LR}
-\frac{3}{2}\,{\rm type2}_{\rm1\Tr,cdiag}^{\rm LR}
+3\,{\rm type3}_{\rm2\Tr,cdiag}^{{\rm LR},l\rm\mathchar`-loop}
\notag\\
& -\frac{3}{2}\,{\rm type3}_{\rm1\Tr,cdiag}^{{\rm LR},l\rm\mathchar`-loop}
+\frac{3}{2}\,{\rm type3}_{\rm2\Tr,cdiag}^{{\rm LR},s\rm\mathchar`-loop}
-\frac{3}{2}\,{\rm type3}_{\rm1\Tr,cdiag}^{{\rm LR},s\rm\mathchar`-loop}
-6\,{\rm type4}_{\rm3\Tr,cdiag}^{{\rm LR},l\rm\mathchar`-loop}
\notag\\
& +3\,{\rm type4}_{\rm2\Tr,cdiag}^{{\rm LR},l\rm\mathchar`-loop}
-3\,{\rm type4}_{\rm3\Tr,cdiag}^{{\rm LR},s\rm\mathchar`-loop}
+3\,{\rm type4}_{\rm2\Tr,cdiag}^{{\rm LR},s\rm\mathchar`-loop}
,\\
-\sqrt{3}\img\, C_{6}^{K\to\pi\pi|_{I=0}}
=& -\frac{3}{2}\,{\rm type1}_{\rm1\Tr,cmix}^{\rm LR}
+3\,{\rm type2}_{\rm2\Tr,cmix}^{\rm LR}
-\frac{3}{2}\,{\rm type2}_{\rm1\Tr,cmix}^{\rm LR}
+3\,{\rm type3}_{\rm2\Tr,cmix}^{{\rm LR},l\rm\mathchar`-loop}
\notag\\
& -\frac{3}{2}\,{\rm type3}_{\rm1\Tr,cmix}^{{\rm LR},l\rm\mathchar`-loop}
+\frac{3}{2}\,{\rm type3}_{\rm2\Tr,cmix}^{{\rm LR},s\rm\mathchar`-loop}
-\frac{3}{2}\,{\rm type3}_{\rm1\Tr,cmix}^{{\rm LR},s\rm\mathchar`-loop}
-6\,{\rm type4}_{\rm3\Tr,cmix}^{{\rm LR},l\rm\mathchar`-loop}
\notag\\
& +3\,{\rm type4}_{\rm2\Tr,cmix}^{{\rm LR},l\rm\mathchar`-loop}
-3\,{\rm type4}_{\rm3\Tr,cmix}^{{\rm LR},s\rm\mathchar`-loop}
+3\,{\rm type4}_{\rm2\Tr,cmix}^{{\rm LR},s\rm\mathchar`-loop}
,\\
-\sqrt{3}\img\, C_{7}^{K\to\pi\pi|_{I=0}}
=& -\frac{3}{4}\,{\rm type1}_{\rm2\Tr,cdiag}^{\rm LR}
-\frac{3}{4}\,{\rm type1}_{\rm1\Tr,cdiag}^{\rm LR}
+\frac{3}{4}\,{\rm type2}_{\rm2\Tr,cdiag}^{\rm LR}
+\frac{3}{4}\,{\rm type2}_{\rm1\Tr,cdiag}^{\rm LR}
\notag\\
& +\frac{3}{4}\,{\rm type3}_{\rm2\Tr,cdiag}^{{\rm LR},l\rm\mathchar`-loop}
+\frac{3}{4}\,{\rm type3}_{\rm1\Tr,cdiag}^{{\rm LR},l\rm\mathchar`-loop}
-\frac{3}{4}\,{\rm type3}_{\rm2\Tr,cdiag}^{{\rm LR},s\rm\mathchar`-loop}
+\frac{3}{4}\,{\rm type3}_{\rm1\Tr,cdiag}^{{\rm LR},s\rm\mathchar`-loop}
\notag\\
& -\frac{3}{2}\,{\rm type4}_{\rm3\Tr,cdiag}^{{\rm LR},l\rm\mathchar`-loop}
-\frac{3}{2}\,{\rm type4}_{\rm2\Tr,cdiag}^{{\rm LR},l\rm\mathchar`-loop}
+\frac{3}{2}\,{\rm type4}_{\rm3\Tr,cdiag}^{{\rm LR},s\rm\mathchar`-loop}
-\frac{3}{2}\,{\rm type4}_{\rm2\Tr,cdiag}^{{\rm LR},s\rm\mathchar`-loop}
,\\
-\sqrt{3}\img\, C_{8}^{K\to\pi\pi|_{I=0}}
=& -\frac{3}{4}\,{\rm type1}_{\rm2\Tr,cmix}^{\rm LR}
-\frac{3}{4}\,{\rm type1}_{\rm1\Tr,cmix}^{\rm LR}
+\frac{3}{4}\,{\rm type2}_{\rm2\Tr,cmix}^{\rm LR}
+\frac{3}{4}\,{\rm type2}_{\rm1\Tr,cmix}^{\rm LR}
\notag\\
& +\frac{3}{4}\,{\rm type3}_{\rm2\Tr,cmix}^{{\rm LR},l\rm\mathchar`-loop}
+\frac{3}{4}\,{\rm type3}_{\rm1\Tr,cmix}^{{\rm LR},l\rm\mathchar`-loop}
-\frac{3}{4}\,{\rm type3}_{\rm2\Tr,cmix}^{{\rm LR},s\rm\mathchar`-loop}
+\frac{3}{4}\,{\rm type3}_{\rm1\Tr,cmix}^{{\rm LR},s\rm\mathchar`-loop}
\notag\\
& -\frac{3}{2}\,{\rm type4}_{\rm3\Tr,cmix}^{{\rm LR},l\rm\mathchar`-loop}
-\frac{3}{2}\,{\rm type4}_{\rm2\Tr,cmix}^{{\rm LR},l\rm\mathchar`-loop}
+\frac{3}{2}\,{\rm type4}_{\rm3\Tr,cmix}^{{\rm LR},s\rm\mathchar`-loop}
-\frac{3}{2}\,{\rm type4}_{\rm2\Tr,cmix}^{{\rm LR},s\rm\mathchar`-loop}
,\\
-\sqrt{3}\img\, C_{9}^{K\to\pi\pi|_{I=0}}
=& -\frac{3}{4}\,{\rm type1}_{\rm2\Tr,cdiag}^{\rm LL}
-\frac{3}{4}\,{\rm type1}_{\rm1\Tr,cdiag}^{\rm LL}
+\frac{3}{4}\,{\rm type2}_{\rm2\Tr,cdiag}^{\rm LL}
+\frac{3}{4}\,{\rm type2}_{\rm1\Tr,cdiag}^{\rm LL}
\notag\\
& +\frac{3}{4}\,{\rm type3}_{\rm2\Tr,cdiag}^{{\rm LL},l\rm\mathchar`-loop}
+\frac{3}{4}\,{\rm type3}_{\rm1\Tr,cdiag}^{{\rm LL},l\rm\mathchar`-loop}
-\frac{3}{4}\,{\rm type3}_{\rm2\Tr,cdiag}^{{\rm LL},s\rm\mathchar`-loop}
+\frac{3}{4}\,{\rm type3}_{\rm1\Tr,cdiag}^{{\rm LL},s\rm\mathchar`-loop}
\notag\\
& -\frac{3}{2}\,{\rm type4}_{\rm3\Tr,cdiag}^{{\rm LL},l\rm\mathchar`-loop}
-\frac{3}{2}\,{\rm type4}_{\rm2\Tr,cdiag}^{{\rm LL},l\rm\mathchar`-loop}
+\frac{3}{2}\,{\rm type4}_{\rm3\Tr,cdiag}^{{\rm LL},s\rm\mathchar`-loop}
-\frac{3}{2}\,{\rm type4}_{\rm2\Tr,cdiag}^{{\rm LL},s\rm\mathchar`-loop}
,\\
-\sqrt{3}\img\, C_{10}^{K\to\pi\pi|_{I=0}}
=& -\frac{3}{4}\,{\rm type1}_{\rm2\Tr,cmix}^{\rm LL}
-\frac{3}{4}\,{\rm type1}_{\rm1\Tr,cmix}^{\rm LL}
+\frac{3}{4}\,{\rm type2}_{\rm2\Tr,cmix}^{\rm LL}
+\frac{3}{4}\,{\rm type2}_{\rm1\Tr,cmix}^{\rm LL}
\notag\\
& +\frac{3}{4}\,{\rm type3}_{\rm2\Tr,cmix}^{{\rm LL},l\rm\mathchar`-loop}
+\frac{3}{4}\,{\rm type3}_{\rm1\Tr,cmix}^{{\rm LL},l\rm\mathchar`-loop}
-\frac{3}{4}\,{\rm type3}_{\rm2\Tr,cmix}^{{\rm LL},s\rm\mathchar`-loop}
+\frac{3}{4}\,{\rm type3}_{\rm1\Tr,cmix}^{{\rm LL},s\rm\mathchar`-loop}
\notag\\
& -\frac{3}{2}\,{\rm type4}_{\rm3\Tr,cmix}^{{\rm LL},l\rm\mathchar`-loop}
-\frac{3}{2}\,{\rm type4}_{\rm2\Tr,cmix}^{{\rm LL},l\rm\mathchar`-loop}
+\frac{3}{2}\,{\rm type4}_{\rm3\Tr,cmix}^{{\rm LL},s\rm\mathchar`-loop}
-\frac{3}{2}\,{\rm type4}_{\rm2\Tr,cmix}^{{\rm LL},s\rm\mathchar`-loop}
\end{align}
for the $\Delta I = 1/2$ channel.

Note that the contractions of {\it type1--3} diagrams given in Eqs.~\eqref{eq:1sttype1}--\eqref{eq:lasttype3} are twice as large as the corresponding contractions $\circledsingle{1}$--$\circled{32}$ given in Ref.~\cite{Blum:2011pu} and consequently the coefficients for these contractions differ by factor of 2.

\subsection{$K\to\sigma$}

All $K\to\sigma$ three-point functions are a linear combination of the following contractions:
\begin{align}
&{\rm type2\sigma}_{\rm1\Tr,cdiag}^{\rm LR/LL}(x_K,x_Q,x_\sigma)
\notag\\*
=& \pm \Tr[
S_l^{\alpha\beta}(x_Q,x_\sigma)
S_l^{\beta\gamma}(x_\sigma,x_Q) \gamma_\mu\gamma_5
S_l^{\gamma\delta}(x_Q,x_K) \gamma_5
S_s^{\delta\alpha}(x_K,x_Q) \gamma_\mu
]
\notag\\
& -\Tr[
S_l^{\alpha\beta}(x_Q,x_\sigma)
S_l^{\beta\gamma}(x_\sigma,x_Q) \gamma_\mu
S_l^{\gamma\delta}(x_Q,x_K) \gamma_5
S_s^{\delta\alpha}(x_K,x_Q) \gamma_\mu\gamma_5
],
\\
&{\rm type2\sigma}_{\rm2\Tr,cdiag}^{\rm LR/LL}(x_K,x_Q,x_\sigma)
\notag\\*
=& \pm \Tr[
S_l^{\alpha\beta}(x_Q,x_\sigma)
S_l^{\beta\alpha}(x_\sigma,x_Q) \gamma_\mu\gamma_5
]\cdot\Tr[
S_l^{\gamma\delta}(x_Q,x_K) \gamma_5
S_s^{\delta\gamma}(x_K,x_Q) \gamma_\mu
]
\notag\\
& -\Tr[
S_l^{\alpha\beta}(x_Q,x_\sigma)
S_l^{\beta\alpha}(x_\sigma,x_Q) \gamma_\mu
]\cdot\Tr[
S_l^{\gamma\delta}(x_Q,x_K) \gamma_5
S_s^{\delta\gamma}(x_K,x_Q) \gamma_\mu\gamma_5
],
\\
&{\rm type2\sigma}_{\rm1\Tr,cmix}^{\rm LR/LL}(x_K,x_Q,x_\sigma)
\notag\\*
=& \pm \Tr[
S_l^{\alpha\beta}(x_Q,x_\sigma)
S_l^{\beta\alpha}(x_\sigma,x_Q) \gamma_\mu\gamma_5
S_l^{\gamma\delta}(x_Q,x_K) \gamma_5
S_s^{\delta\gamma}(x_K,x_Q) \gamma_\mu
]
\notag\\
& -\Tr[
S_l^{\alpha\beta}(x_Q,x_\sigma)
S_l^{\beta\alpha}(x_\sigma,x_Q) \gamma_\mu
S_l^{\gamma\delta}(x_Q,x_K) \gamma_5
S_s^{\delta\gamma}(x_K,x_Q) \gamma_\mu\gamma_5
],
\\
&{\rm type2\sigma}_{\rm2\Tr,cmix}^{\rm LR/LL}(x_K,x_Q,x_\sigma)
\notag\\*
=& \pm \Tr[
S_l^{\alpha\beta}(x_Q,x_\sigma)
S_l^{\beta\gamma}(x_\sigma,x_Q) \gamma_\mu\gamma_5
]\cdot\Tr[
S_l^{\gamma\delta}(x_Q,x_K) \gamma_5
S_s^{\delta\alpha}(x_K,x_Q) \gamma_\mu
]
\notag\\
& -\Tr[
S_l^{\alpha\beta}(x_Q,x_\sigma)
S_l^{\beta\gamma}(x_\sigma,x_Q) \gamma_\mu
]\cdot\Tr[
S_l^{\gamma\delta}(x_Q,x_K) \gamma_5
S_s^{\delta\alpha}(x_K,x_Q) \gamma_\mu\gamma_5
],
\\
&{\rm type3\sigma}_{\rm1\Tr,cdiag}^{{\rm LR/LL},l\rm\mathchar`-loop}(x_K,x_Q,x_\sigma)
\notag\\*
=& \pm \Tr[
S_l^{\alpha\beta}(x_Q,x_Q) \gamma_\mu\gamma_5
S_l^{\beta\gamma}(x_Q,x_\sigma)
S_l^{\gamma\delta}(x_\sigma,x_K) \gamma_5
S_s^{\delta\alpha}(x_K,x_Q) \gamma_\mu
]
\notag\\
& -\Tr[
S_l^{\alpha\beta}(x_Q,x_Q) \gamma_\mu
S_l^{\beta\gamma}(x_Q,x_\sigma)
S_l^{\gamma\delta}(x_\sigma,x_K) \gamma_5
S_s^{\delta\alpha}(x_K,x_Q) \gamma_\mu\gamma_5
],
\\
&{\rm type3\sigma}_{\rm1\Tr,cdiag}^{{\rm LR/LL},s\rm\mathchar`-loop}(x_K,x_Q,x_\sigma)
\notag\\*
=& - \Tr[
S_s^{\alpha\beta}(x_Q,x_Q) \gamma_\mu\gamma_5
S_l^{\beta\gamma}(x_Q,x_\sigma)
S_l^{\gamma\delta}(x_\sigma,x_K) \gamma_5
S_s^{\delta\alpha}(x_K,x_Q) \gamma_\mu
]
\notag\\
& \pm\Tr[
S_s^{\alpha\beta}(x_Q,x_Q) \gamma_\mu
S_l^{\beta\gamma}(x_Q,x_\sigma)
S_l^{\gamma\delta}(x_\sigma,x_K) \gamma_5
S_s^{\delta\alpha}(x_K,x_Q) \gamma_\mu\gamma_5
],
\\
&{\rm type3\sigma}_{\rm2\Tr,cdiag}^{{\rm LR/LL},q\rm\mathchar`-loop}(x_K,x_Q,x_\sigma)
\notag\\*
=& \pm \Tr[
S_q^{\alpha\alpha}(x_Q,x_Q) \gamma_\mu\gamma_5
]\cdot\Tr[
S_l^{\beta\gamma}(x_Q,x_\sigma)
S_l^{\gamma\delta}(x_\sigma,x_K) \gamma_5
S_s^{\delta\beta}(x_K,x_Q) \gamma_\mu
]
\notag\\
& -\Tr[
S_q^{\alpha\alpha}(x_Q,x_Q) \gamma_\mu
]\cdot\Tr[
S_l^{\beta\gamma}(x_Q,x_\sigma)
S_l^{\gamma\delta}(x_\sigma,x_K) \gamma_5
S_s^{\delta\beta}(x_K,x_Q) \gamma_\mu\gamma_5
],
\\
&{\rm type3\sigma}_{\rm1\Tr,cmix}^{{\rm LR/LL},l\rm\mathchar`-loop}(x_K,x_Q,x_\sigma)
\notag\\*
=& \pm \Tr[
S_l^{\alpha\alpha}(x_Q,x_Q) \gamma_\mu\gamma_5
S_l^{\beta\gamma}(x_Q,x_\sigma)
S_l^{\gamma\delta}(x_\sigma,x_K) \gamma_5
S_s^{\delta\beta}(x_K,x_Q) \gamma_\mu
]
\notag\\
& -\Tr[
S_l^{\alpha\alpha}(x_Q,x_Q) \gamma_\mu
S_l^{\beta\gamma}(x_Q,x_\sigma)
S_l^{\gamma\delta}(x_\sigma,x_K) \gamma_5
S_s^{\delta\beta}(x_K,x_Q) \gamma_\mu\gamma_5
],
\\
&{\rm type3\sigma}_{\rm1\Tr,cmix}^{{\rm LR/LL},s\rm\mathchar`-loop}(x_K,x_Q,x_\sigma)
\notag\\*
=& - \Tr[
S_s^{\alpha\alpha}(x_Q,x_Q) \gamma_\mu\gamma_5
S_l^{\beta\gamma}(x_Q,x_\sigma)
S_l^{\gamma\delta}(x_\sigma,x_K) \gamma_5
S_s^{\delta\beta}(x_K,x_Q) \gamma_\mu
]
\notag\\
& \pm\Tr[
S_s^{\alpha\alpha}(x_Q,x_Q) \gamma_\mu
S_l^{\beta\gamma}(x_Q,x_\sigma)
S_l^{\gamma\delta}(x_\sigma,x_K) \gamma_5
S_s^{\delta\beta}(x_K,x_Q) \gamma_\mu\gamma_5
],
\\
&{\rm type3\sigma}_{\rm2\Tr,cmix}^{{\rm LR/LL},q\rm\mathchar`-loop}(x_K,x_Q,x_\sigma)
\notag\\*
=& \pm \Tr[
S_q^{\alpha\beta}(x_Q,x_Q) \gamma_\mu\gamma_5
]\cdot\Tr[
S_l^{\beta\gamma}(x_Q,x_\sigma)
S_l^{\gamma\delta}(x_\sigma,x_K) \gamma_5
S_s^{\delta\alpha}(x_K,x_Q) \gamma_\mu
]
\notag\\
& -\Tr[
S_q^{\alpha\beta}(x_Q,x_Q) \gamma_\mu
]\cdot\Tr[
S_l^{\beta\gamma}(x_Q,x_\sigma)
S_l^{\gamma\delta}(x_\sigma,x_K) \gamma_5
S_s^{\delta\alpha}(x_K,x_Q) \gamma_\mu\gamma_5
],
\\
&{\rm type4\sigma}_{\rm2\Tr,cdiag}^{{\rm LR/LL},l\rm\mathchar`-loop}(x_K,x_Q,x_\sigma)
\notag\\*
=& \pm \Tr[
S_l^{\alpha\beta}(x_Q,x_Q) \gamma_\mu\gamma_5
S_l^{\beta\gamma}(x_Q,x_K)\gamma_5
S_s^{\gamma\alpha}(x_K,x_Q) \gamma_\mu
]\cdot\Tr[
S_l^{\delta\delta}(x_\sigma,x_\sigma)
]
\notag\\
& -\Tr[
S_l^{\alpha\beta}(x_Q,x_Q) \gamma_\mu
S_l^{\beta\gamma}(x_Q,x_K)\gamma_5
S_s^{\gamma\alpha}(x_K,x_Q) \gamma_\mu\gamma_5
]\cdot\Tr[
S_l^{\delta\delta}(x_\sigma,x_\sigma)
],
\\
&{\rm type4\sigma}_{\rm2\Tr,cdiag}^{{\rm LR/LL},s\rm\mathchar`-loop}(x_K,x_Q,x_\sigma)
\notag\\*
=& - \Tr[
S_s^{\alpha\beta}(x_Q,x_Q) \gamma_\mu\gamma_5
S_l^{\beta\gamma}(x_Q,x_K)\gamma_5
S_s^{\gamma\alpha}(x_K,x_Q) \gamma_\mu
]\cdot\Tr[
S_l^{\delta\delta}(x_\sigma,x_\sigma)
]
\notag\\
& \pm\Tr[
S_s^{\alpha\beta}(x_Q,x_Q) \gamma_\mu
S_l^{\beta\gamma}(x_Q,x_K)\gamma_5
S_s^{\gamma\alpha}(x_K,x_Q) \gamma_\mu\gamma_5
]\cdot\Tr[
S_l^{\delta\delta}(x_\sigma,x_\sigma)
],
\\
&{\rm type4\sigma}_{\rm3\Tr,cdiag}^{{\rm LR/LL},q\rm\mathchar`-loop}(x_K,x_Q,x_\sigma)
\notag\\*
=& \pm \Tr[
S_q^{\alpha\alpha}(x_Q,x_Q) \gamma_\mu\gamma_5
]\cdot\Tr[
S_l^{\beta\gamma}(x_Q,x_K)\gamma_5
S_s^{\gamma\beta}(x_K,x_Q) \gamma_\mu
]\cdot\Tr[
S_l^{\delta\delta}(x_\sigma,x_\sigma)
]
\notag\\
& -\Tr[
S_q^{\alpha\alpha}(x_Q,x_Q) \gamma_\mu
]\cdot\Tr[
S_l^{\beta\gamma}(x_Q,x_K)\gamma_5
S_s^{\gamma\beta}(x_K,x_Q) \gamma_\mu\gamma_5
]\cdot\Tr[
S_l^{\delta\delta}(x_\sigma,x_\sigma)
],
\\
&{\rm type4\sigma}_{\rm2\Tr,cmix}^{{\rm LR/LL},l\rm\mathchar`-loop}(x_K,x_Q,x_\sigma)
\notag\\*
=& \pm \Tr[
S_l^{\alpha\alpha}(x_Q,x_Q) \gamma_\mu\gamma_5
S_l^{\beta\gamma}(x_Q,x_K)\gamma_5
S_s^{\gamma\beta}(x_K,x_Q) \gamma_\mu
]\cdot\Tr[
S_l^{\delta\delta}(x_\sigma,x_\sigma)
]
\notag\\
& -\Tr[
S_l^{\alpha\alpha}(x_Q,x_Q) \gamma_\mu
S_l^{\beta\gamma}(x_Q,x_K)\gamma_5
S_s^{\gamma\beta}(x_K,x_Q) \gamma_\mu\gamma_5
]\cdot\Tr[
S_l^{\delta\delta}(x_\sigma,x_\sigma)
],
\\
&{\rm type4\sigma}_{\rm2\Tr,cmix}^{{\rm LR/LL},s\rm\mathchar`-loop}(x_K,x_Q,x_\sigma)
\notag\\*
=& - \Tr[
S_s^{\alpha\alpha}(x_Q,x_Q) \gamma_\mu\gamma_5
S_l^{\beta\gamma}(x_Q,x_K)\gamma_5
S_s^{\gamma\beta}(x_K,x_Q) \gamma_\mu
]\cdot\Tr[
S_l^{\delta\delta}(x_\sigma,x_\sigma)
]
\notag\\
& \pm\Tr[
S_s^{\alpha\alpha}(x_Q,x_Q) \gamma_\mu
S_l^{\beta\gamma}(x_Q,x_K)\gamma_5
S_s^{\gamma\beta}(x_K,x_Q) \gamma_\mu\gamma_5
]\cdot\Tr[
S_l^{\delta\delta}(x_\sigma,x_\sigma)
],
\\
&{\rm type4\sigma}_{\rm3\Tr,cmix}^{{\rm LR/LL},q\rm\mathchar`-loop}(x_K,x_Q,x_\sigma)
\notag\\*
=& \pm \Tr[
S_q^{\alpha\beta}(x_Q,x_Q) \gamma_\mu\gamma_5
]\cdot\Tr[
S_l^{\beta\gamma}(x_Q,x_K)\gamma_5
S_s^{\gamma\alpha}(x_K,x_Q) \gamma_\mu
]\cdot\Tr[
S_l^{\delta\delta}(x_\sigma,x_\sigma)
]
\notag\\
& -\Tr[
S_q^{\alpha\beta}(x_Q,x_Q) \gamma_\mu
]\cdot\Tr[
S_l^{\beta\gamma}(x_Q,x_K)\gamma_5
S_s^{\gamma\alpha}(x_K,x_Q) \gamma_\mu\gamma_5
]\cdot\Tr[
S_l^{\delta\delta}(x_\sigma,x_\sigma)
].
\end{align}
We have only $\Delta I = 1/2$ channel of the $K\to\sigma$ three-point functions, which is defined as $C_i^{K\to\sigma} = \langle \sigma(x_\sigma)Q_i(x_Q)O^K(x_K)^\dag\rangle$ and given as follows:
\begin{align}
\sqrt{2}\img\, C_{1}^{K\to\sigma}
=& +\,{\rm type2\sigma}_{\rm2\Tr,cdiag}^{\rm LL}
+\,{\rm type3\sigma}_{\rm2\Tr,cdiag}^{{\rm LL},l\rm\mathchar`-loop}
-2\,{\rm type4\sigma}_{\rm3\Tr,cdiag}^{{\rm LL},l\rm\mathchar`-loop}
,\\
\sqrt{2}\img\, C_{2}^{K\to\sigma}
=& +\,{\rm type2\sigma}_{\rm2\Tr,cmix}^{\rm LL}
+\,{\rm type3\sigma}_{\rm2\Tr,cmix}^{{\rm LL},l\rm\mathchar`-loop}
-2\,{\rm type4\sigma}_{\rm3\Tr,cmix}^{{\rm LL},l\rm\mathchar`-loop}
,\\
\sqrt{2}\img\, C_{3}^{K\to\sigma}
=& +2\,{\rm type2\sigma}_{\rm2\Tr,cdiag}^{\rm LL}
-\,{\rm type2\sigma}_{\rm1\Tr,cdiag}^{\rm LL}
+2\,{\rm type3\sigma}_{\rm2\Tr,cdiag}^{{\rm LL},l\rm\mathchar`-loop}
-\,{\rm type3\sigma}_{\rm1\Tr,cdiag}^{{\rm LL},l\rm\mathchar`-loop}
\notag\\
& +\,{\rm type3\sigma}_{\rm2\Tr,cdiag}^{{\rm LL},s\rm\mathchar`-loop}
-\,{\rm type3\sigma}_{\rm1\Tr,cdiag}^{{\rm LL},s\rm\mathchar`-loop}
-4\,{\rm type4\sigma}_{\rm3\Tr,cdiag}^{{\rm LL},l\rm\mathchar`-loop}
+2\,{\rm type4\sigma}_{\rm2\Tr,cdiag}^{{\rm LL},l\rm\mathchar`-loop}
\notag\\
& -2\,{\rm type4\sigma}_{\rm3\Tr,cdiag}^{{\rm LL},s\rm\mathchar`-loop}
+2\,{\rm type4\sigma}_{\rm2\Tr,cdiag}^{{\rm LL},s\rm\mathchar`-loop}
,\\
\sqrt{2}\img\, C_{4}^{K\to\sigma}
=& +2\,{\rm type2\sigma}_{\rm2\Tr,cmix}^{\rm LL}
-\,{\rm type2\sigma}_{\rm1\Tr,cmix}^{\rm LL}
+2\,{\rm type3\sigma}_{\rm2\Tr,cmix}^{{\rm LL},l\rm\mathchar`-loop}
-\,{\rm type3\sigma}_{\rm1\Tr,cmix}^{{\rm LL},l\rm\mathchar`-loop}
\notag\\
& +\,{\rm type3\sigma}_{\rm2\Tr,cmix}^{{\rm LL},s\rm\mathchar`-loop}
-\,{\rm type3\sigma}_{\rm1\Tr,cmix}^{{\rm LL},s\rm\mathchar`-loop}
-4\,{\rm type4\sigma}_{\rm3\Tr,cmix}^{{\rm LL},l\rm\mathchar`-loop}
+2\,{\rm type4\sigma}_{\rm2\Tr,cmix}^{{\rm LL},l\rm\mathchar`-loop}
\notag\\
& -2\,{\rm type4\sigma}_{\rm3\Tr,cmix}^{{\rm LL},s\rm\mathchar`-loop}
+2\,{\rm type4\sigma}_{\rm2\Tr,cmix}^{{\rm LL},s\rm\mathchar`-loop}
,\\
\sqrt{2}\img\, C_{5}^{K\to\sigma}
=& +2\,{\rm type2\sigma}_{\rm2\Tr,cdiag}^{\rm LR}
-\,{\rm type2\sigma}_{\rm1\Tr,cdiag}^{\rm LR}
+2\,{\rm type3\sigma}_{\rm2\Tr,cdiag}^{{\rm LR},l\rm\mathchar`-loop}
-\,{\rm type3\sigma}_{\rm1\Tr,cdiag}^{{\rm LR},l\rm\mathchar`-loop}
\notag\\
& +\,{\rm type3\sigma}_{\rm2\Tr,cdiag}^{{\rm LR},s\rm\mathchar`-loop}
-\,{\rm type3\sigma}_{\rm1\Tr,cdiag}^{{\rm LR},s\rm\mathchar`-loop}
-4\,{\rm type4\sigma}_{\rm3\Tr,cdiag}^{{\rm LR},l\rm\mathchar`-loop}
+2\,{\rm type4\sigma}_{\rm2\Tr,cdiag}^{{\rm LR},l\rm\mathchar`-loop}
\notag\\
& -2\,{\rm type4\sigma}_{\rm3\Tr,cdiag}^{{\rm LR},s\rm\mathchar`-loop}
+2\,{\rm type4\sigma}_{\rm2\Tr,cdiag}^{{\rm LR},s\rm\mathchar`-loop}
,\\
\sqrt{2}\img\, C_{6}^{K\to\sigma}
=& +2\,{\rm type2\sigma}_{\rm2\Tr,cmix}^{\rm LR}
-\,{\rm type2\sigma}_{\rm1\Tr,cmix}^{\rm LR}
+2\,{\rm type3\sigma}_{\rm2\Tr,cmix}^{{\rm LR},l\rm\mathchar`-loop}
-\,{\rm type3\sigma}_{\rm1\Tr,cmix}^{{\rm LR},l\rm\mathchar`-loop}
\notag\\
& +\,{\rm type3\sigma}_{\rm2\Tr,cmix}^{{\rm LR},s\rm\mathchar`-loop}
-\,{\rm type3\sigma}_{\rm1\Tr,cmix}^{{\rm LR},s\rm\mathchar`-loop}
-4\,{\rm type4\sigma}_{\rm3\Tr,cmix}^{{\rm LR},l\rm\mathchar`-loop}
+2\,{\rm type4\sigma}_{\rm2\Tr,cmix}^{{\rm LR},l\rm\mathchar`-loop}
\notag\\
& -2\,{\rm type4\sigma}_{\rm3\Tr,cmix}^{{\rm LR},s\rm\mathchar`-loop}
+2\,{\rm type4\sigma}_{\rm2\Tr,cmix}^{{\rm LR},s\rm\mathchar`-loop}
,\\
\sqrt{2}\img\, C_{7}^{K\to\sigma}
=& +\frac{1}{2}\,{\rm type2\sigma}_{\rm2\Tr,cdiag}^{\rm LR}
+\frac{1}{2}\,{\rm type2\sigma}_{\rm1\Tr,cdiag}^{\rm LR}
+\frac{1}{2}\,{\rm type3\sigma}_{\rm2\Tr,cdiag}^{{\rm LR},l\rm\mathchar`-loop}
+\frac{1}{2}\,{\rm type3\sigma}_{\rm1\Tr,cdiag}^{{\rm LR},l\rm\mathchar`-loop}
\notag\\
& -\frac{1}{2}\,{\rm type3\sigma}_{\rm2\Tr,cdiag}^{{\rm LR},s\rm\mathchar`-loop}
+\frac{1}{2}\,{\rm type3\sigma}_{\rm1\Tr,cdiag}^{{\rm LR},s\rm\mathchar`-loop}
-\,{\rm type4\sigma}_{\rm3\Tr,cdiag}^{{\rm LR},l\rm\mathchar`-loop}
-\,{\rm type4\sigma}_{\rm2\Tr,cdiag}^{{\rm LR},l\rm\mathchar`-loop}
\notag\\
& +\,{\rm type4\sigma}_{\rm3\Tr,cdiag}^{{\rm LR},s\rm\mathchar`-loop}
-\,{\rm type4\sigma}_{\rm2\Tr,cdiag}^{{\rm LR},s\rm\mathchar`-loop}
,\\
\sqrt{2}\img\, C_{8}^{K\to\sigma}
=& +\frac{1}{2}\,{\rm type2\sigma}_{\rm2\Tr,cmix}^{\rm LR}
+\frac{1}{2}\,{\rm type2\sigma}_{\rm1\Tr,cmix}^{\rm LR}
+\frac{1}{2}\,{\rm type3\sigma}_{\rm2\Tr,cmix}^{{\rm LR},l\rm\mathchar`-loop}
+\frac{1}{2}\,{\rm type3\sigma}_{\rm1\Tr,cmix}^{{\rm LR},l\rm\mathchar`-loop}
\notag\\
& -\frac{1}{2}\,{\rm type3\sigma}_{\rm2\Tr,cmix}^{{\rm LR},s\rm\mathchar`-loop}
+\frac{1}{2}\,{\rm type3\sigma}_{\rm1\Tr,cmix}^{{\rm LR},s\rm\mathchar`-loop}
-\,{\rm type4\sigma}_{\rm3\Tr,cmix}^{{\rm LR},l\rm\mathchar`-loop}
-\,{\rm type4\sigma}_{\rm2\Tr,cmix}^{{\rm LR},l\rm\mathchar`-loop}
\notag\\
& +\,{\rm type4\sigma}_{\rm3\Tr,cmix}^{{\rm LR},s\rm\mathchar`-loop}
-\,{\rm type4\sigma}_{\rm2\Tr,cmix}^{{\rm LR},s\rm\mathchar`-loop}
,\\
\sqrt{2}\img\, C_{9}^{K\to\sigma}
=& +\frac{1}{2}\,{\rm type2\sigma}_{\rm2\Tr,cdiag}^{\rm LL}
+\frac{1}{2}\,{\rm type2\sigma}_{\rm1\Tr,cdiag}^{\rm LL}
+\frac{1}{2}\,{\rm type3\sigma}_{\rm2\Tr,cdiag}^{{\rm LL},l\rm\mathchar`-loop}
+\frac{1}{2}\,{\rm type3\sigma}_{\rm1\Tr,cdiag}^{{\rm LL},l\rm\mathchar`-loop}
\notag\\
& -\frac{1}{2}\,{\rm type3\sigma}_{\rm2\Tr,cdiag}^{{\rm LL},s\rm\mathchar`-loop}
+\frac{1}{2}\,{\rm type3\sigma}_{\rm1\Tr,cdiag}^{{\rm LL},s\rm\mathchar`-loop}
-\,{\rm type4\sigma}_{\rm3\Tr,cdiag}^{{\rm LL},l\rm\mathchar`-loop}
-\,{\rm type4\sigma}_{\rm2\Tr,cdiag}^{{\rm LL},l\rm\mathchar`-loop}
\notag\\
& +\,{\rm type4\sigma}_{\rm3\Tr,cdiag}^{{\rm LL},s\rm\mathchar`-loop}
-\,{\rm type4\sigma}_{\rm2\Tr,cdiag}^{{\rm LL},s\rm\mathchar`-loop}
,\\
\sqrt{2}\img\, C_{10}^{K\to\sigma}
=& +\frac{1}{2}\,{\rm type2\sigma}_{\rm2\Tr,cmix}^{\rm LL}
+\frac{1}{2}\,{\rm type2\sigma}_{\rm1\Tr,cmix}^{\rm LL}
+\frac{1}{2}\,{\rm type3\sigma}_{\rm2\Tr,cmix}^{{\rm LL},l\rm\mathchar`-loop}
+\frac{1}{2}\,{\rm type3\sigma}_{\rm1\Tr,cmix}^{{\rm LL},l\rm\mathchar`-loop}
\notag\\
& -\frac{1}{2}\,{\rm type3\sigma}_{\rm2\Tr,cmix}^{{\rm LL},s\rm\mathchar`-loop}
+\frac{1}{2}\,{\rm type3\sigma}_{\rm1\Tr,cmix}^{{\rm LL},s\rm\mathchar`-loop}
-\,{\rm type4\sigma}_{\rm3\Tr,cmix}^{{\rm LL},l\rm\mathchar`-loop}
-\,{\rm type4\sigma}_{\rm2\Tr,cmix}^{{\rm LL},l\rm\mathchar`-loop}
\notag\\
& +\,{\rm type4\sigma}_{\rm3\Tr,cmix}^{{\rm LL},s\rm\mathchar`-loop}
-\,{\rm type4\sigma}_{\rm2\Tr,cmix}^{{\rm LL},s\rm\mathchar`-loop}.
\end{align}

\subsection{$K$ to vacuum}

We use the following fundamental contractions for the $K$ to vacuum correlation functions:
\begin{align}
{\rm k2vac}_{\rm1\Tr,cdiag}^{{\rm LR/LL},l\rm\mathchar`-loop}(x_K,x_Q)
=& \pm \Tr[
S_l^{\alpha\beta}(x_Q,x_Q) \gamma_\mu\gamma_5
S_l^{\beta\gamma}(x_Q,x_K)\gamma_5
S_s^{\gamma\alpha}(x_K,x_Q) \gamma_\mu
]
\notag\\
& -\Tr[
S_l^{\alpha\beta}(x_Q,x_Q) \gamma_\mu
S_l^{\beta\gamma}(x_Q,x_K)\gamma_5
S_s^{\gamma\alpha}(x_K,x_Q) \gamma_\mu\gamma_5
],
\\
{\rm k2vac}_{\rm1\Tr,cdiag}^{{\rm LR/LL},s\rm\mathchar`-loop}(x_K,x_Q)
=& - \Tr[
S_s^{\alpha\beta}(x_Q,x_Q) \gamma_\mu\gamma_5
S_l^{\beta\gamma}(x_Q,x_K)\gamma_5
S_s^{\gamma\alpha}(x_K,x_Q) \gamma_\mu
]
\notag\\
& \pm\Tr[
S_s^{\alpha\beta}(x_Q,x_Q) \gamma_\mu
S_l^{\beta\gamma}(x_Q,x_K)\gamma_5
S_s^{\gamma\alpha}(x_K,x_Q) \gamma_\mu\gamma_5
],
\\
{\rm k2vac}_{\rm2\Tr,cdiag}^{{\rm LR/LL},q\rm\mathchar`-loop}(x_K,x_Q)
=& \pm \Tr[
S_q^{\alpha\alpha}(x_Q,x_Q) \gamma_\mu\gamma_5
]\cdot\Tr[
S_l^{\beta\gamma}(x_Q,x_K)\gamma_5
S_s^{\gamma\beta}(x_K,x_Q) \gamma_\mu
]
\notag\\
& -\Tr[
S_q^{\alpha\alpha}(x_Q,x_Q) \gamma_\mu
]\cdot\Tr[
S_l^{\beta\gamma}(x_Q,x_K)\gamma_5
S_s^{\gamma\beta}(x_K,x_Q) \gamma_\mu\gamma_5
],
\\
{\rm k2vac}_{\rm1\Tr,cmix}^{{\rm LR/LL},l\rm\mathchar`-loop}(x_K,x_Q)
=& \pm \Tr[
S_l^{\alpha\alpha}(x_Q,x_Q) \gamma_\mu\gamma_5
S_l^{\beta\gamma}(x_Q,x_K)\gamma_5
S_s^{\gamma\beta}(x_K,x_Q) \gamma_\mu
]
\notag\\
& -\Tr[
S_l^{\alpha\alpha}(x_Q,x_Q) \gamma_\mu
S_l^{\beta\gamma}(x_Q,x_K)\gamma_5
S_s^{\gamma\beta}(x_K,x_Q) \gamma_\mu\gamma_5
],
\\
{\rm k2vac}_{\rm1\Tr,cmix}^{{\rm LR/LL},s\rm\mathchar`-loop}(x_K,x_Q)
=& - \Tr[
S_s^{\alpha\alpha}(x_Q,x_Q) \gamma_\mu\gamma_5
S_l^{\beta\gamma}(x_Q,x_K)\gamma_5
S_s^{\gamma\beta}(x_K,x_Q) \gamma_\mu
]
\notag\\
& \pm\Tr[
S_s^{\alpha\alpha}(x_Q,x_Q) \gamma_\mu
S_l^{\beta\gamma}(x_Q,x_K)\gamma_5
S_s^{\gamma\beta}(x_K,x_Q) \gamma_\mu\gamma_5
],
\\
{\rm k2vac}_{\rm2\Tr,cmix}^{{\rm LR/LL},q\rm\mathchar`-loop}(x_K,x_Q)
=& \pm \Tr[
S_q^{\alpha\beta}(x_Q,x_Q) \gamma_\mu\gamma_5
]\cdot\Tr[
S_l^{\beta\gamma}(x_Q,x_K)\gamma_5
S_s^{\gamma\alpha}(x_K,x_Q) \gamma_\mu
]
\notag\\
& -\Tr[
S_q^{\alpha\beta}(x_Q,x_Q) \gamma_\mu
]\cdot\Tr[
S_l^{\beta\gamma}(x_Q,x_K)\gamma_5
S_s^{\gamma\alpha}(x_K,x_Q) \gamma_\mu\gamma_5
].
\end{align}
The $K$ to vacuum correlation functions $C_i^{K\to|0\rangle}=\langle Q_i(x_Q)O^K(x_K)^\dag\rangle$ read
\begin{align}
\img\, C_{1}^{K\to|0\rangle}
=\,& \,{\rm k2vac}_{\rm2\Tr,cdiag}^{{\rm LL},l\rm\mathchar`-loop}
,\\
\img\, C_{2}^{K\to|0\rangle}
=\,& \,{\rm k2vac}_{\rm2\Tr,cmix}^{{\rm LL},l\rm\mathchar`-loop}
,\\
\img\, C_{3}^{K\to|0\rangle}
=\,& 2\,{\rm k2vac}_{\rm2\Tr,cdiag}^{{\rm LL},l\rm\mathchar`-loop}
-\,{\rm k2vac}_{\rm1\Tr,cdiag}^{{\rm LL},l\rm\mathchar`-loop}
+\,{\rm k2vac}_{\rm2\Tr,cdiag}^{{\rm LL},s\rm\mathchar`-loop}
-\,{\rm k2vac}_{\rm1\Tr,cdiag}^{{\rm LL},s\rm\mathchar`-loop}
,\\
\img\, C_{4}^{K\to|0\rangle}
=\,& 2\,{\rm k2vac}_{\rm2\Tr,cmix}^{{\rm LL},l\rm\mathchar`-loop}
-\,{\rm k2vac}_{\rm1\Tr,cmix}^{{\rm LL},l\rm\mathchar`-loop}
+\,{\rm k2vac}_{\rm2\Tr,cmix}^{{\rm LL},s\rm\mathchar`-loop}
-\,{\rm k2vac}_{\rm1\Tr,cmix}^{{\rm LL},s\rm\mathchar`-loop}
,\\
\img\, C_{5}^{K\to|0\rangle}
=\,& 2\,{\rm k2vac}_{\rm2\Tr,cdiag}^{{\rm LR},l\rm\mathchar`-loop}
-\,{\rm k2vac}_{\rm1\Tr,cdiag}^{{\rm LR},l\rm\mathchar`-loop}
+\,{\rm k2vac}_{\rm2\Tr,cdiag}^{{\rm LR},s\rm\mathchar`-loop}
-\,{\rm k2vac}_{\rm1\Tr,cdiag}^{{\rm LR},s\rm\mathchar`-loop}
,\\
\img\, C_{6}^{K\to|0\rangle}
=\,& 2\,{\rm k2vac}_{\rm2\Tr,cmix}^{{\rm LR},l\rm\mathchar`-loop}
-\,{\rm k2vac}_{\rm1\Tr,cmix}^{{\rm LR},l\rm\mathchar`-loop}
+\,{\rm k2vac}_{\rm2\Tr,cmix}^{{\rm LR},s\rm\mathchar`-loop}
-\,{\rm k2vac}_{\rm1\Tr,cmix}^{{\rm LR},s\rm\mathchar`-loop}
,\\
\img\, C_{7}^{K\to|0\rangle}
=\,& \frac{1}{2}\,{\rm k2vac}_{\rm2\Tr,cdiag}^{{\rm LR},l\rm\mathchar`-loop}
+\frac{1}{2}\,{\rm k2vac}_{\rm1\Tr,cdiag}^{{\rm LR},l\rm\mathchar`-loop}
-\frac{1}{2}\,{\rm k2vac}_{\rm2\Tr,cdiag}^{{\rm LR},s\rm\mathchar`-loop}
+\frac{1}{2}\,{\rm k2vac}_{\rm1\Tr,cdiag}^{{\rm LR},s\rm\mathchar`-loop}
,\\
\img\, C_{8}^{K\to|0\rangle}
=\,& \frac{1}{2}\,{\rm k2vac}_{\rm2\Tr,cmix}^{{\rm LR},l\rm\mathchar`-loop}
+\frac{1}{2}\,{\rm k2vac}_{\rm1\Tr,cmix}^{{\rm LR},l\rm\mathchar`-loop}
-\frac{1}{2}\,{\rm k2vac}_{\rm2\Tr,cmix}^{{\rm LR},s\rm\mathchar`-loop}
+\frac{1}{2}\,{\rm k2vac}_{\rm1\Tr,cmix}^{{\rm LR},s\rm\mathchar`-loop}
,\\
\img\, C_{9}^{K\to|0\rangle}
=\,& \frac{1}{2}\,{\rm k2vac}_{\rm2\Tr,cdiag}^{{\rm LL},l\rm\mathchar`-loop}
+\frac{1}{2}\,{\rm k2vac}_{\rm1\Tr,cdiag}^{{\rm LL},l\rm\mathchar`-loop}
-\frac{1}{2}\,{\rm k2vac}_{\rm2\Tr,cdiag}^{{\rm LL},s\rm\mathchar`-loop}
+\frac{1}{2}\,{\rm k2vac}_{\rm1\Tr,cdiag}^{{\rm LL},s\rm\mathchar`-loop}
,\\
\img\, C_{10}^{K\to|0\rangle}
=\,& \frac{1}{2}\,{\rm k2vac}_{\rm2\Tr,cmix}^{{\rm LL},l\rm\mathchar`-loop}
+\frac{1}{2}\,{\rm k2vac}_{\rm1\Tr,cmix}^{{\rm LL},l\rm\mathchar`-loop}
-\frac{1}{2}\,{\rm k2vac}_{\rm2\Tr,cmix}^{{\rm LL},s\rm\mathchar`-loop}
+\frac{1}{2}\,{\rm k2vac}_{\rm1\Tr,cmix}^{{\rm LL},s\rm\mathchar`-loop}.
\end{align}

\subsection{Contractions for power-divergence subtractions}

We define fundamental contractions as follows:
\begin{align}
{\rm type3}^{\gamma_5}(x_K,x_Q,x_1,x_2)
&= \Tr[
S_l(x_Q,x_1)\gamma_5
S_l(x_1,x_2)\gamma_5
S_l(x_2,x_K)\gamma_5
S_s(x_K,x_Q)\gamma_5
]
\notag\\
&+\Tr[
S_l(x_Q,x_2)\gamma_5
S_l(x_2,x_1)\gamma_5
S_l(x_1,x_K)\gamma_5
S_s(x_K,x_Q)\gamma_5
],
\\
{\rm type3\sigma}^{\gamma_5}(x_K,x_Q,x_\sigma)
&= \Tr[
S_l(x_Q,x_\sigma)
S_l(x_\sigma,x_K)\gamma_5
S_s(x_K,x_Q)\gamma_5
],
\\
{\rm type4}^{\gamma_5}(x_K,x_Q,x_1,x_2)
&= \Tr[
S_l(x_Q,x_K)\gamma_5
S_s(x_K,x_Q)\gamma_5
]\cdot\Tr[
S_l(x_1,x_2)\gamma_5
S_l(x_2,x_1)\gamma_5
],
\\
{\rm type4\sigma}^{\gamma_5}(x_K,x_Q,x_\sigma)
&= \Tr[
S_l(x_Q,x_K)\gamma_5
S_s(x_K,x_Q)\gamma_5
]\cdot\Tr[
S_l(x_\sigma,x_\sigma)
],
\\
{\rm k2vac}^{\gamma_5}(x_K,x_Q)
&= \Tr[
S_l(x_Q,x_K)\gamma_5
S_s(x_K,x_Q)\gamma_5
],
\end{align}
where we omit the color indices since they are all contracted in the same way as the spin indices. 
The three- and two-point functions used for the subtractions read
\begin{align}
-\sqrt3\img\, C_{\gamma_5}^{K\to\pi\pi}(x_K,x_Q,x_1,x_2)
&= \langle\pi\pi^{I=0}(x_1,x_2)\bar s\gamma_5d(x_Q)O^K(x_K)^\dag\rangle
=-\frac{3}{2}\,{\rm type3}^{\gamma_5} +3\,{\rm type4}^{\gamma_5},
\\
\sqrt2\img\, C_{\gamma_5}^{K\to\sigma}(x_K,x_Q,x_\sigma)
&= \langle\sigma(x_\sigma)\bar s\gamma_5d(x_Q)O^K(x_K)^\dag\rangle
=-\,{\rm type3\sigma}^{\gamma_5} +2\,{\rm type4\sigma}^{\gamma_5},
\\
\img\, C_{\gamma_5}^{K\to|0\rangle}(x_K,x_Q)
&= \langle\bar s\gamma_5d(x_Q)O^K(x_K)^\dag\rangle
= - {\rm k2vac}^{\gamma_5}.
\end{align}

\section{Systematic error on the amplitudes due to the interpolation}
\label{sec:syserr_onshell_limit}

In this appendix we describe how we estimate the systematic error due to the interpolation of the matrix elements to the physical kinematics.  Throughout this appendix the matrix elements in the SMOM($\slashchar q,\slashchar q$) scheme under consideration are the ones after a step scaling from $\mu_1$ to $\mu_h$.

\begin{table}[tp]
\centering
\begin{tabular}{ccc}
\hline
$j$ & $I=2$ & $I=0$ \\
\hline
1 & $0.070(10)$ & $0.038(30)$ \\
2 & --- & $0.0203(59)$ \\
3 & --- & $0.0216(67)$ \\
5 & --- & $0.0191(59)$ \\
6 & --- & $0.0220(35)$ \\
7 & $-0.0279(13)$ & $0.0009(11)$ \\
8 & $-0.0199(11)$ & $0.00191(92)$ \\
\hline
\end{tabular}
\caption{Values of $\Delta M_{I,j}^{\prime\,{\rm SMOM}({\scriptsize\slashchar q},{\scriptsize\slashchar q})}(\mu_h)$ divided by the central value of the corresponding $M_{I,j}^{\prime\,{\rm SMOM}({\scriptsize\slashchar q},{\scriptsize\slashchar q}),\rm lin}(\mu_h)$.  Errors in the parentheses are statistical only.}
\label{tab:DeltaMERI}
\end{table}

We define the systematic error on the matrix elements due to the interpolation as the difference
\begin{equation}
\Delta M_{I,j}^{\prime\,{\rm SMOM}({\scriptsize\slashchar q},{\scriptsize\slashchar q})}(\mu_h)
= M_{I,j}^{\prime\,{\rm SMOM}({\scriptsize\slashchar q},{\scriptsize\slashchar q}),{\rm lin}}(\mu_h) - M_{I,j}^{\prime\,{\rm SMOM}({\scriptsize\slashchar q},{\scriptsize\slashchar q}),{\rm quad}}(\mu_h),
\end{equation}
whose ratios to $M_{I,j}^{\prime\,{\rm SMOM}({\scriptsize\slashchar q},{\scriptsize\slashchar q}),\rm lin}(\mu_h)$ are tabulated in Table~\ref{tab:DeltaMERI}.  The table indicates that the systematic error due to the interpolation is well resolved and maximally a few percent except for $j=1$, where the systematic error for $I=2$ is about 7\% and that for $I=0$ is not resolved but fairly less than 10\%.

In order to propagate these systematic errors to the $K\to\pi\pi$ amplitudes, we consider the summation form
\begin{equation}
A_I = \sum_{j\in\cal C} C_j^{\prime\,{\rm SMOM}({\scriptsize \slashchar q},{\scriptsize \slashchar q})}(\mu_h) M_{I,j}^{\prime\,{\rm SMOM}({\scriptsize \slashchar q},{\scriptsize \slashchar q})}(\mu_h),
\end{equation}
where
\begin{equation}
C_j^{\prime\,{\rm SMOM}({\scriptsize \slashchar q},{\scriptsize \slashchar q})}(\mu_h) = \sum_{i=1}^{10}\sum_{k\in\cal C}C_i^{\rm\overline{MS}}(\mu_h)\left(T_{ik}+\Delta T_{ik}^{\rm\overline{MS}}\right)
R_{kj}^{{\rm\overline{MS}\leftarrow SMOM}({\scriptsize \slashchar q},{\scriptsize \slashchar q})}(\mu_h)
\end{equation}
is calculated perturbatively.
We estimate the systematic error on the amplitudes as
\begin{equation}
\Delta A_I =
\sqrt{\sum_{j\in\cal C}\left(C_j^{\prime\,{\rm SMOM}({\scriptsize \slashchar q},{\scriptsize \slashchar q})}(\mu_h)\Delta M_{I,j}^{\prime\,{\rm SMOM}({\scriptsize \slashchar q},{\scriptsize \slashchar q})}(\mu_h)\right)^2}.
\end{equation}
We find
\begin{align}
{\rm Re}(\Delta A_2)/{\rm Re}(\overline A_2) &= 0.071(10),\\
{\rm Im}(\Delta A_2)/{\rm Im}(\overline A_2) &= 0.0328(18),\\
{\rm Re}(\Delta A_0)/{\rm Re}(\overline A_0) &= 0.0184(52),\\
{\rm Im}(\Delta A_0)/{\rm Im}(\overline A_0) &= 0.0254(39),
\end{align}
where $\overline A_I$ stands for the central value of $A_I$.
The errors from `On-shell limit' shown in Table~\ref{tab:syserrAI} are from these values.

\section{Supplemental figures and tables}
\label{sec:supplemental}

\begin{figure}[tbp]
\begin{center}
\begin{tabular}{c}
\includegraphics[width=80mm]{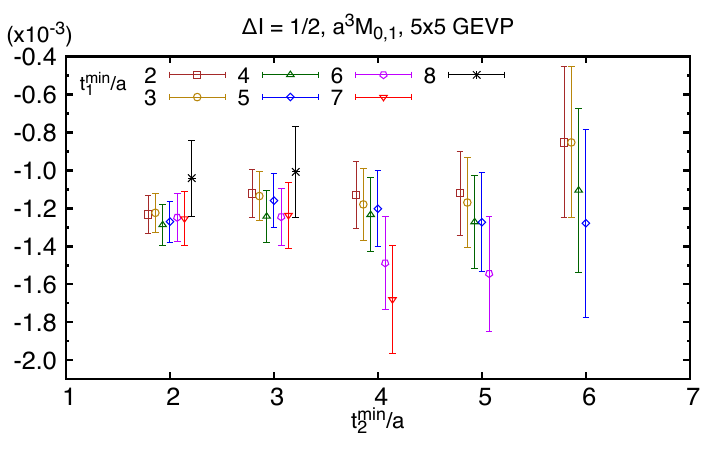}
\end{tabular}
\hfill
\begin{tabular}{c}
\includegraphics[width=80mm]{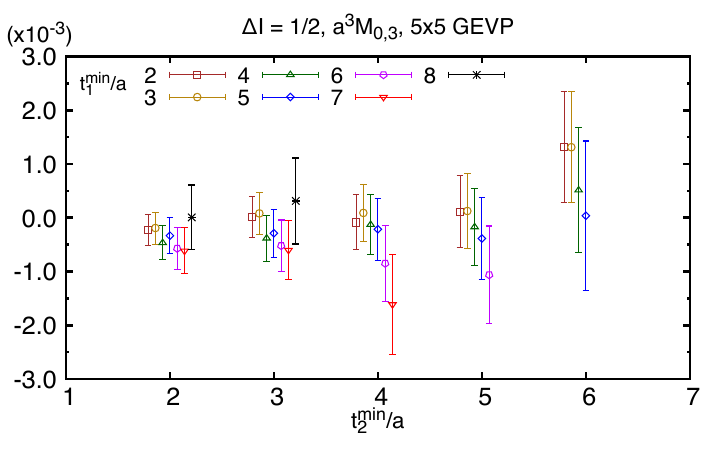}
\end{tabular}
\begin{tabular}{c}
\includegraphics[width=80mm]{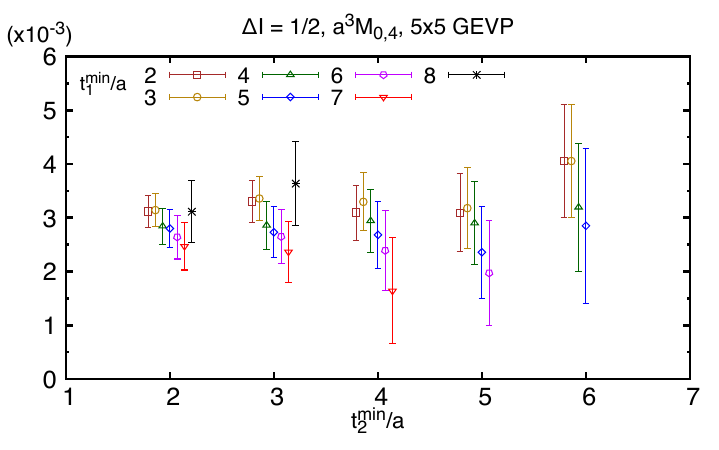}
\end{tabular}
\hfill
\begin{tabular}{c}
\includegraphics[width=80mm]{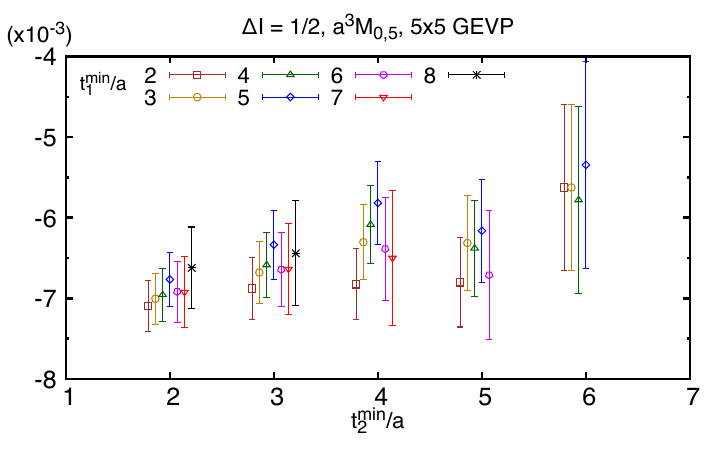}
\end{tabular}
\begin{tabular}{c}
\includegraphics[width=80mm]{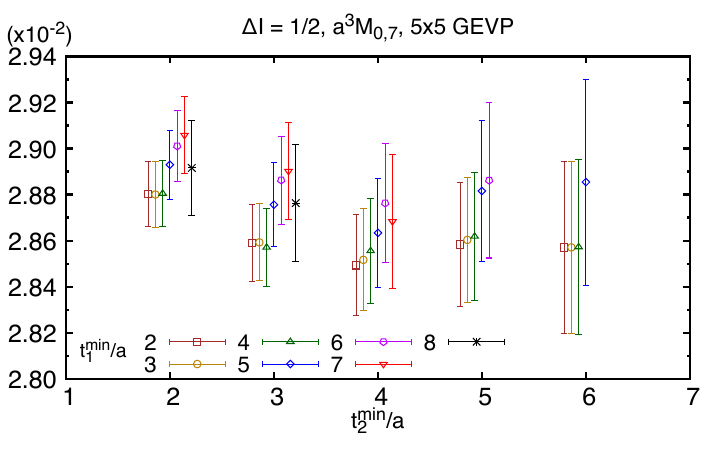}
\end{tabular}
\hfill
\begin{tabular}{c}
\includegraphics[width=80mm]{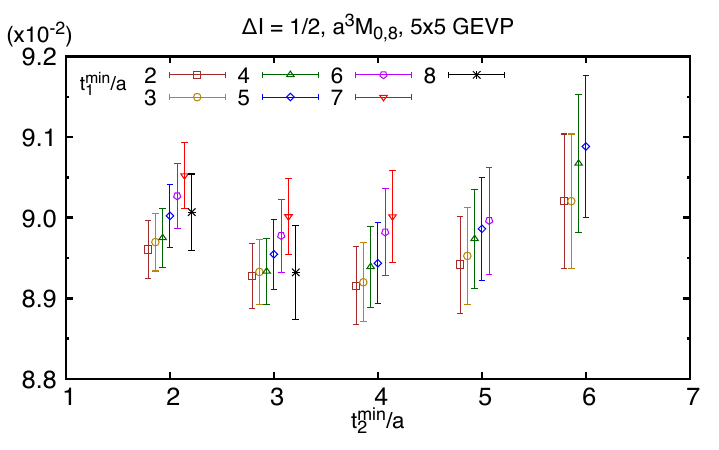}
\end{tabular}
\begin{tabular}{c}
\includegraphics[width=80mm]{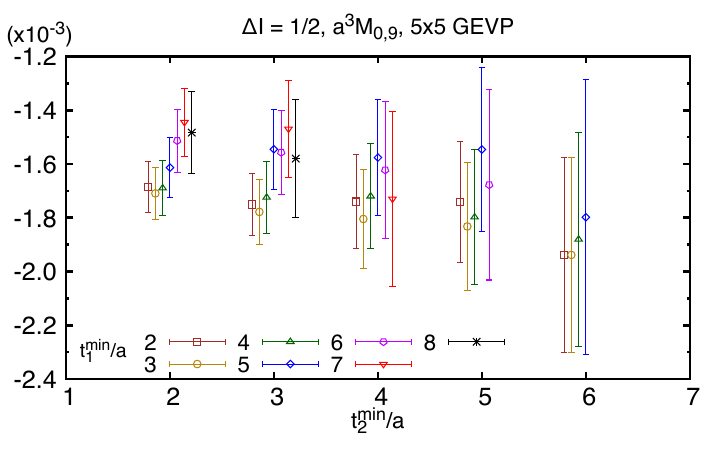}
\end{tabular}
\hfill
\begin{tabular}{c}
\includegraphics[width=80mm]{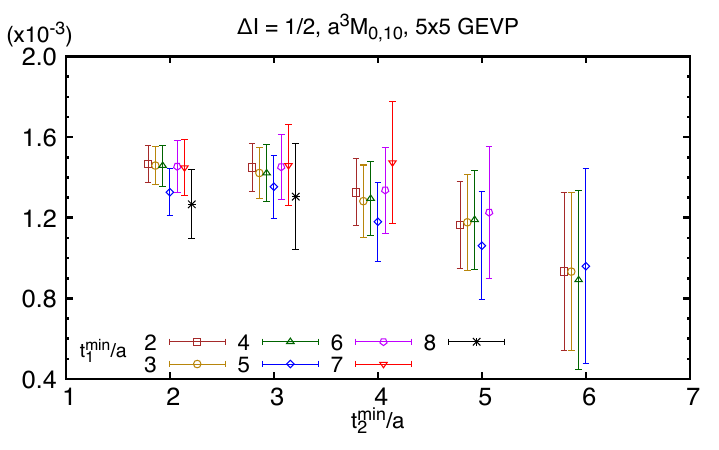}
\end{tabular}
\caption{$\Delta I=1/2$ channel of $K\to\pi\pi$ matrix elements of the four-quark operators $Q_{1,3,4,5,7,8,9,10}$ with the ground two-pion final state.  The result for those of $Q_{2,6}$ are shown on the left panels of Figure~\ref{fig:range_dps_I0_n0}.}
\label{fig:range_dps_I0_n0_suppl}
\end{center}
\end{figure}

\begin{figure}[tbp]
\begin{center}
\begin{tabular}{c}
\includegraphics[width=80mm]{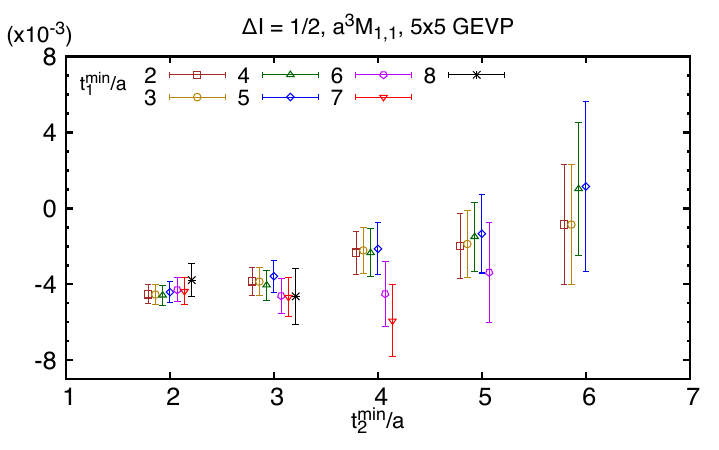}
\end{tabular}
\hfill
\begin{tabular}{c}
\includegraphics[width=80mm]{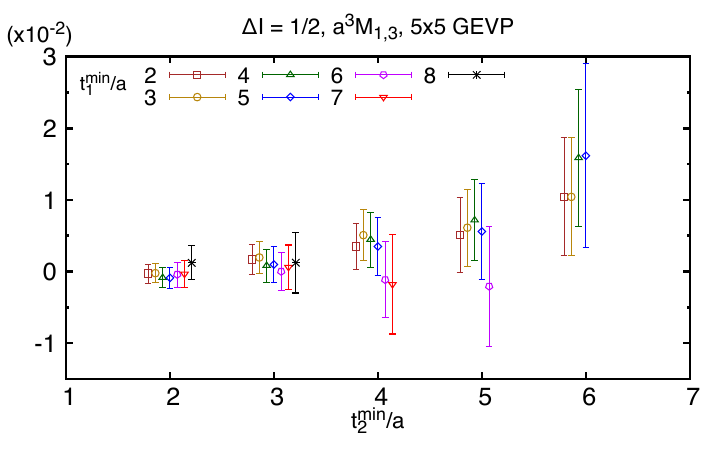}
\end{tabular}
\begin{tabular}{c}
\includegraphics[width=80mm]{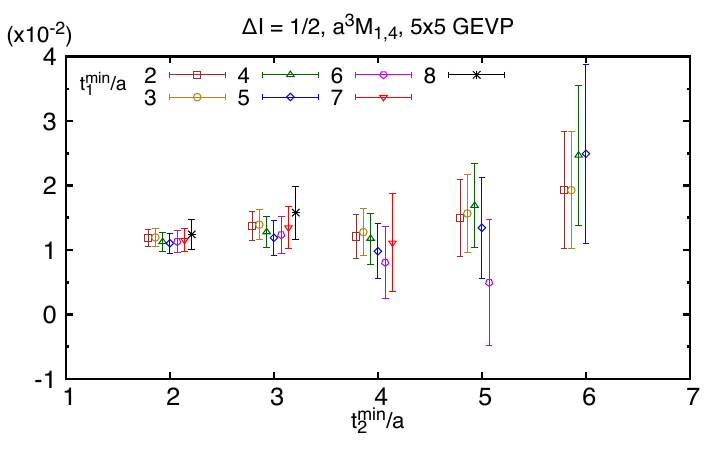}
\end{tabular}
\hfill
\begin{tabular}{c}
\includegraphics[width=80mm]{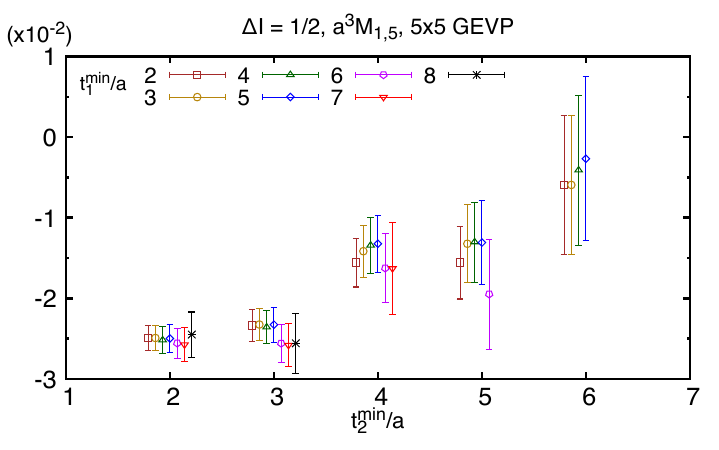}
\end{tabular}
\begin{tabular}{c}
\includegraphics[width=80mm]{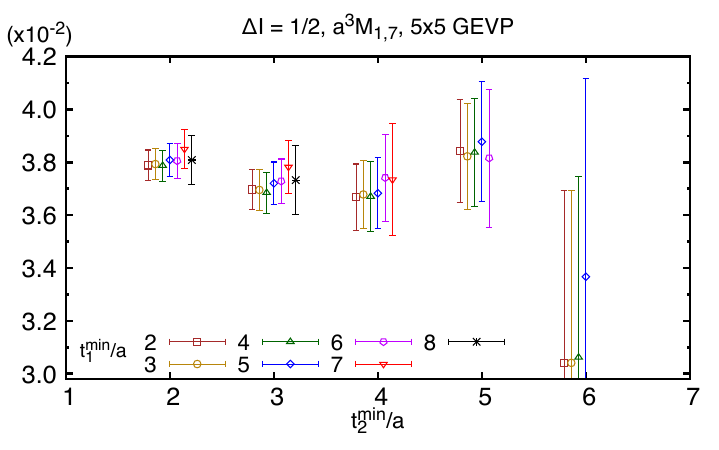}
\end{tabular}
\hfill
\begin{tabular}{c}
\includegraphics[width=80mm]{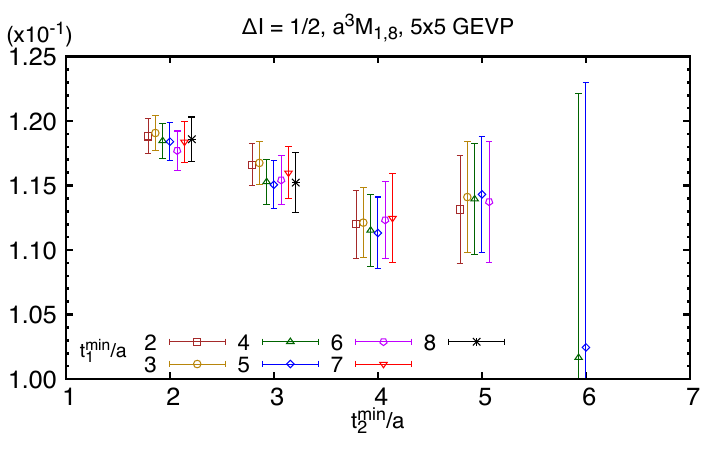}
\end{tabular}
\begin{tabular}{c}
\includegraphics[width=80mm]{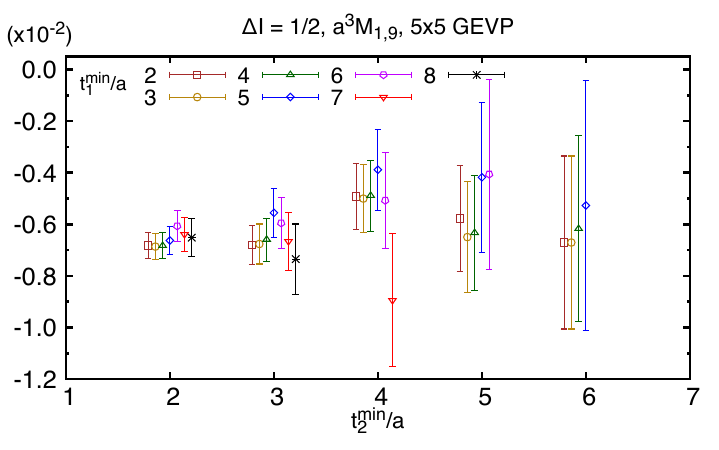}
\end{tabular}
\hfill
\begin{tabular}{c}
\includegraphics[width=80mm]{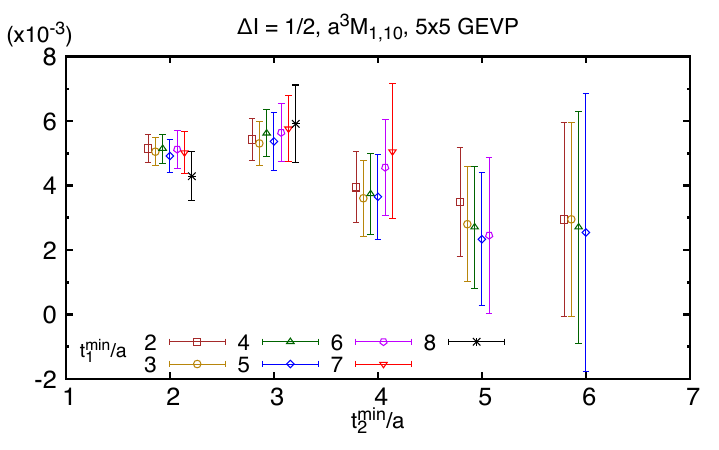}
\end{tabular}
\caption{$\Delta I=1/2$ channel of $K\to\pi\pi$ matrix elements of the four-quark operators $Q_{1,3,4,5,7,8,9,10}$ with the first-excited two-pion final state.  The result for those of $Q_{2,6}$ are shown on the left panels of Figure~\ref{fig:range_dps_I0_n1}.}
\label{fig:range_dps_I0_n1_suppl}
\end{center}
\end{figure}

\begin{table*}[tp]
\centering
\begin{tabular}{|c|ccccccc|}
\hline
 & 1 & 2 & 3 & 5 & 6 & 7 & 8\\
\hline
1 & $0.47461(14)$ & $0$ & $0$ & $0$ & $0$ & $0$ & $0$\\
2 & $0$ & $0.5384(62)$ & $-0.0780(49)$ & $-0.0025(26)$ & $-0.0024(14)$ & $0$ & $0$\\
3 & $0$ & $-0.0634(26)$ & $0.5542(26)$ & $-0.0030(10)$ & $0.00563(75)$ & $0$ & $0$\\
5 & $0$ & $-0.048(20)$ & $-0.043(16)$ & $0.5672(72)$ & $-0.1102(43)$ & $0$ & $0$\\
6 & $0$ & $0.0029(82)$ & $0.0277(80)$ & $-0.0437(31)$ & $0.3854(22)$ & $0$ & $0$\\
7 & $0$ & $0$ & $0$ & $0$ & $0$ & $0.55930(17)$ & $-0.103918(38)$\\
8 & $0$ & $0$ & $0$ & $0$ & $0$ & $-0.036095(23)$ & $0.36730(13)$\\
\hline
\hline
1 & $0.42042(13)$ & $0$ & $0$ & $0$ & $0$ & $0$ & $0$\\
2 & $0$ & $0.4970(78)$ & $-0.0682(70)$ & $0.0034(27)$ & $-0.0061(18)$ & $0$ & $0$\\
3 & $0$ & $-0.0561(34)$ & $0.5080(37)$ & $-0.0026(15)$ & $0.0065(12)$ & $0$ & $0$\\
5 & $0$ & $-0.006(24)$ & $-0.013(21)$ & $0.5188(83)$ & $-0.0941(58)$ & $0$ & $0$\\
6 & $0$ & $0.000(13)$ & $0.040(14)$ & $-0.0358(51)$ & $0.3829(41)$ & $0$ & $0$\\
7 & $0$ & $0$ & $0$ & $0$ & $0$ & $0.49986(16)$ & $-0.087247(35)$\\
8 & $0$ & $0$ & $0$ & $0$ & $0$ & $-0.037058(27)$ & $0.36232(13)$\\
\hline
\hline
1 & $0.56764(21)$ & $0$ & $0$ & $0$ & $0$ & $0$ & $0$\\
2 & $0$ & $0.4542(91)$ & $-0.385(11)$ & $-0.0053(34)$ & $0.0012(23)$ & $0$ & $0$\\
3 & $0$ & $0.164(17)$ & $1.089(18)$ & $0.0064(65)$ & $-0.0105(46)$ & $0$ & $0$\\
5 & $0$ & $-0.067(44)$ & $-0.051(45)$ & $0.632(16)$ & $-0.106(11)$ & $0$ & $0$\\
6 & $0$ & $-0.075(26)$ & $-0.090(34)$ & $-0.078(10)$ & $0.5613(92)$ & $0$ & $0$\\
7 & $0$ & $0$ & $0$ & $0$ & $0$ & $0.62059(22)$ & $-0.108774(48)$\\
8 & $0$ & $0$ & $0$ & $0$ & $0$ & $-0.093300(83)$ & $0.56162(28)$\\
\hline
\end{tabular}
\caption{Renormalization matrices with the chiral basis on the $24^3$ ensemble for SMOM$(\gamma_\mu,\gamma_\mu)$ at $\mu=\mu_1$ (upper), SMOM$(\gamma_\mu,\gamma_\mu)$ at $\mu=\mu_2$ (middle) and SMOM$(\slashchar{q},\slashchar{q})$ at $\mu=\mu_2$ (lower).  The result for SMOM$(\slashchar{q},\slashchar{q})$ at $\mu=\mu_1$ is shown in Table~\ref{tab:NPR_1p3GeV_qslash_qslash_odd.bs1.L24}.}
\label{tab:NPR_rest.L24}
\end{table*}

\begin{table*}[tp]
\centering
\begin{tabular}{|c|ccccccc|}
\hline
 & 1 & 2 & 3 & 5 & 6 & 7 & 8\\
\hline
1 & $0.94625(42)$ & $0$ & $0$ & $0$ & $0$ & $0$ & $0$\\
2 & $0$ & $0.955(93)$ & $-0.262(85)$ & $0.021(33)$ & $0.051(26)$ & $0$ & $0$\\
3 & $0$ & $-0.121(35)$ & $1.133(34)$ & $0.005(11)$ & $-0.003(12)$ & $0$ & $0$\\
5 & $0$ & $-0.15(27)$ & $-0.17(25)$ & $0.959(94)$ & $0.449(80)$ & $0$ & $0$\\
6 & $0$ & $0.11(14)$ & $0.07(14)$ & $0.007(48)$ & $2.224(48)$ & $0$ & $0$\\
7 & $0$ & $0$ & $0$ & $0$ & $0$ & $0.97575(56)$ & $0.3671(13)$\\
8 & $0$ & $0$ & $0$ & $0$ & $0$ & $0.0165(11)$ & $2.3154(52)$\\
\hline
\hline
1 & $0.95422(29)$ & $0$ & $0$ & $0$ & $0$ & $0$ & $0$\\
2 & $0$ & $0.985(49)$ & $-0.157(42)$ & $-0.022(19)$ & $0.023(15)$ & $0$ & $0$\\
3 & $0$ & $-0.106(20)$ & $1.065(19)$ & $-0.0131(70)$ & $-0.0062(65)$ & $0$ & $0$\\
5 & $0$ & $-0.10(15)$ & $-0.13(12)$ & $0.864(56)$ & $0.245(47)$ & $0$ & $0$\\
6 & $0$ & $0.001(76)$ & $-0.058(71)$ & $-0.032(28)$ & $1.744(24)$ & $0$ & $0$\\
7 & $0$ & $0$ & $0$ & $0$ & $0$ & $0.97529(28)$ & $0.23528(65)$\\
8 & $0$ & $0$ & $0$ & $0$ & $0$ & $0.00920(46)$ & $1.8345(26)$\\
\hline
\hline
1 & $0.90212(23)$ & $0$ & $0$ & $0$ & $0$ & $0$ & $0$\\
2 & $0$ & $1.115(77)$ & $0.100(62)$ & $-0.002(24)$ & $0.023(17)$ & $0$ & $0$\\
3 & $0$ & $-0.488(49)$ & $0.534(42)$ & $-0.027(16)$ & $0.009(10)$ & $0$ & $0$\\
5 & $0$ & $-0.19(28)$ & $-0.20(22)$ & $0.847(82)$ & $0.243(58)$ & $0$ & $0$\\
6 & $0$ & $0.09(12)$ & $0.09(10)$ & $-0.027(40)$ & $1.603(33)$ & $0$ & $0$\\
7 & $0$ & $0$ & $0$ & $0$ & $0$ & $0.97601(23)$ & $0.18587(60)$\\
8 & $0$ & $0$ & $0$ & $0$ & $0$ & $0.07499(83)$ & $1.6123(32)$\\
\hline
\end{tabular}
\caption{Step scaling matrices with the chiral basis on the 32Ifine ensemble from the low scale $\mu$ to the high scale $\mu_h\approx4$~GeV.  Results for SMOM$(\gamma_\mu,\gamma_\mu)$ at $\mu=\mu_1$ (upper), SMOM$(\gamma_\mu,\gamma_\mu)$ at $\mu=\mu_2$ (middle) and SMOM$(\slashchar{q},\slashchar{q})$ at $\mu=\mu_2$ (lower) are shown.  The result for SMOM$(\slashchar{q},\slashchar{q})$ at $\mu=\mu_1$ is shown in Table~\ref{tab:NPR_1p3GeV_ss_qslash_qslash_odd.bs1}.}
\label{tab:ss_rest}
\end{table*}

\begin{table*}[tp]
\centering
\begin{tabular}{|c|ccccccc|}
\hline
 & 1 & 2 & 3 & 5 & 6 & 7 & 8\\
\hline
1 & $0.44910(24)$ & $0$ & $0$ & $0$ & $0$ & $0$ & $0$\\
2 & $0$ & $0.530(46)$ & $-0.219(41)$ & $0.008(19)$ & $0.014(12)$ & $0$ & $0$\\
3 & $0$ & $-0.137(17)$ & $0.637(17)$ & $-0.0001(69)$ & $0.0049(55)$ & $0$ & $0$\\
5 & $0$ & $-0.11(13)$ & $-0.11(12)$ & $0.525(55)$ & $0.067(37)$ & $0$ & $0$\\
6 & $0$ & $0.063(72)$ & $0.089(71)$ & $-0.094(29)$ & $0.856(22)$ & $0$ & $0$\\
7 & $0$ & $0$ & $0$ & $0$ & $0$ & $0.53248(35)$ & $0.03344(49)$\\
8 & $0$ & $0$ & $0$ & $0$ & $0$ & $-0.07433(66)$ & $0.8487(20)$\\
\hline
\hline
1 & $0.40117(17)$ & $0$ & $0$ & $0$ & $0$ & $0$ & $0$\\
2 & $0$ & $0.498(24)$ & $-0.146(20)$ & $-0.008(11)$ & $0.0039(72)$ & $0$ & $0$\\
3 & $0$ & $-0.1121(96)$ & $0.5480(93)$ & $-0.0097(40)$ & $0.0065(31)$ & $0$ & $0$\\
5 & $0$ & $-0.048(72)$ & $-0.061(58)$ & $0.439(31)$ & $0.012(22)$ & $0$ & $0$\\
6 & $0$ & $0.005(41)$ & $0.040(40)$ & $-0.079(17)$ & $0.671(13)$ & $0$ & $0$\\
7 & $0$ & $0$ & $0$ & $0$ & $0$ & $0.47879(21)$ & $0.00016(24)$\\
8 & $0$ & $0$ & $0$ & $0$ & $0$ & $-0.06338(28)$ & $0.66386(98)$\\
\hline
\hline
1 & $0.51209(23)$ & $0$ & $0$ & $0$ & $0$ & $0$ & $0$\\
2 & $0$ & $0.521(45)$ & $-0.322(45)$ & $-0.009(16)$ & $0.013(11)$ & $0$ & $0$\\
3 & $0$ & $-0.133(29)$ & $0.770(31)$ & $-0.011(11)$ & $0.0016(70)$ & $0$ & $0$\\
5 & $0$ & $-0.19(16)$ & $-0.21(15)$ & $0.516(54)$ & $0.048(37)$ & $0$ & $0$\\
6 & $0$ & $-0.065(82)$ & $-0.082(93)$ & $-0.142(31)$ & $0.902(25)$ & $0$ & $0$\\
7 & $0$ & $0$ & $0$ & $0$ & $0$ & $0.58836(26)$ & $-0.00177(35)$\\
8 & $0$ & $0$ & $0$ & $0$ & $0$ & $-0.10389(73)$ & $0.8973(19)$\\
\hline
\end{tabular}
\caption{Renormalization matrices with step scaling performed from the scale $\mu$ to $\mu_h\approx4$~GeV.  Results for SMOM$(\gamma_\mu,\gamma_\mu)$ at $\mu=\mu_1$ (upper), SMOM$(\gamma_\mu,\gamma_\mu)$ at $\mu=\mu_2$ (middle) and SMOM$(\slashchar{q},\slashchar{q})$ at $\mu=\mu_2$ (lower) are shown.  The result for SMOM$(\slashchar{q},\slashchar{q})$ at $\mu=\mu_1$ is shown in Table~\ref{tab:NPR_1p3GeV_qslash_qslash_odd_4GeV.bs1.L24}.}
\label{tab:ssZ_rest}
\end{table*}

\bibliographystyle{unsrt}
\bibliography{main.bib}
\end{document}